\documentclass[aps,pre,twocolumn,showpacs,superscriptaddress,floatfix]{revtex4-1}
\usepackage{graphicx}
\usepackage{amssymb}
\usepackage{color}
\usepackage{amsmath}
\usepackage{ulem}
\usepackage{mathrsfs}

\newcommand{\up}{\uparrow}

\newcommand{\be}{\begin{equation}}
\newcommand{\ee}{\end{equation}}

\begin{document}

\title{Random-Singlet Phase in Disordered Two-Dimensional Quantum Magnets} 
  
\author{Lu Liu} 
\affiliation{Department of Physics, Beijing Normal University, Beijing 100875, China}

\author{Hui Shao} 
\affiliation{Beijing Computational Science Research Center, Beijing 100193, China}
\affiliation{Department of Physics, Boston University, 590 Commonwealth Avenue, Boston, Massachusetts 02215, USA}

\author{Yu-Cheng Lin}
\email{yc.lin@nccu.edu.tw}
\affiliation{Graduate Institute of Applied Physics, National Chengchi University, Taipei, Taiwan}

\author{Wenan Guo} 
\email{waguo@bnu.edu.cn}
\affiliation{Department of Physics, Beijing Normal University, Beijing 100875, China}
\affiliation{Beijing Computational Science Research Center, Beijing 100193, China}

\author{Anders W. Sandvik}
\email{sandvik@bu.edu}
\affiliation{Beijing National Laboratory for Condensed Matter Physics and Institute of Physics, Chinese Academy of Sciences, Beijing 100190, China}
\affiliation{Department of Physics, Boston University, 590 Commonwealth Avenue, Boston, Massachusetts 02215, USA}
\affiliation{Beijing Computational Science Research Center, Beijing 100193, China}

\begin{abstract}
  We study effects of disorder (quenched randomness) in a two-dimensional square-lattice $S=1/2$ quantum spin system, the $J$-$Q$ model with a multi-spin
  interaction $Q$ supplementing the Heisenberg exchange $J$. In the absence of disorder the system hosts antiferromagnetic (AFM) and columnar
  valence-bond-solid (VBS) ground states. The VBS breaks $Z_4$ symmetry spontaneously, and in the presence of arbitrarily weak disorder
  it forms domains. Using quantum Monte Carlo simulations, we demonstrate two different kinds
  of such disordered VBS states. Upon dilution, a removed site in one sublattice forces a left-over localized spin in the opposite sublattice.
  These spins interact through the host system and always form AFM order. In the case of random $J$ or $Q$ interactions in the intact lattice,
  we find a different, spin-liquid-like state with no magnetic or VBS order but with algebraically decaying mean correlations. Here we identify
  localized spinons at the nexus of domain walls separating regions with the four different VBS patterns. These spinons form correlated groups
  with the same number of spinons and antispinons. Within such a group, we argue that there is a strong tendency to singlet formation, because of
  the native pairing and relatively strong spinon-spinon interactions mediated by the domain walls. Thus, the spinon groups are effectively
  isolated from each other and
  no long-range AFM order forms. The mean spin correlations decay as $r^{-2}$ as a function of distance $r$. We propose that this state
  is a two-dimensional analogue of the well-known random singlet (RS) state in one dimension, though, in contrast to the latter, the
  dynamic exponent $z$ here is finite. By studying quantum-critical scaling of the magnetic susceptibility, we find that $z$ varies, taking
  the value $z=2$ at the AFM--RS phase boundary and growing upon moving into the RS phase (thus causing a power-law
  divergent susceptibility). The RS state discovered here in a system without geometric frustration may correspond to the same fixed point
  as the RS state recently proposed for frustrated systems, and the ability to study it without Monte Carlo sign problems opens up opportunities
  for further detailed characterization of its static and dynamic properties. We also discuss experimental evidence of the RS phase in the
  quasi-two-dimensional square-lattice random-exchange quantum magnets Sr$_2$CuTe$_{1-x}$W$_x$O$_6$ for $x$ in the range $0.2-0.5$. 
\end{abstract}

\date{\today}

\maketitle

\section{Introduction}
\label{sec:intro}

In the quest to classify and characterize ground states and excitations of quantum many-body systems, 
disorder (quenched randomness) plays a central role. Beyond the fundamental scientific interest in understanding the interplay 
between quantum fluctuations and intrinsic randomness, there are also potential practical implications: In the same way 
as pure crystalline states of matter are often not optimal for achieving desired properties of materials, e.g., in the case
of metals hardened by limiting the size of crystal grains, it is likely that quantum technologies will emerge that exploit
disorder effects. For example, random spin chains have been proposed as key elements for
memories \cite{nandkishore15,smith16} and state transfer channels \cite{yao11} in quantum computing. Two-dimensional (2D)
quantum spin systems, which we consider here, is another natural setting for exploring novel disorder-induced states.

Recent experimental efforts have been devoted to searches for quantum spin liquids in quasi-2D insulators. Several
candidate systems showing the qualitative signatures of spin liquids have been identified, e.g., in a series of
organic salts where the spins reside on triangular lattices \cite{shimizu03,yamashita08,manna10,pratt11} and in the
kagome-lattice herbertsmithite \cite{lee07,vries09,helton10,han12}. It has so far not been possible to unambiguously match the
properties of these systems to theoretically proposed spin liquids, however, and it has been proposed that disorder
effects are crucial for understanding the observed behaviors \cite{singh10}. In a more extreme interpretation put forward
recently \cite{watanabe14,kawamura14,shimokawa,uematsu17,kimchi17,kimchi18,wu18}, disorder is even responsible for realizing a certain
spin liquid, the random singlet (RS) state, in some triangular, kagome, and frustrated honeycomb lattice systems, e.g., the triangular
YbMgGaO$_4$ \cite{li15,li17} where disorder is present in the form of random occupation of Mg and Ga ions on equivalent lattice
sites between the magnetic layers. While such a state has not yet been observed in systems without geometric frustration, there
is recent experimental evidence for an RS state in a square lattice system; the double perovskite Sr$_2$CuTe$_{1-x}$W$_{x}$O$_6$.
Here the disorder is in the form of random Te$\leftrightarrow$W substitutions relative to the isostructural compounds Sr$_2$CuTeO$_6$
and Sr$_2$CuWO$_6$, which have dominant nearest- and next-nearest-neighbor spin interactions, respectively
\cite{mustonen18a,mustonen18b,watanabe18}.

We will here show that disorder can induce a spin-liquid-like state---a gapless state with algebraic correlation functions---in a
2D quantum spin system on the square lattice even without geometric frustration. We will refer to this state as an RS state, for reasons
to be discussed further in Sec.~\ref{sec:background}. To demonstrate the existence of the RS state and to characterize its properties,
we carry our large-scale quantum Monte Carlo (QMC) simulations of an $S=1/2$ quantum spin model, the $J$-$Q$ model, which in the absence
of disorder hosts both a N\'eel antiferromagnetic (AFM) and a spontaneously singlet-dimerized valence-bond solid (VBS)
ground state. The transition between these states is driven by enhancing
the formation of correlated singlets by increasing the multi-spin (here six-spin) interaction $Q$, which competes with the Heisenberg exchange $J$.
We show that randomness in the coupling constants leads to the formation of domains in the four-fold degenerate VBS state, with different
realizations of the bond order and with domain walls of the type expected \cite{levin04} to lead to localized spinons at each nexus of four
domain walls. These spinons form in correlated groups of even numbers, as a consequence of the domain-wall topology. We will show evidence
for domain-wall mediated enhanced spinon-spinon interactions, which leads to singlet formation within the groups and no residual AFM ordering
of the spinons. As a contrast, we also consider a site-diluted system, in which the remnant local moments associated with removed sites are not
spatially strongly correlated; thus residual AFM order forms and there is no RS phase.

We will present a broad survey of the phase diagrams, quantum phase transitions, and basic ground state and temperature $T>0$ properties of the
2D RS phase in different versions of the random $J$-$Q$ model. The spin and bond correlations at $T=0$ decay as power laws, likely as a consequence
of rare events in the form of singlet formation over large distances. At $T>0$, using lattices sufficiently large to reach the thermodynamic
limit, we find power-law scaling in $T$ of the magnetic susceptibility. This behavior allows us to extract the dynamic exponent $z$, which
we find is varying inside the RS phase.

It is possible that the RS state we identify here is the same one, in the sense of renormalization group (RG) fixed points, as the one proposed recently
to arise out of a VBS on the triangular lattice in the presence of random couplings \cite{kimchi17}. It may then also be a realization of the unusual
magnetic states observed in YbMgGaO$_4$ and Sr$_2$CuTe$_{1-x}$W$_{x}$O$_6$, and possibly in many other disordered spin liquid candidates as well. The
possibility of creating this state with a ``designer Hamiltonian'' within the $J$-$Q$ family of models is very significant, as this unfrustrated (in
the geometric sense) system is amenable to large-scale QMC studies without the ``sign problems'' plaguing simulations of models with frustration.
Thus the RS state in these systems can be characterized essentially completely---far beyond the analytical calculations in Ref.~\onlinecite{kimchi17} and
the exact diagonalization (ED) numerics on small frustrated Heisenberg lattices in Refs.~\onlinecite{watanabe14,kawamura14,shimokawa,uematsu17}, and on slightly
larger triangular lattices by density-matrix renormalization group (DMRG) calculations in Ref.~\onlinecite{wu18}. In particular, we are able to reliably
study the AFM--RS quantum phase transition.

The paper is organized as follows: In Sec.~\ref{sec:background} we discuss the broader context of our work and provide specifics of the models
considered. In addition to the main focus on different kinds of disorder in the $J$-$Q$ model, we will also discuss a simpler case as a point
of reference: the statically columnar-dimerized Heisenberg model in which localized moments different from the VBS spinons form in the neighborhood
of removed sites. In Sec.~\ref{sec:spinons}, in order to aid in the presentation and interpretation of the extensive QMC results in the later sections, 
we discuss qualitatively the phenomena and mechanisms we have identified as responsible for the RS state, specifically the pairing of localized spinons
and the role of VBS domain walls in mediating effective magnetic interactions. In Sec.~\ref{sec:ground} we present results of ground-state projector
QMC calculations of static properties of all the models considered, with the main focus on the order parameters and correlation
functions in the RS phase in the cases where this state is attained. We demonstrate the existence of a universal continuous AFM--RS quantum phase
transition. In Sec.~\ref{sec:tfinite} we discuss susceptibility results at $T>0$ which allow us to extract the dynamic exponent at the AFM--RS transition and
in the RS phase. In Sec.~\ref{sec:jeff} we provide evidence for the mechanism underlying the formation of the RS state; spinon interactions mediated by
VBS domain walls. We conclude in Sec.~\ref{sec:discussion} with a brief summary and further discussion of our results and their significance
in the context of both theory and experiments. We re-analyze the recent susceptibility data for Sr$_2$CuTe$_{1-x}$W$_{x}$O$_6$ with $x$ in the range
$0.2-0.5$ \cite{watanabe18} and demonstrate that the divergence at low $T$ is slower than $1/T$, consistent with what we found for the RS phase.

\section{Background and Models}
\label{sec:background}

\subsection{Infinite-randomness fixed points and the random singlet phase}

Theoretically, when randomness is a relevant perturbation under RG transformations, fixed points 
corresponding to ground state phases and critical points appear beyond those realized in pure, translationally-invariant 
systems \cite{vojta10,vojta13}. In some cases the RG flow converges to non-zero but finite disorder, e.g., at critical
points in many quantum spin glasses \cite{read95,guo94,guo96,rieger96}, boson systems with random potentials \cite{fisher89}
or random hopping \cite{iyer12}, and Heisenberg antiferromagnets \cite{melin00,lin03,lin06}. However, the randomness can also increase
without bounds in the RG flow, leading to an infinite-randomness fixed point (IRFP). This broad class of fixed points has been
extensively studied using strong-disorder RG (SDRG) methods in quantum systems in one 
\cite{dasgupta80,fisher94,melin02,refael02,refael02,refael04,hoyosl08,pielawa13,shu16}
and higher dimensions \cite{bhatt82,pich98,motrunich00,lin06b,lin07} (in addition to many applications in classical statistical physics \cite{igloi05}). 
The most striking general property of the IRFPs is an infinite dynamic exponent $z$, i.e., the scaling 
relationship between energy ($\epsilon$) and length ($l$) scales is exponential instead of the conventional power-law 
relation $\epsilon \sim l^{-z}$. Moreover, rare instances of long-distance entangled spins (or particles) lead to
different behaviors of the mean and typical correlation functions versus distance $r$, decaying, respectively, 
as a power law and exponentially.

An important example of an IRFP is the 1D RS phase realized in the antiferromagnetic Heisenberg chain with random couplings
\cite{dasgupta80,fisher94,hoyosl08,shu16}. Here the SDRG procedure gives the ground state as a single ``frozen'' configuration of
valence bonds (singlet spin pairs), with a characteristic bond-length distribution. The long long-distance entangled spins (long valence bonds)
lead to the mean spin correlations decaying as $r^{-2}$ \cite{fisher94,shu16} (while the typical correlations decay exponentially) and the entanglement
entropy diverging logarithmically with the system size \cite{refael04}. 

IRFPs have been identified also in 2D systems, primarily in transverse-field Ising models \cite{motrunich00,lin06b,lin07}
but also in experiments on the superconductor--metal transition in Ga thin films \cite{xing15}. However, no convincing case of 
such a phase or critical point has been reported in 2D quantum magnets with SU(2) spin-isotropic interactions, such as the standard
Heisenberg exchange, as far as we are aware. If an RS state exists in such systems, one would expect it to have algebraically
decaying mean correlation functions, as in the 1D case. We are not aware of any strict definition of an RS state in 2D and we here
simply use this term for a non-uniform singlet state without any long-range order, but with power-law decaying correlation functions.
Such a state should roughly correspond to a product of frozen singlets pairs as in the 1D case, perhaps with some other distribution
of valence-bond lengths and non-trivial spatial bond correlations.

If the 2D RS state also corresponds to an IRFP, the dynamic exponent should presumably be infinite as well.
However, an RS state can also in principle exist which has finite $z$,
although such a state corresponding to an RG fixed point at finite disorder strength does not exist in random Heisenberg chains. Finite-disorder
fixed points have been obtained in SDRG calculations on the 2D Heisenberg model with various types of disorder \cite{melin00,lin03},
but it is not clear whether the SDRG method, by its construction and underlying assumptions, produces the correct fixed point when
it does not flow to infinite disorder strength.

As mentioned in the Introduction, Sec.~\ref{sec:intro}, there are some experimental indications of 2D disorder-induced spin liquids
with finite $z$ in frustrated quantum magnets, according to interpretations supported by numerical studies of the $S=1/2$ Heisenberg
antiferromagnet with random couplings on the triangular and kagome lattices \cite{watanabe14,kawamura14,shimokawa,wu18}, and also on
the honeycomb lattice with frustrated interactions \cite{uematsu17}. These may very well be realizations of an RS
state, as proposed. However, a full characterization of the putative RS ground state and its low-temperature thermodynamic properties
(i.e., the form of the asymptotic long-distance correlations and the value of the dynamic exponent) was not possible, because of
the limited lattice sizes accessible to ED \cite{watanabe14,kawamura14,shimokawa,uematsu17} and DMRG \cite{wu18} calculations.
The recently developed theory of the RS state arising out of a VBS on the frustrated triangular lattice \cite{kimchi17} contains
ingredients---VBS domains and localized spinons---that were not discussed in the context of the numerical works. 

Here we consider a class of $S=1/2$ quantum spin models on the 2D square lattice, with no geometric frustration but with interactions leading to
weakened AFM order or nonmagnetic VBS states on uniform lattices. In systems with random couplings, the dynamic exponent is finite and varying
throughout the RS phase, which is a clear indication of a class of finite disorder RG fixed points. Our results suggest a mechanism of pairing of
localized spinons, which leads to the RS state instead of a weakly ordered AFM state (which had been regarded as the most likely state forming in
the random VBS in the absence of frustrated interactions \cite{kimchi17}). Importantly, this RS state in an unfrustrated, bipartite system can be
induced also in cases where the pure host system is not yet in the VBS state (though not in the standard Heisenberg model with random
nearest-neighbor couplings \cite{laflorencie06}), because local VBS domains are still created in response to the disorder. This observation,
along with other considerations, suggests a possible universal scenario that connects our square-lattice RS state directly to the above mentioned
works on frustrated models with various host states \cite{watanabe14,kawamura14,shimokawa,uematsu17,wu18,kimchi17}. To definitely confirm this
universality would require more detailed work on the frustrated systems, however, since the frustrated systems have not been characterized
to the extent that we are able to do here for the random $J$-$Q$ model.

\subsection{Random singlet state in the 2D J-Q model}
\label{sec:randomjq}

\begin{figure}[t]
\includegraphics[width=80mm, clip]{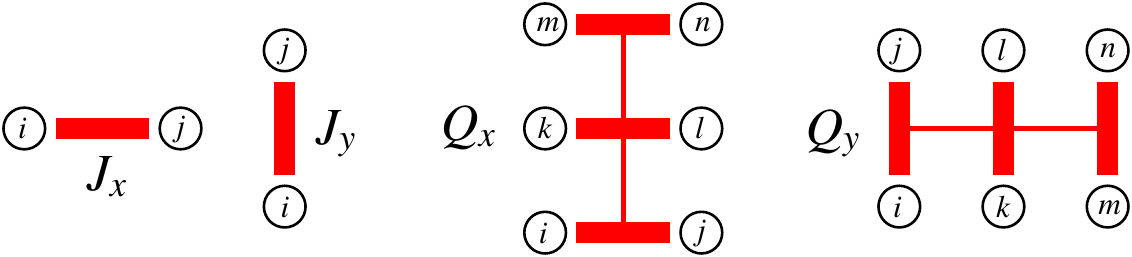}
\vskip-2mm
\caption{Illustration of the terms of the $J$-$Q$ model used in this work. The circles are sites on the square lattice, labeled
in accordance with the Hamiltonian, Eq.~(\ref{jqham}). The red bars connecting two sites are the singlet projectors, and connected
bars in the $Q$ terms indicate products.}
\label{terms}
\end{figure}

We will study a square-lattice Heisenberg antiferromagnet with nearest-neighbor exchange $J$ augmented with certain multi-spin 
interactions of strength $Q$ (the $J$-$Q$ model). The unadulterated translationally invariant model is defined by the 
Hamiltonian \cite{lou09,sandvik12}
\begin{equation}
\label{jqham}
H = -J \sum \limits_{\langle ij\rangle} P_{ij} - 
Q \hskip-3mm \sum \limits_{\langle ijklmn\rangle} \hskip-2.5mm P_{ij}P_{kl}P_{mn}, 
\end{equation}
where $P_{ij}$ is the singlet projector for two $S=1/2$ spins,
\begin{equation}
P_{ij} = \frac{1}{4} - {\bf S}_i \cdot {\bf S}_j.
\end{equation}
In the sums in Eq.~(\ref{jqham}),
${\langle ij\rangle}$ indicates nearest-neighbor sites, and the index pairs $ij$, $kl$, and $mn$ in ${\langle ijklmn\rangle}$ are
neighbors forming a horizontal or vertical column, as illustrated in Fig.~\ref{terms}. The summations are over all pairs and
columns, so that the Hamiltonian respects all the symmetries of the square lattice, including the 90$^\circ$ rotation symmetry
when $J_x=J_y=J$ and $Q_x=Q_y=Q$ as we have assumed in Eq.~(\ref{jqham}); we will not consider the more general cases
$J_x \not= J_y$ or $Q_x \not= Q_y$ here. We will introduce various forms of disorder in the model,
including site dilution and random $J$ and $Q$ couplings drawn from suitable distributions; detailed definitions of the
different cases are presented in Sec.~\ref{sec:ground}.

In the uniform system the $Q$ interactions compete against the exchange terms $J$, disfavoring the strong AFM order
present for $Q=0$ (the standard 2D Heisenberg model \cite{manousakis91}) by producing correlated local singlets. The interactions are not
frustrated in the standard (geometric) sense, however, and the model is amenable to large-scale QMC simulations for all positive values of
the ratio $g=Q/J$ (with $J\ge 0$, $Q\ge 0$ being of primary interest) \cite{sandvik10a}. The ground state is long-range AFM ordered for
$g < g_c$, with $g_c \approx 1.50$ \cite{lou09}, and is a spontaneously dimerized VBS for $g>g_c$. In the VBS phase the $Z_4$ symmetry
of four degenerate columnar dimer patterns is broken when $L \to \infty$.

A columnar VBS state and an AFM--VBS transition is also realized if the $Q$-interaction (often called $Q_3$) in Eq.~(\ref{jqham}) is
replaced by a simpler one with only two singlet projectors (or $Q_2$) \cite{sandvik07}. The critical coupling ratio $g_c$ is then much larger, $g_c \approx 22$,
and the  VBS order is much weaker throughout the phase. A much larger number of studies have been devoted to the issue of deconfined quantum criticality
within this model \cite{sandvik07,melko08,jiang08,sandvik10c,harada13,chen13,block13,pujari15}. Disorder effects on the VBS state are easier to study with
the more extended $Q_3$ term  in Eq.~(\ref{jqham}), and we will here demonstrate RS behavior for a significant range of coupling ratios when either
the $J$ or the $Q$ interactions are random. We expect these disorder effects to be generic for VBS phases on bipartite lattices.

Before the advent of the $J$-$Q$ model, VBS physics was normally associated with geometric frustration, in models such as the $J$-$J'$
Heisenberg model with nearest- ($J$) and next-nearest-neighbor ($J'$) couplings. These systems are not amenable to large-scale QMC
studies because of mixed-sign sampling weights (the sign problem), except at the variational level when sampling suitably parametrized and
optimized wave functions \cite{hu13,morita15}. While great progress on frustrated models has been made in the last several years with DMRG and
methods based on tensor product states (see e.g., the recent papers \cite{gong14,wang17,wang16,haghshenas17} for applications
to the $J$-$J'$ Heisenberg model), various convergence issues or limited system sizes still make it impossible to carry out calculations as
reliable as QMC simulations of sign-problem free models.

The $J$-$Q$ models exhibit many of the phenomena of long-standing interest in the context of frustrated quantum magnetism, in particular the AFM--VBS
transition \cite{shao16}, which appears to realize the exotic deconfined quantum-critical (DQC) point scenario \cite{senthil04a,senthil04b}. It is
presently not clear whether exactly this transition is also realized in non-bipartite quantum magnets, e.g., in the square-lattice Heisenberg model with
first and second neighbor interactions---there may instead be an extended algebraic spin liquid phase between the AFM and VBS phases
\cite{gong14,morita15,wang17}. The DQC phenomenon has nevertheless attracted a great deal of interest as it is a prominent example of a quantum phase
transition beyond the standard Landau-Ginzburg-Wilson framework. The $J$-$Q$ models offer unique opportunities to study the emergent degrees of
freedom---spinons and gauge fields---that are the ingredients of the field-theory description of the DQC point and the VBS phase. A very interesting
question is how these degrees of freedom respond to quenched disorder; this issue is one aspect of the work presented here.

By the Imry-Ma argument \cite{imry75}, in the presence of even an infinitesimal degree of randomness in the local interactions, the VBS can no longer
exist as a long-range ordered state, due to different columnar dimerization patterns being energetically favored in different parts of the lattice.
Thus, the uniform VBS breaks up into finite domains of different VBS patterns. An extreme case (in the sense of very small VBS domains) of such a
disordered dimer state has been dubbed the valence-bond glass \cite{tarzia08}. It essentially consists of a random arrangement of short valence bonds and
has been discussed in the experimental context of herbertsmithite \cite{lee07,vries09} and in certain 3D frustrated quantum
magnets \cite{vries10,carlo11}. The kagome spin $S=1/2$ lattice of herbertsmithite 
is to some degree diluted with non-magnetic impurities, and these also liberate spinons from the singlet ground state \cite{singh10}.
It was argued that these spinons interact and form a gapless critical RS state. In this case the spinons can be regarded as a byproduct of the
dilution, and in the original picture of the valence-bond glass without dilution \cite{tarzia08} there were no such spinons.

In analogy with one dimensional spin chains with VBS ground states \cite{lavarelo13,shu16}, and considering the nature of the elementary domain walls
in 2D VBS states \cite{levin04}, one should expect a VBS broken up into domains to also have localized spinons at the nexus of domain walls. Therefore,
interesting magnetic properties due to local moments can arise even without the explicit introduction of moments by dilution. Indeed, it was very recently
argued \cite{kimchi17} that a spin-liquid-like state (referred to as an RS state) arises in this way on the triangular lattice when the pristine host
system is a VBS. The RS state there is formed as a direct consequence of the randomly interacting localized spinons at the nexus of domain walls. Though
spinons do not appear in the scenario discussed in the context of the ED \cite{watanabe14,kawamura14,shimokawa,uematsu17} and DMRG \cite{wu18} studies of
frustrated systems, localized spinons may still give rise to the physical properties observed in these numerical calculations, but they were
not studied explicitly (which would also not be easy with the very small lattices considered). On the square lattice with bipartite
interactions, this kind of state has not been previously expected, however, and it was argued that the most likely scenario for systems like the
random $J$-$Q$ model is that the liberated spinons form a subsystem with AFM order, instead of a fully disordered RS state \cite{kimchi17}.

\begin{figure}[t]
\includegraphics[width=48mm, clip]{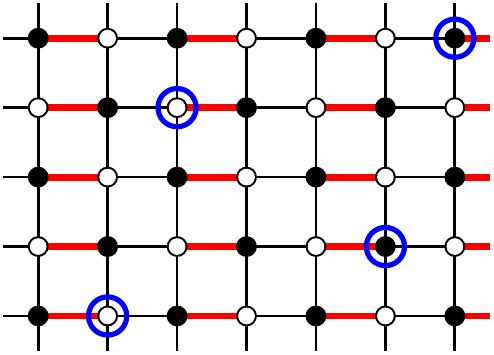}
\vskip-2mm
\caption{The statically dimerized $J_1$-$J_2$ Heisenberg model, with thin black bonds and thick red bonds representing exchange couplings
  ${\bf S}_i \cdot {\bf S}_j$ of strength $J_1$ and $J_2$, respectively, between $S=1/2$ spins. The A and B sublattices are indicated with solid
  and open black circles. The larger blue circles indicate randomly removed sites. For the intact system with $j_2=J_2/J_1$ larger than 
  $j_{2c} \approx 1.91$ \cite{singh88,matsumoto01,sandvik10a}, the ground state is approximately a product of singlets on the strong bonds,
  and upon dilution the 'dangling spins' remaining at the 'broken dimer' adjacent to each removed spin will constitute localized magnetic
  $S=1/2$ moments.}
\label{j1j2fig}
\end{figure}

An example, illustrated in Fig.~\ref{j1j2fig}, of a well understood system in which residual AFM forms among impurity spins is the diluted columnar
dimerized Heisenberg model, which we will later use as a bench-mark case for our numerical analysis techniques. For sufficiently large ratio $j_2=J_2/J_1$
of the intra- to inter-dimer couplings, in the quantum paramagnetic phase, the removed sites leave behind 'dangling' spins at the sites near the broken
dimers, and these form a subsystem with AFM order due to effective bipartite interactions mediated by the inert spin-gapped dimer host \cite{nagaosa96}.
Thus, the quantum phase transition out of the AFM ground state, at $j_2 \approx 1.91$ in the intact system \cite{singh88,matsumoto01,sandvik10a}, is destroyed
and replaced by a cross-over from strong to weak AFM order \cite{yasuda01,santos17}. In a disordered VBS on the square lattice, one might imagine that
the disorder induced spinons should be subject to a similar ordering mechanism \cite{kimchi17}. However, our results and arguments suggest that the correlated
nature of spinon-antispinon pairs (and larger complexes of even numbers of spinons) was not taken fully into account previously. In
particular, we argue that a key missing ingredient in the analysis of bipartite systems Kimchi et al.~\cite{kimchi17} is that the VBS domain walls act as
channels of enhanced spinon-spinon interactions within the groups of even numbers of spinons, thus leading to stronger than expected tendency to local
singlet formation and, apparently, no residual AFM ordering.

The RS state proposed on the triangular lattice may eventually be unstable to the formation of a quantum spin glass (a state characterized by randomly frozen
moments instead of singlets), according to the arguments by Kimchi et al.~\cite{kimchi17}. RS physics could then still appear on long length scales and be
experimentally relevant, although the asymptotic properties of the system, e.g., the thermodynamics at very low temperatures, would be different. This kind of
cross-over, with distinct RS behavior up to some length scale or down to some energy scale, may also be expected in the event that the bipartite RS would be unstable
to AFM ordering. Here we find RS physics in the random $J$-$Q$ model and no signs of cross-over into weak AFM order up to the largest lattices studied, $64\times 64$
sites. We also find non-trivial low-temperature properties that we associate with the RS state. Moreover, we find a distinct transition point with universal critical
exponents separating the RS and AFM states. Thus, the RS state appears to be stable.

Though it is not immediately clear whether the RS phase that we identify and characterize here corresponds to the same fixed point as the state identified
on the triangular lattice by Kimchi et al.~\cite{kimchi17}, this would be the simplest scenario. We show here that the RS state can also form in some
cases even though the bipartite host system is not yet VBS ordered but still in the AFM state, as long as there are sufficient interactions (here $Q$ terms)
favoring the formation of some local VBS domains. This role should also be played by standard frustrated interactions, in systems with VBS states as well
as other states, such as spin liquids or weakly ordered AFMs. The RS state in the disordered $J$-$Q$ model could then indeed correspond to the same RG fixed point
as the states discussed previously in the context of a variety of frustrated host systems,
including  ED studies \cite{watanabe14,kawamura14,shimokawa,uematsu17} and DMRG calculations \cite{wu18}. In the numerical works, the physical picture
presented for the nature of the RS state was different, however, with an emphasis in Refs.~\onlinecite{watanabe14,kawamura14,shimokawa,uematsu17} put on
the singlet pairs (Anderson localization of singlets) \cite{shimokawa} and no reference to the localized spinons and VBS domains emphasized in our work
here and in Ref.~\onlinecite{kimchi17}.

In Secs.~\ref{sec:ground} and \ref{sec:tfinite} we will present $T=0$ and $T>0$ QMC
results for the Hamiltonian Eq.~(\ref{jqham}) with random $J$ and random $Q$ couplings,
as well as for a site diluted system with no randomness in the remaining $J$ and $Q$ interactions. For reference we also present results for the diluted
$J_1$-$J_2$ Heisenberg model in Fig.~\ref{j1j2fig}.
To characterize the ground states of these systems in an unbiased way, we use a ground-state projector QMC method formulated
in the valence-bond basis \cite{sandvik05,sandvik10b}, and to obtain properties at $T>0$ we use the stochastic series expansion (SSE) QMC method \cite{sandvik99}.
To make the results sections more accessible and concise, in Sec.~\ref{sec:spinons} we first outline the physical scenario that arises out of the many different
calculations reported in the subsequent sections.

\section{Domain walls and spinons in the disordered valence-bond solid}
\label{sec:spinons}

On the 2D square lattice and with the bipartite nature of a model such as the $J$-$Q$ model, the main question regarding the disordered VBS state
is whether the spinons localizing at each nexus of four domain walls \cite{levin04} will form long-range AFM order or some other collective state
with only short-range or algebraic spin-spin correlations. As already discussed in Sec.~\ref{sec:randomjq}, one might suspect \cite{kimchi17} that AFM
order should exist for all values of $g=Q/J$, in analogy with the fate of the quantum paramagnet and N\'eel--paramagnetic quantum phase transition
in Heisenberg models with static dimerization when spins are randomly diluted (Fig.~\ref{j1j2fig}). This picture neglects important spatial
correlations among the localized spinons, however, as well as the nature of the VBS domain walls that connect the spinons.

\begin{figure}[t]
\centering
\includegraphics[width=82mm, clip]{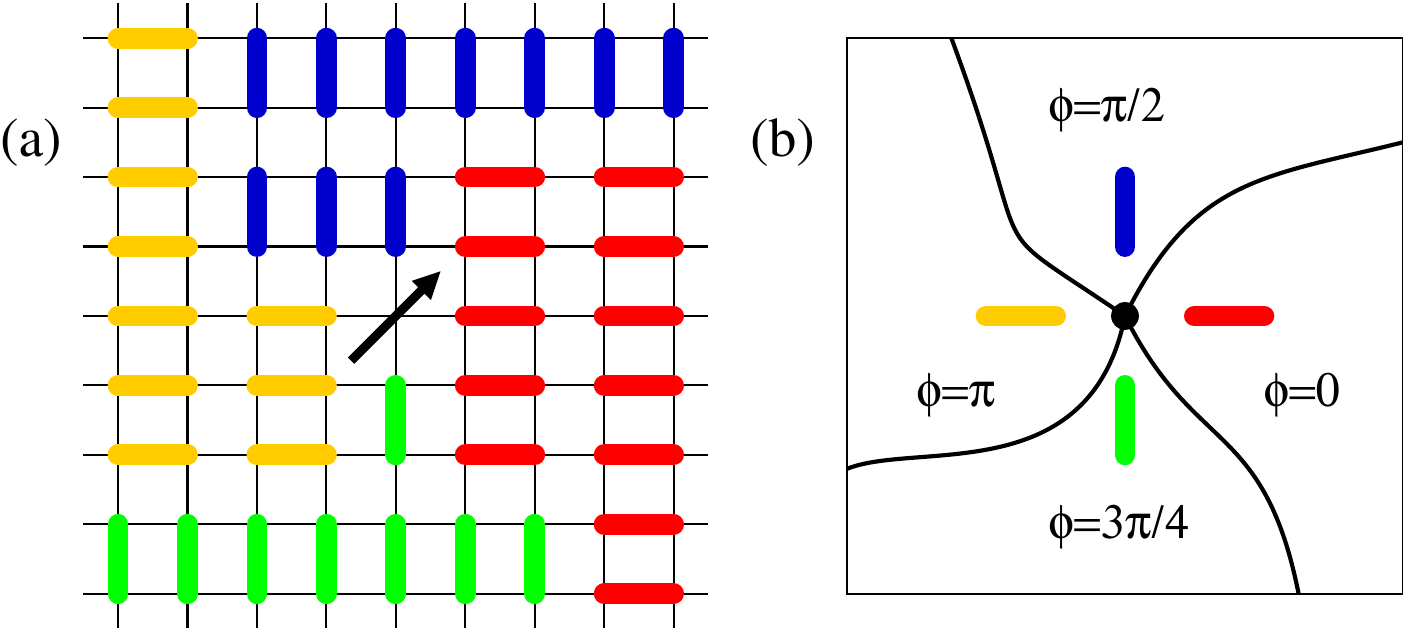}
\vskip-2mm
\caption{(a) Illustration of a spinon forming at an unpaired spin at the nexus of domain walls separating the four different columnar 
VBS patterns on the square lattice. (b) The four VBS patterns associated with angles $\phi$ in a simplified view of the VBS vortex, with 
only the bonds closest to the core shown in the same color coding as in (a). A similar vortex can also be constructed starting from 
core bonds rotated by $90^\circ$ relative to those shown here.}
\label{spinon}
\end{figure}

\subsection{Bound spinons as excitations of the pure VBS}

To understand the spatial spinon correlations, consider first an individual, localized spinon created by a topological defect in the VBS (in a pure
or random system). As illustrated in Fig.~\ref{spinon}, the four lattice bonds pointing out from the site of an unpaired spin (the core of the spinon)
correspond to the four different VBS patterns. Another bond arrangement at the unpaired spin can also form, with the bonds rotated by 90$^\circ$ relative to
those in the figure \cite{levin04}, but our simulations of the $J$-$Q$ model often show the 'star' configuration at the spinon (but this local arrangement
should not change the properties of the domain walls discussed in Ref.~\cite{levin04}). The four bonds and the corresponding extended VBS domains can be
associated with angles $\phi$ as indicated. Note that the energetically favored domain walls correspond to a $\pi/2$ phase twist \cite{levin04}, while walls
with $\pi$ phase change are unstable and break up into two $\pi/2$ walls (as shown explicitly in Ref.~\cite{shao15}). This is the origin of the proper
classification of the symmetry of the VBS as the cyclic $Z_4$ group, or 'clock' symmetry \cite{senthil04a,levin04}. Within a domain wall, the angle $\phi$
(properly defined by coarse graining and averaging over fluctuations) changes continuously, and it is clear that this kind of detect is a vortex-like 
topological defect of the VBS. Such a vortex forming around a vacancy has been studied with the $J$-$Q$ model and a field-theoretical description \cite{kaul08}.
A spinon should be considered as a composite object of the VBS vortex with the unpaired spin at its core.

Note that a spinon can be associated with either sublattice A or B, and the way in which the angle $\phi$ changes, increasing or decreasing, when
going around the spinon in a given direction depends on the sublattice. Thus, we can also refer to the two cases as vortices (sublattice A) or antivortices
(sublattice B), or spinon and antispinon. This classification remains valid also in the presence of longer valence bonds, as long as only bonds connecting the 
two sublattices are allowed. This is exactly the case with bipartite interactions, where bonds connecting sites on the same sublattice are always eliminated
when a state written in the valence-bond basis is time evolved \cite{bondnote}. Note that, fluctuations of the VBS vortices involving longer bonds also
lead to the unpaired spin fluctuating around the vortex core, instead of being completely centered at the core (and of course the core itself becomes
a more extended object).

\begin{figure}[t]
\centering
\includegraphics[width=40mm,angle=90,clip]{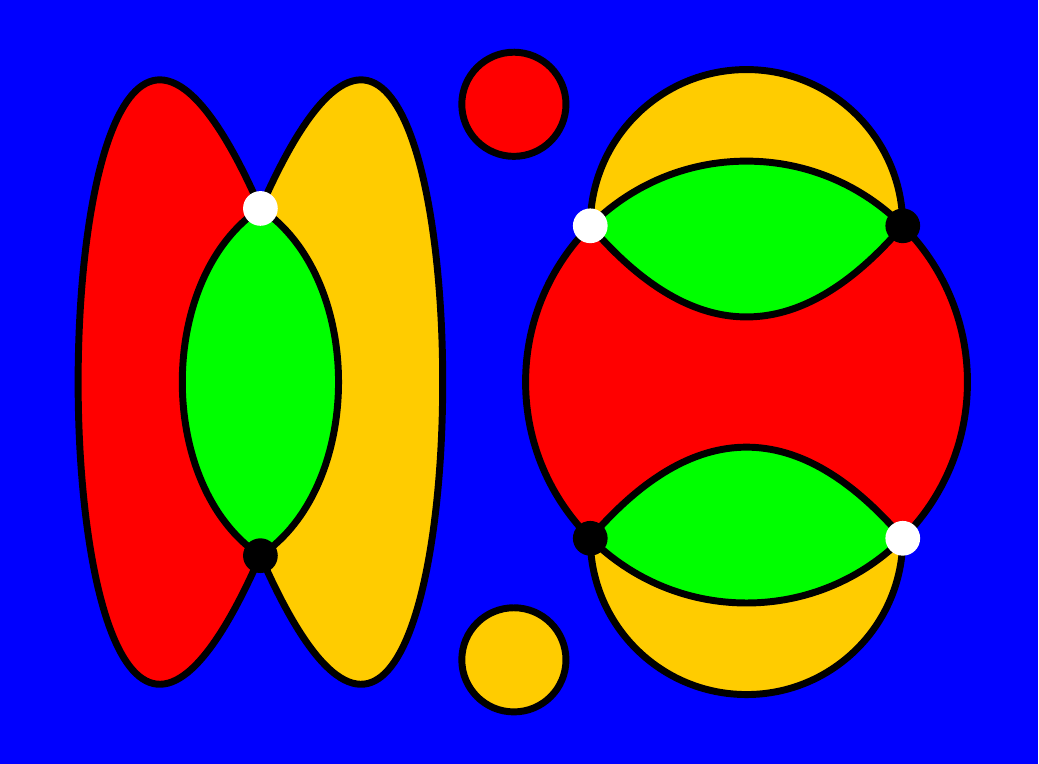}
\vskip-2mm
\caption{Illustration of multi-spinon complexes; a spinon pair (left), with the spinon and antispinon marked as black and white circles,
  respectively, and a quadruplet (right) consisting of two spinons and two antispinons. Two trivial domains, the yellow and red circles, are
  also shown. The color coding of the VBS domains is as in Fig.~\ref{spinon}, and all domain walls are of the elementary type where the VBS
  angle twists by $\Delta\phi=\pi/2$.}
\label{domains}
\end{figure}

When exciting the uniform VBS singlet ground state into its low-lying $S=1$ states, spinons always have to be introduced as pairs of spinons and antispinons,
and these remain bound to each other as dispersing gapped ``triplons''. In a simplified static picture,
when separating the two members of a triplon, domains form such that each spinon
is connected to all four types of domains as in Fig.~\ref{spinon}. As shown in Fig.~\ref{domains}, this leads to a four-stranded confining string, akin to the
(more complicated) quark-confining strings in quantum chromodynamics \cite{sulejman17}. Here we have not shown the details of the bonds within the domains,
only the colors corresponding to the coding in Fig.~\ref{spinon}. As already mentioned, in principle there will also be valence bonds of length greater than
one lattice spacing, but the pictures remain valid as long as the probability of longer bonds decays sufficiently rapidly with the bond length. If we consider
the total-spin singlet state of the two spinons (an $S=0$ excitation of the VBS), there will also be a bond connecting the spinon and the antispinon sites.
Such a long bond corresponds to a small gap between the singlet and triplet excitations (vanishing in the limit of large separation). In the non-random VBS,
the spinons can not actually be far separated in this way, because other spinons can be excited from the VBS ground state as the string energy
becomes sufficiently high. The confining string will then break, thus limiting the number of bound spinon-antispinon states; again analogous to the case of
quark confinement (mesons).

\subsection{Localized spinons in the disordered VBS}

In a system with random couplings, different VBS angles $\phi \in \{ 0,\pi/2,\pi,3\pi/2\}$ will be favored in different parts of the system and
the domain size will be governed by the competition of the energy cost of the domain walls and the energy gains due to the disorder. In classical
systems, according to the Imry-Ma argument \cite{imry75}, this always leads to domain formation at $T=0$ in dimensionality $D<2$, while for $D>2$ the
uniform state is stable in the presence of weak disorder. Considering entropy effects, the uniform state is also unstable at $T>0$ in $D=2$. Similarly
one can expect quantum fluctuations to also always lead to domain formation in systems with two spatial dimensions at $T=0$ \cite{kimchi17}. At least for
weak disorder, the domain walls should still be of the $\pi/2$-twist type. These domain walls can meet in various ways without breaking bonds,
but the case of a nexus of four different domains is special and requires the breaking of bonds into unpaired spinons, as in Fig.~\ref{spinon}.

As in the uniform VBS state discussed above, spinons forming in a VBS broken up into domains must also always appear in groups of
an even number---half of the spinons and half of them antispinons. In Fig.~\ref{domains}, a quadruplet is shown along with the spinon pair already discussed.
It is this inherent correlation among spinons and, importantly, the tendency to singlet formation within the groups, that we believe prohibit the formation
of AFM order in the random VBS arising out of the columnar VBS in the J-Q model. The effective interactions between the
spinons should be mediated through the domain walls (and we will show explicit evidence for this), because they have much smaller local mass gaps than the
bulk of the VBS domains (through which interactions between different spinon groups have to be mediated). We will also later comment on this picture
in the context of SDRG theory.

\subsection{Basic properties of the RS state}

According to our findings reported in Sec.~\ref{sec:ground}, the above described disordered VBS state in the $J$-$Q$ model with random couplings (either random $J$
or random $Q$, both of which we will study, or all random, which we have not considered) should be classified as an RS state, a non-uniform spin liquid
with mean spin correlations decaying with distance as $r^{-2}$. The form of the spin correlation function is, thus, the same as in the 1D RS phase, and the
dimer (bond singlet) correlations decay with a higher power, likely $r^{-4}$, which again would be the same as in 1D \cite{shu16}. Unlike the
1D RS state, we do not find a divergent dynamic exponent, however. By investigating the temperature dependence of the uniform magnetic susceptibility
we find $z=2$ ($T$ independent susceptibility) at the AFM--RS phase boundary and $z>2$ (power-law divergent susceptibility) inside the RS phase.

\begin{figure}[t]
\includegraphics[width=50mm,clip]{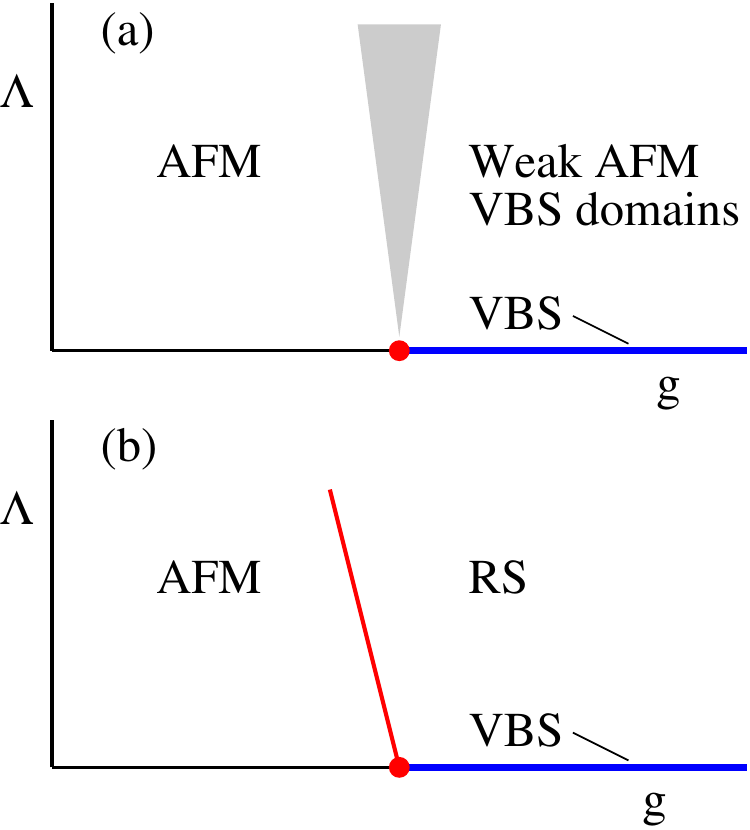}
\vskip-3mm
\caption{Schematic ground state phase diagrams of models such as the $J$-$Q$ model in the presence of disorder. Here $\Lambda$
  represents a disorder strength and $g$ is a tuning parameter existing also in the pure system (e.g., $g=Q/J$
  in the $J$-$Q$ model). In the pure model, $\Lambda=0$, there is a DQC point (red circles) separating the AFM and VBS phases. The VBS is destroyed,
  breaking up into domains, for any $\Lambda>0$. In (a), which applies to the model with site dilution, there is no phase transition vs $g$ when
  $\Lambda > 0$, only a cross-over (indicated by the wedge) between the standard AFM state and a state with finite VBS domains in which weak
  AFM order forms among localized effective moments. In (b), which applies to the case of random coupling constants, there is a true continuous quantum phase
  transition between the AFM and RS phases for at least some range of $\Lambda > 0$.}
\label{phases}
\end{figure}

In further support of a disordered VBS state with no AFM order, we also compare the model with random couplings with a site-diluted $J$-$Q$ model.
Here, like in the diluted $J_1$-$J_2$ model in Fig.~\ref{j1j2fig}, there will be effective moments associated with the removed sites. Thus, while there
may also be localized pair-correlated spinons associated with the meeting points of four domain walls, now there are also moments at random locations
without any intrinsic pairing of A and B sublattice moments.
The vacancies should also lead to topological defects similar to those discussed above, but, since there is no constraint on their sublattice occupation, it
will typically not be possible to pair all the released moments up into spinon-antispinon singlets. The picture of weakly interacting singlet pairs
leading to the RS state is then inapplicable. Indeed, in this case we find a VBS broken up into domains and weak AFM order, and no RS state exists in
the ground state phase diagram.

In Fig.~\ref{phases} we sketch generic phase diagrams expected based on our findings for the $J$-$Q$ model in the presence of the different types of
disorder discussed in this paper. Here we have used a disorder strength denoted by $\Lambda$ on the vertical axis and outlined phases and phase boundaries
in the plane $(g,\Lambda)$, where $g$ is the parameter driving the AFM--VBS transition in the clean system $(\Lambda=0)$. In our actual calculations we
will vary $g=Q/J$ and study several examples of disorder in $J$ or $Q$, but we will not draw full phase diagrams, merely detect the relevant phase transitions
and study the properties of the phases in certain cases to demonstrate their existence. We expect the phase diagrams in Fig.~\ref{phases} to be generic
for disordered 2D quantum magnets that host AFM--VBS quantum phase transitions in the absence of disorder. 

Note the way the AFM--RS phase boundary has been drawn in Fig.~\ref{phases} as tilted into the AFM phase, i.e.,
one can reach the RS state not only from the VBS phase of the pure system but also (for some types of disorder) from the AFM state even when it is quite far
from the AFM--VBS transition. This is interpreted as the tendency to local VBS domain formation in the presence of disorder. On the square lattice the
Heisenberg model with only nearest-neighbor couplings $J$, disorder in the form of random unfrustrated $J$s does not induce an RS phase \cite{laflorencie06},
and a critical strength of frustrated interactions is presumably required, like in the other frustrated systems, to induce
it \cite{watanabe14,kawamura14,shimokawa,uematsu17,kimchi17,kimchi18,wu18}. The $Q$ interactions of the $J$-$Q$ model explicitly
favor local correlated singlets and apparently mimic the effects of geometrically frustrated interactions in their ability to generate the RS state.

\section{Ground state properties}
\label{sec:ground}

We here present QMC results for the $J$-$Q$ model defined in Eq.~(\ref{jqham}) in the presence of disorder in the form of random $J$ or random $Q$.
In most cases we will use a bimodal distribution of couplings, $J_{ij} \in \{0,\Lambda\}$ or $Q_{ijklmn} \in \{0,\Lambda\}$, with equal probability for the
two values, but in some cases we will also consider uniform distributions with the couplings bounded by the above values. To contrast
random couplings and site dilution, we also consider the $J$-$Q$ model where a given fraction of the sites, randomly selected, are missing. All operators
in Eq.~(\ref{jqham}) touching one or several missing sites are then removed from the Hamiltonian.

To bench-mark our calculations for the $J$-$Q$ model against
a case where it is known that site dilution induces AFM order in a quantum paramagnetic host, we also consider the diluted statically dimerized Heisenberg
model illustrated in Fig.~\ref{j1j2fig}. In all cases, we average QMC results over a large number of independent realization of the disorder (hundreds to
thousands) on square lattices with $N=L\times L$ sites and periodic boundary conditions.

Below, in Sec.~\ref{sec:vbprojector} we will first briefly describe the QMC algorithm used in the ground state calculations and also introduce the main
observables we use to characterize the systems. In the following subsections, we present results for all the models; the diluted $J_1$-$J_2$ model in
Sec.~\ref{sec:j1j2}, the diluted $J$-$Q$ model in Sec.~\ref{sec:jqdil}, and the random $J$ and random $Q$ systems in Sec.~\ref{sec:jrand} and
Sec.~\ref{sec:qrand}, respectively.

\subsection{Ground state projector method}
\label{sec:vbprojector}

The QMC method we use here projects out the ground state from a trial wave function $|\Psi(0)\rangle$ written in the valence bond basis consisting of all
possible tilings of the square lattice into bipartite singlet bonds. Acting with $(-H)^m$ on this state, we obtain an un-normalized state $|\Psi(m)\rangle$;
thus expectation values of operators $A$ are evaluated in the form
\begin{equation}
\langle A\rangle = \frac{\langle \Psi(m)|A|\Psi(m)\rangle}{\langle \Psi(m)|\Psi(m)\rangle},
\label{expvala}
\end{equation}
for sufficiently large $m$.
The different propagation paths contributing to $|\Psi(m)\rangle$ are sampled by expressing $H$ as a sum over the $J$ and $Q$ terms in Eq.~(\ref{jqham}) and
carrying out Monte Carlo updates on the corresponding strings (products) of $m$ such operators acting on $|\Psi(0)\rangle$. In this process, the
spin degrees of freedom are put back in by also sampling the $\uparrow\downarrow$ and $\downarrow\uparrow$ contributions to each valence bond (where
one can show that the signs associated with the singlet always cancel out for systems with bipartite interactions) \cite{sandvik10a}. This way, the
projector QMC method in practice becomes very similar to the finite-temperature SSE method \cite{sandvik99}, with the main difference being that the
periodic imaginary-time boundary conditions in the SSE method are replaced by boundary conditions given by the trial state $|\Psi(0)\rangle$. The
exact choice of this state is not critical, though a good variational state can improve the convergence rate in $m$ significantly.

The advantage of the projector approach
relative to taking the limit $T \to 0$ in SSE calculations is that the valence bonds restrict the system to the singlet sector (and other sectors can also
be accessed by simple modifications). Thus, low-lyings $S>0$ states that require very low temperatures to be filtered out in $T>0$ calculations are
excluded from the outset. For further technical details on the method we refer to the literature \cite{sandvik10a,sandvik10b}.

In the valence bond basis, expectation values are expressed using transition graphs \cite{liang88,sutherland88}
obtained by superimposing the bond configuration from the left
and right projected states in Eq.~(\ref{expvala}). Spin-rotationally averaged quantities can be expressed using the loops of the transition graphs, e.g.,
the spin-spin correlation function between two sites $i$ and $j$ vanishes if the two sites are in different loops and is $\pm 3/4$ for sites in
the same loop (with the plus and minus sign corresponding to sites on the same and different sublattices, respectively). Higher-order correlation
functions involve more complicated expressions with the transition graph loops \cite{beach06}.

\subsubsection{Order parameters and correlations}

Here we will focus on the order parameters of the AFM and VBS phases. The former is the conventional sublattice (staggered) magnetization
\begin{equation}
{\bf M} = \frac{1}{N} \sum_{i=1}^N (-1)^{x_i+y_i}{\bf S}_i,
\label{mag1}
\end{equation}
where the coordinates $x_i,y_i \in \{0,L-1\}$. Since the simulations do not break the spin-rotation symmetry we evaluate the expectation value of the
squared order parameter, $\langle M^2\rangle$, which has a simple loop expression. The VBS order can form with horizontal or vertical bonds, and these are
captured by the bond-order parameters
\begin{subequations}
\begin{align}
 D_x = \frac{1}{N} \sum_{x,y} (-1)^x {\bf S}_{x,y}\cdot {\bf S}_{x+1,y}, \label{dimx} \\
 D_y = \frac{1}{N} \sum_{x,y} (-1)^y {\bf S}_{x,y}\cdot {\bf S}_{x,y+1}, \label{dimy}
\end{align}
where, for convenience, we have switched to a notation where the double subscripts on ${\bf S}_{x,y}$ refer to the integer coordinates $\{0,\ldots,L-1\}$
on the square lattice.
\label{dim1}
\end{subequations}
In this case as well we need the squared order parameter, $\langle D^2\rangle = \langle D_x^2\rangle + \langle D_y^2\rangle$, which has a reasonably
simple direct transition-graph loop estimator \cite{beach06}.

With the above order parameters we can also define the corresponding Binder cumulants. In the case of the O(3) symmetric AFM order the proper
definition of the cumulant is
\begin{equation}
U_M = \frac{5}{2} \left ( 1 - \frac{3}{5}\frac{\langle M^4\rangle}{\langle M^2\rangle^2}\right ),
\label{umdef}
\end{equation}
where the coefficients are chosen such that, with increasing system size, $U_m \to 1$ in the AFM phase and $U_m \to 0$ if there is no AFM order.
For $\langle M^4\rangle$ as well there is a simple direct loop expression \cite{beach06}. In the case of VBS order, the coefficients of the
cumulant should be chosen as appropriate for a two-component U(1) symmetric vector order parameter, thus
\begin{equation}
U_D = 2 - \frac{\langle D^4\rangle}{\langle D^2\rangle^2}.
\label{uddef}
\end{equation}
Here $\langle D^4\rangle$ involves eight-spin correlation functions that in practice are too difficult to compute efficiently \cite{beach06}.
We therefore invoke an approximation in Eq.~(\ref{uddef}) that does not impact the scaling properties of the cumulant; we simply evaluate $(D_x,D_y)$
using the loop estimator for the two-point operators (\ref{dimx}) and (\ref{dimy}), and then use these classical numbers to evaluate $D^2$ and $D^4$.
While the expectation values entering in Eq.~(\ref{uddef}) are then  not strictly the correct quantum-mechanical expectation values, they still reflect
perfectly the absence or presence of VBS order in the system \cite{udnote}, and $U_D$ maintains the desired properties discussed above.

In addition to the squared order parameters $\langle M^2\rangle$ and $\langle D^2\rangle$ evaluated on the full lattice as described above,
we will also consider the distance dependent spin and dimer correlation functions,
\begin{subequations}
\begin{align}
&C_s(\mathbf{r}) = \langle {\bf S}_{x,y} \cdot {\bf S}_{x+r_x,y+r_y}\rangle, \label{csdef} \\
&C_d(\mathbf{r}) = \langle ({\bf S}_{x,y}\cdot {\bf S}_{x+1,y})({\bf S}_{x+r_x,y+r_y}\cdot {\bf S}_{x+1+r_x,y+r_y})\rangle \nonumber \\
&~~~~~~~~~~~~~~  - \langle {\bf S}_{x,y}\cdot {\bf S}_{x+1,y}\rangle^2 \label{cddef},
\end{align}
where we spatially average over the reference coordinates $(x,y)$ for each disorder sample. In the case of the spin correlations we will also
consider the probability distribution of values without averaging over $(x,y)$ or disorder realizations. The spin correlations have a staggered sign
$(-1)^{r_x+r_y}$, while the sign of the dimer correlator with $x$ oriented bond as above is $(-1)^{r_x}$ (and we take the proper average with
the $y$-oriented ones). When presenting results we remove these signs. In $C_d(r)$ it is sometimes better to use the difference between even and
odd distances instead of removing the squared mean value.
\label{cordef}
\end{subequations}

\subsubsection{Spinon strings}
\label{sec:spinonstring}

\begin{figure}[t]
\includegraphics[width=50mm, clip]{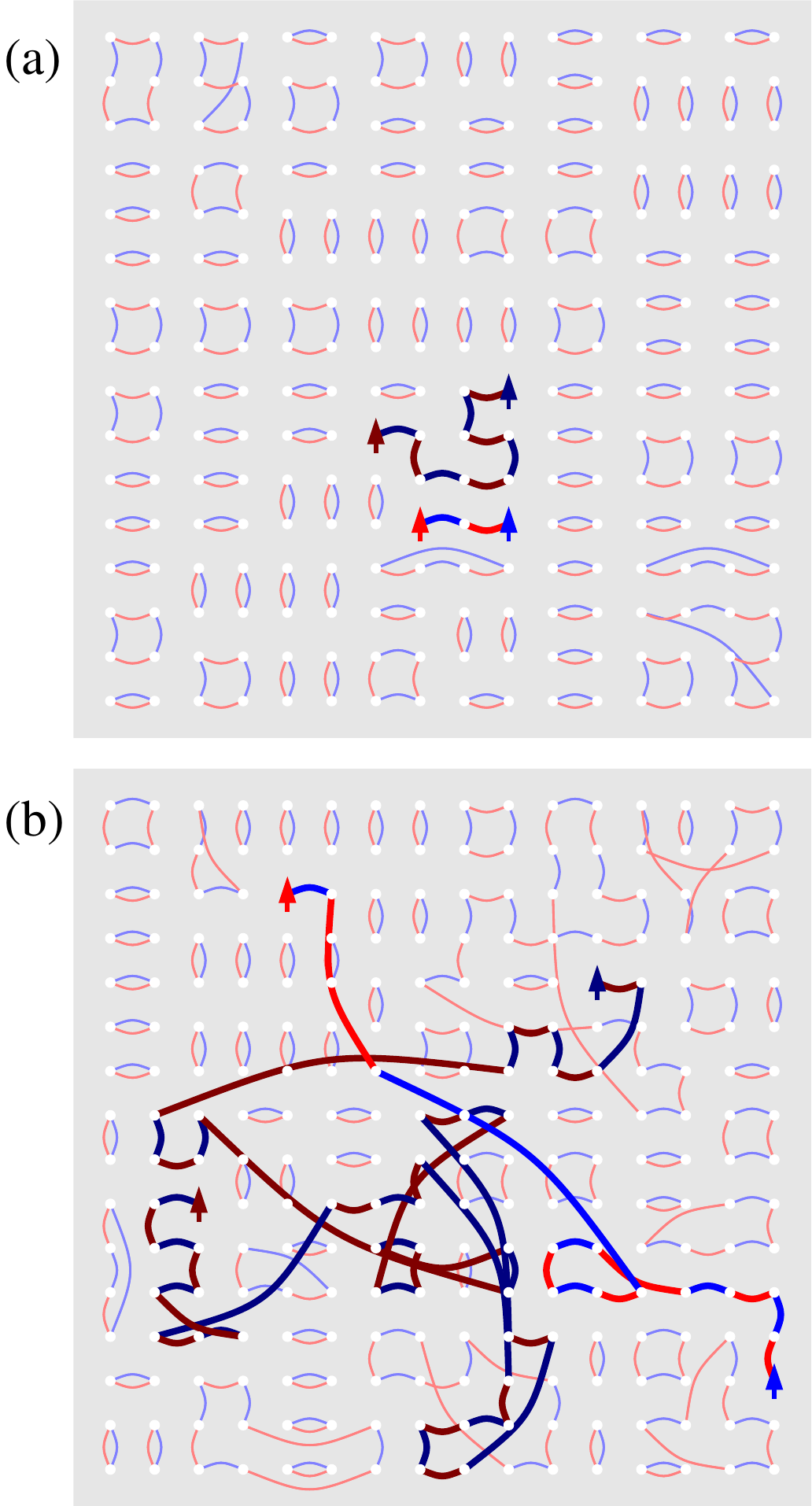}
\vskip-1mm
\caption{Transition graphs in the $S=1$ ground state of an $L=16$ $J$-$Q$ system without disorder in the VBS (a) and AFM (b) phase. For clarity,
  open boundaries are used here to avoid bonds crossing the boundaries. Red and blue arches correspond to bra and ket 
  valence bonds, with thicker bonds representing the two open strings (depicted in different color shades for clarity) that terminate at unpaired spins 
  (one in the bra and one in the ket state). These end spins are always on sublattice A in one of the strings and B in the other one. In (a), defects in 
  the columnar symmetry-broken VBS pattern originate both from the presence of the spinons and by the intrinsic VBS fluctuations. Formation of a 
  clearly columnar state can only be observed for much larger system sizes \cite{lou09}.}
  \label{strings}
\end{figure}

In addition to the physical observables in the singlet sector discussed above, it is also useful to consider the lowest state with total spin $S=1$,
in which some aspects of spinons can be probed directly. In the valence-bond basis, an $S=1$ state can be expressed with a ``broken bond'', e.g.,
with one bond replaced by two $\up$ spins, one each on sublattice A and B (or with one bond treated as a triplet) \cite{sandvik05,wang10,tang11}.
These unpaired spins will propagate under the action of the Hamiltonian, and one can characterize their collective nature as bound or unbound,
and, in the former case, quantify the size of the bound state \cite{tang11,shao16}. Reflecting the non-orthogonality of the valence-bond states,
when forming a transition graph out of bra and ket states, the spinons do not have to occupy the same sites in the two states. Referring to the bra
and ket sites occupied by the unpaired spins as $a,a'$ and $b,b'$ on sublattice A and B, respectively, open strings of valence bonds will form in the
transition graph between $a$ and $a'$ and between $b$ and $b'$, as illustrated in Fig.~\ref{strings}. The extended nature of the strings reflect
the intrinsic size of the spinons \cite{tang11}.

Here we will characterize an $S=1$ state by simply using the number of sites involved in the spinon strings. As we will see in Sec.~\ref{sec:qrand}, 
the mean number of sites in the strings scales very differently in the AFM and RS states, and this provides a way, along with other methods, to locate 
the phase transition between these two states. In addition, in some cases (in Sec.~\ref{sec:jeff}) we will also use the difference in ground state energy 
between the $S=1$ and $S=0$ sectors to extract the spin gap. The spatial distribution of the spinon strings can also give information on the structure of 
the lowest $S=1$ wave function; this will be investigated in Sec.~\ref{sec:jeff}. For technical details on how to carry out the simulations with
broken valence bonds we refer to Refs.~\onlinecite{sandvik05,wang10,tang11,shao15}.

\subsection{Site diluted $J_1$-$J_2$ static-dimer model}
\label{sec:j1j2}

We begin our discussion of QMC results with a brief study of a statically dimerized system, where in the uniform system there is a quantum phase transition
from an AFM to a trivial quantum paramagnet due to singlet formation at the stronger bonds. In the case of the columnar model illustrated in Fig.~\ref{j1j2fig},
the critical coupling ratio $j_{2c} \approx 1.91$ \cite{singh88,matsumoto01,sandvik10a}. For $j_2 > j_{2c}$, it is well known that effective $S=1/2$ moments
localize around diluted sites in such a system, and that these moments interact with each other by non-frustrated effective couplings mediated by the gapped host
system \cite{nagaosa96}, thus inducing AFM order also in the previously quantum-disordered phase \cite{yasuda01,santos17}. Here we use this system as a means
of illustrating how this weak dilution induced AFM order is manifested in the quantities that we will later study in the more interesting models. For these
illustrations we take the vacancy fraction $p=1/32$, with a canonical ensemble such that exactly $N/32$ sites are removed, with equal numbers on
the two sublattices. This density of vacancies is far below the classical percolation threshold, $p_c \approx 0.407$, beyond which no long-range
order can exist.

\begin{figure}[t]
\includegraphics[width=65mm, clip]{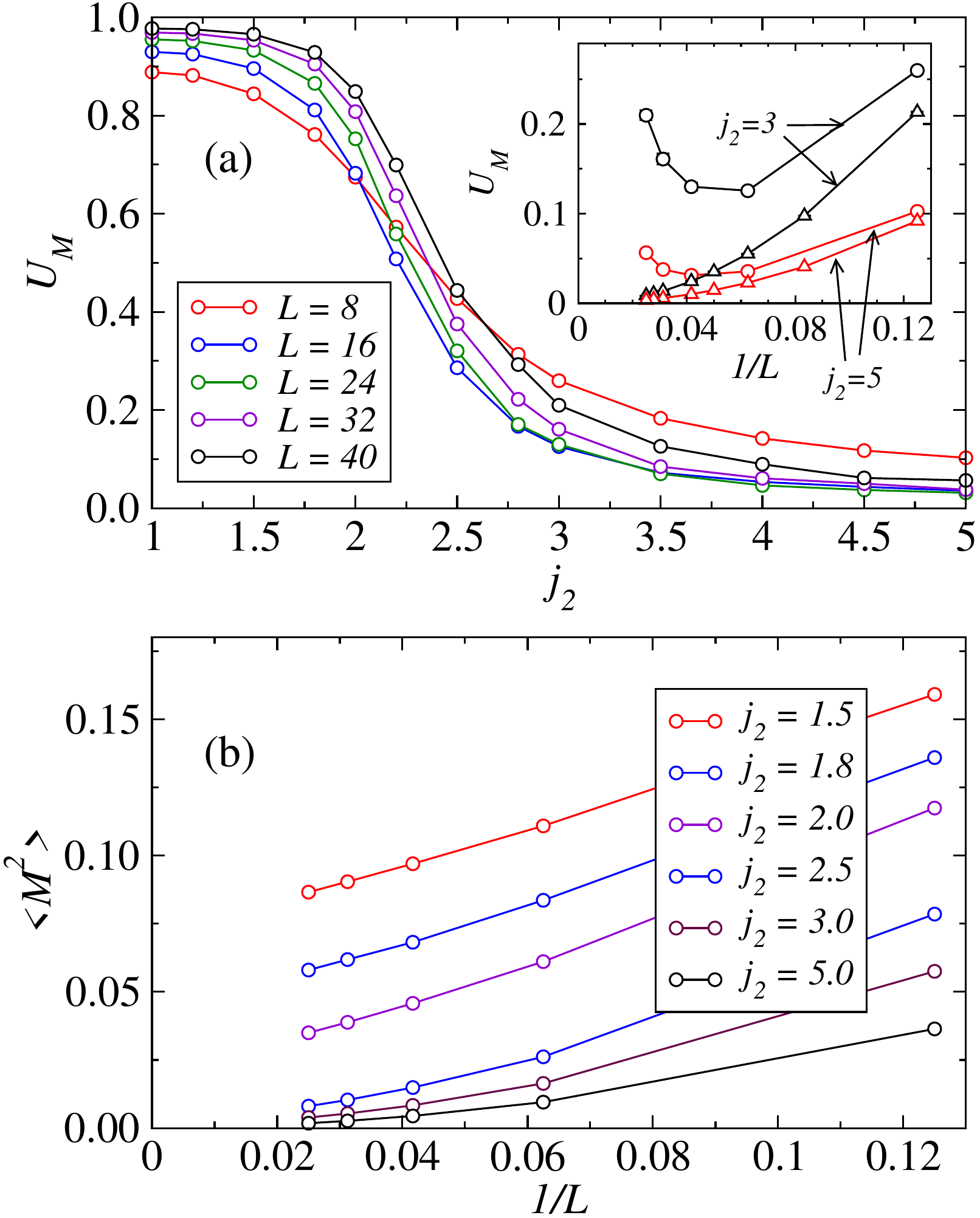}
\vskip-2mm
\caption{Results for the diluted $J_1$-$J_2$ model at vacancy fraction $p=1/32$. (a) AFM Binder cumulant vs the coupling ratio for different
system sizes. The inset shows the size dependence for $j_2=3$ and $5$ for both diluted (circles) and intact (triangles) systems. (b) The size
dependence of the squared sublattice magnetization for several values of the coupling ratio. Error bars are smaller than the symbol size
in all cases.}
\label{j1j2dil}
\end{figure}

Fig.~\ref{j1j2dil} shows results for both the squared sublattice magnetization and its Binder cumulant. The latter turns out to be a more sensitive
quantity for detecting weak order. If there is a critical point separating the AFM phase from a non-AFM phase, the cumulants for two different system
sizes, graphed versus the control parameter, should cross each other at a point that drifts toward the critical point with increasing $L$. However, as
shown in Fig.~\ref{j1j2dil}(a), the crossing points in this case drift rapidly toward higher $j_2$ values and no convergence with increasing $L$ to a critical
coupling can be found. In the inset of Fig.~\ref{j1j2dil}(a), the size dependence at two values of the coupling ratio deep inside the quantum paramagnet
are shown. Here one can observe non-monotonic behaviors indicating asymptotic flows toward the value $U_M=1$ expected for long-range ordered AFM states. This
behavior can be seen even though the order parameter itself, shown in Fig.~\ref{j1j2dil}(b), is very small. Here all the curves for different $j_2$ should
extrapolate to $\langle M^2\rangle > 0$ when $L \to \infty$, but for large $j_2$ the values are very small and not easy to extract precisely. With the
behavior of the Binder cumulants, we can nevertheless confirm that there is long-range order at least up to $j_2=5$, and there is no reason to expect
any other phase for still larger $j_2$.

The reason for the decreasing AFM order with increasing coupling ratio $j_2$ deserves some discussion. This behavior can have more than one
source and the most important should be: (i) The localized moments induce some AFM order in their vicinity and so each diluted site can contribute effectively
more than one unit of staggered magnetization. This effect decreases with increasing $j_2$ as the host becomes less susceptible to induced order.
(ii) Some of the local moments will form singlets and do not contribute (or contribute very little) to the overall AFM ordering. This effect may also
increase with increasing $j_2$, as the effective interactions among moments at fixed distance becomes weaker and the distribution of couplings becomes
broader. Therefore, some  moment pairs will become more specifically coupled to each other than to other more distant spins in their surroundings.
The AFM order cannot be destroyed by these effects, however, as there will always be unpairable moments on sufficiently large length scales, which
is supported by previous numerical studies \cite{yasuda01,santos17}.

\subsection{Random $Q$ model}
\label{sec:jrand}

We next consider the intact lattice with randomness in the $Q$ interactions, using an extreme case of bimodal coupling distribution where 
each $Q$ term in Eq.~(\ref{jqham}) is either absent or present (with equal probability). Here we take the strength of the present six-spin
couplings as $2Q$, so that the parameter $Q$ is the average six-spin coupling. As $Q$ increases, the effective value of the
disorder strength, $\Lambda=Q$ in Fig.~\ref{phases}(b), also increases when defined in relation to the constant $J$ coupling. We will
demonstrate a quantum phase transition between the AFM phase and the phase that we characterize as an RS phase as the coupling ratio $Q/J$
increases. We will argue that the phase diagram is of the type schematically illustrated in Fig.~\ref{phases}(b), though we will not 
consider the full phase  boundary versus $\Lambda$. We will demonstrate the existence of a quantum critical point separating the two
phases along one path in parameter space and also characterize the ground state properties of the RS phase in various ways.

\subsubsection{VBS domains and apparent lack of AFM order}

\begin{figure}[t]
\includegraphics[width=75mm, clip]{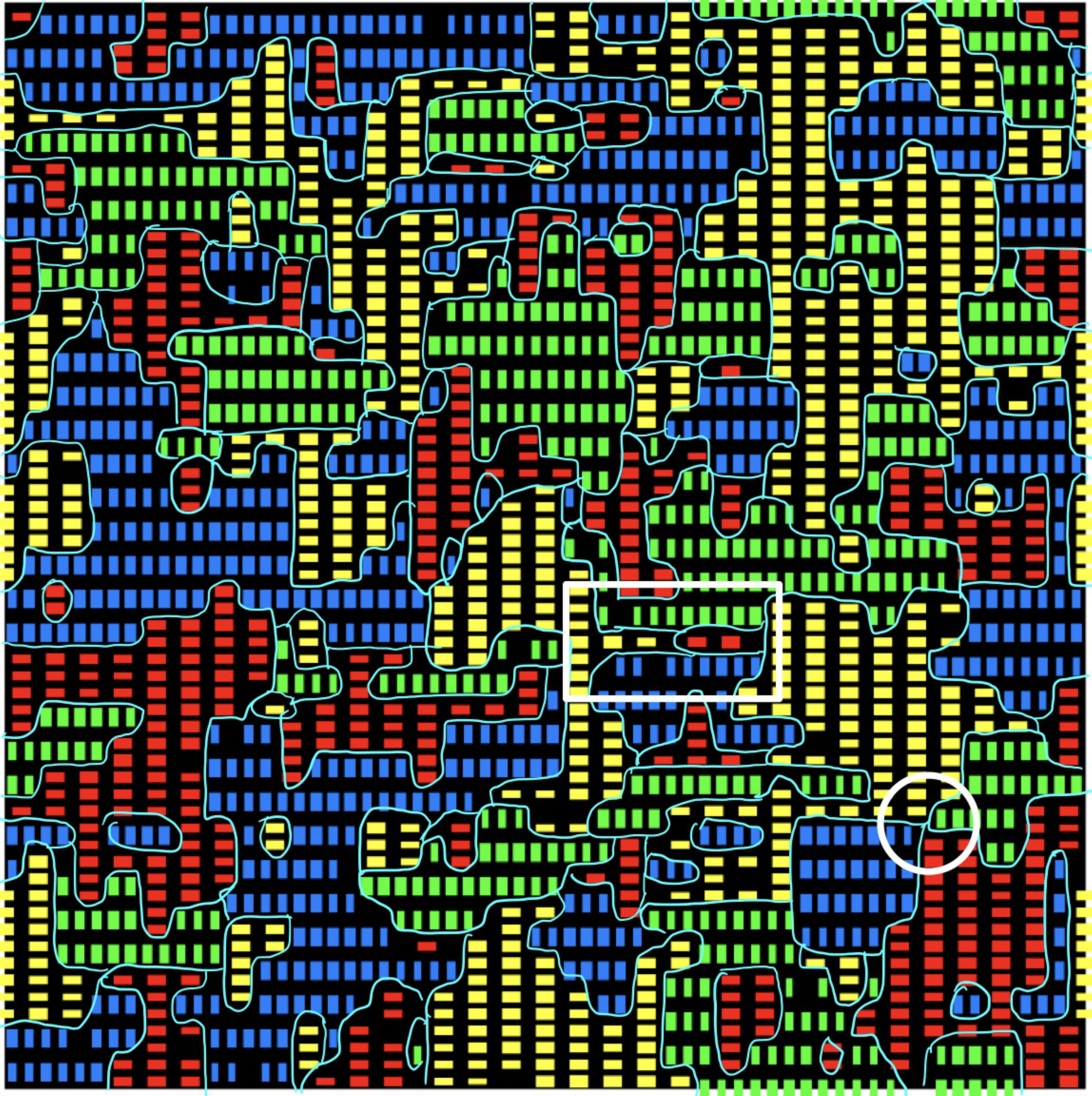}
\vskip-1mm
\caption{Visualization of the VBS pattern in the $J$-$Q$ model generated in simulations with one realization of random bimodal couplings
  $Q \in \{0,1\}$ and $J=0$ on a periodic $64\times 64$ lattice.
  The colored links visualize the corresponding correlations $\langle {\bf S}_i  \cdot {\bf S}_j\rangle$ between the spins $i$
  and $j$ connected by the link, with the line thickness indicating the magnitude of the correlation. A given link is drawn only
  if it is the strongest link for both spins $i$ and $j$, and the color coding corresponds to the convention defined in Fig.~\ref{spinon}.
  The domain boundaries are drawn by hand (turquoise curves). The circle indicates an example of a spinon, at which four domain walls meet
  (as in Fig.~\ref{spinon}, but with a different bond arrangement at the nexus of the domains).
  The rectangle encloses a segment of a $\pi$ domain wall, in which two spinons are located.}
\label{vbsdom}
\end{figure}

First, in Fig.~\ref{vbsdom}, we visualize the VBS domains forming in this kind of system for large $Q/J$, where the pure system is deep inside the
VBS phase. Here we observe several instances of meeting points of four domain walls, where spinons are expected to be localized. The clearest
example of such a spinon region is indicated by a circle in the low-right corner in Fig.~\ref{vbsdom}. Note that the static dimer pattern,
which in Fig.~\ref{vbsdom} is represented by the nearest-neighbor spin correlations, can be misleading due to the fact that it
does not convey completely the quantum fluctuations. A thin line or the absence of any line on a given site implies large fluctuations of the
associated spins, as further explained in the caption of Fig.~\ref{vbsdom}, but the nature of those fluctuations is not apparent. Later, in
Sec.~\ref{sec:jeff}, we will also visualize the local spin fluctuations and demonstrate that they are small within the bulk of VBS domains and
large at regions corresponding to spinons and domain walls. Despite possible
shortcomings of this type of visualization, it nevertheless makes clear the typical domain size and the manner in which domains meet. A notable feature
is that there are mainly domain walls of the type where the angle $\phi$ (Fig.~\ref{spinon}) changes by $\pi/2$, as would be expected according to
the discussion in Sec.~\ref{sec:spinons}. Some very short segments of $\pi$ domain walls can also be seen, with a line of bonds oriented
perpendicularly to those of the adjacent domains located in the gap between those domains. The $\pi$ domain walls in a pure system with a two-fold degenerate VBS
are gapless with deconfined spinons \cite{sulejman17}, and in a disordered system with a pinned $\pi$ domain wall one can expect localized spinons to form
pairwise as well. These spinons can also be regarded as meeting points of four domains, with two of the domains being extremely narrow (chain-like).
Examples of local VBS patterns indicative of such spinons can also be seen in Fig.~\ref{vbsdom}, in the form of $\pi$ phase shifts between the VBS
patterns of chain segments between two domains. One such domain wall is enclosed by a rectangle in the figure.

\begin{figure}[t]
\includegraphics[width=65mm, clip]{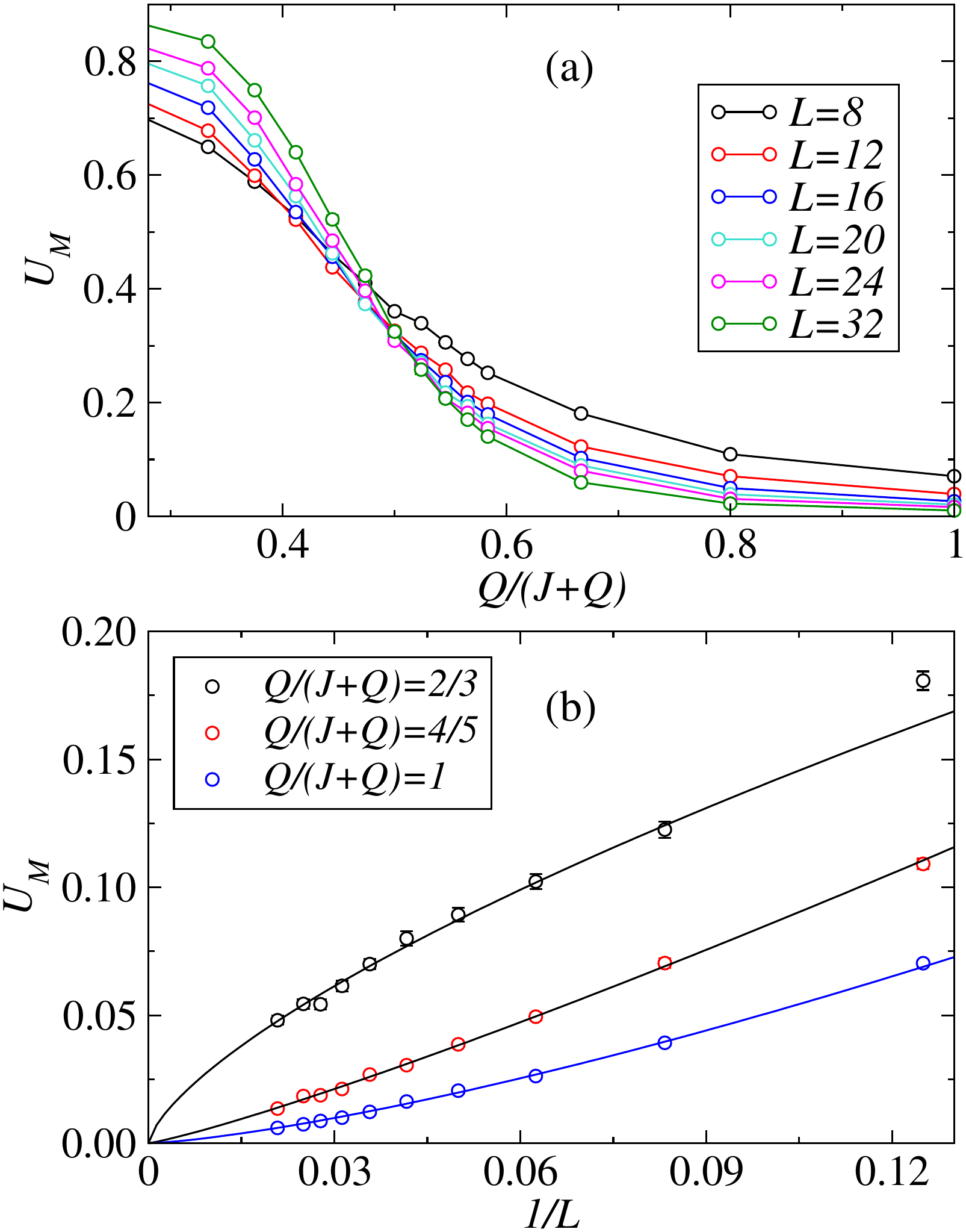}
\vskip-1mm
\caption{AFM Binder cumulant of the random $Q$ model. In (a), results for several different system
sizes are graphed versus the coupling ratio $Q/(J+Q)$, and in (b) results for three different cases inside the RS phase are
graphed vs the inverse system size along with power-law fits.}
\label{qdisum}
\end{figure}

The main question now is whether AFM order is induced among the localized spinons that presumably exist in the random VBS environment. We again study
the AFM Binder cumulant, Eq.~(\ref{umdef}), as a function of the $Q$ interaction. For convenience, to span the full range of interactions,
we graph $U_M$ versus $Q/(J+Q)$ in Fig.~\ref{qdisum}(a). Interestingly, unlike the diluted $J_1$-$J_2$ model, Fig.~\ref{j1j2dil} (and also
the diluted $J$-$Q$ model to be discussed later in Sec.~\ref{sec:jqdil}), in this case it appears that the cumulants for different system sizes 
develop a common crossing point as $L$ increases; the standard signal of a quantum phase
transition of the AFM state. Furthermore, as shown in Fig.~\ref{qdisum}(b), for values of $Q/J$ larger than the apparent asymptotic crossing point, the
cumulants decrease steadily toward zero and there are no indications of any upturn expected if the state has weak AFM order. One could of course wonder
whether the turning point might occur only for even larger system sizes, but the very different behaviors of the crossing points between the diluted models,
where they drift strongly as the system size increases (as shown in Fig.~\ref{j1j2dil} in the case of the $J_1$-$J_2$ model) suggests that the phase
diagrams really are different.

\subsubsection{Existence of a phase transition}

\begin{figure}[t]
\includegraphics[width=65mm, clip]{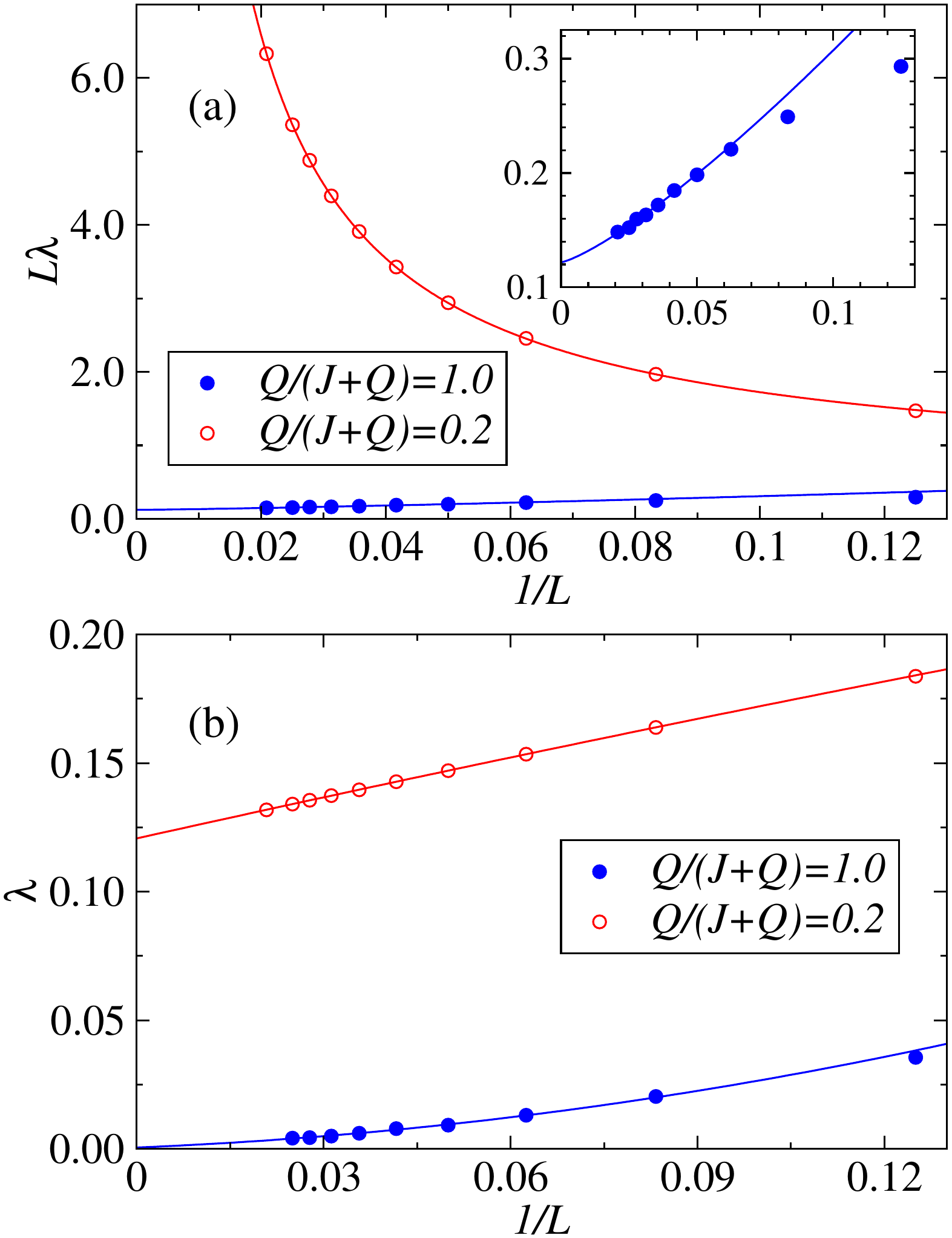}
\vskip-2mm
\caption{Finite-size scaling properties of the string fraction $\lambda$ of the random $Q$ model in the AFM state, at $Q/(J+Q)=0.2$, and deep
  in the RS phase, at $Q/(J+Q)=1$. In (a) $\lambda$ is scaled by $L$ to demonstrate $\lambda \propto L^{-1}$ in the RS phase (the inset shows
  the results on a more detailed scale). The results in (b) illustrate the expected size independent string fraction in the AFM phase.
  Error bars are smaller than the symbols.}
\label{lambda}
\end{figure}

The possibility of AFM order for large $Q/J$ in the random $Q$ model is excluded if we can convincingly establish the existence of a quantum critical point
where the AFM order parameter and related quantities exhibit critical scaling. 
To this end, we will analyze the drift with $L$ of the cumulant crossing points, and also
consider an alternative way of locating the critical point.

As discussed in Sec.~\ref{sec:vbprojector}, QMC simulations in the valence-bond basis allow also for studies of the lowest triplet state, 
which is associated with strings representing spinons in the sampled transition graphs (see Fig.~\ref{strings}). In an AFM state one can expect
the spinon strings to cover a finite fraction of the system (and then the spinons are not well-defined particles \cite{tang11}). We therefore
define the string fraction $\lambda$ as the mean fraction of sites covered by one of the spinon strings. In Fig.~\ref{lambda} we demonstrate that,
indeed, $\lambda$ approaches a constant when $L$  increases inside the AFM phase, while in the RS phase $\lambda \propto L^{-1}$. We do not have a
rigorous explanation for the latter behavior, but it appears to be a very robust feature of the RS phase. Superficially, it would seem to indicate
that the spinons are not completely localized but involve of the order of $L$ spins. However, it should be noted that many spinons can be involved
in forming the lowest triplet, and the spinon strings will migrate during the simulations between all of them. The strings then also
partially occupy the domain walls (see further discussion of this issue in Sec.~\ref{subsec:strings}), and the mean string fraction is not just probing
an individual localized spinon. The precise meaning of the length of the spinon strings in disordered systems should be further investigated; here we merely
exploit the apparent utility of $\lambda$ for locating the AFM--RS transition.

\begin{figure}[t]
\includegraphics[width=65mm, clip]{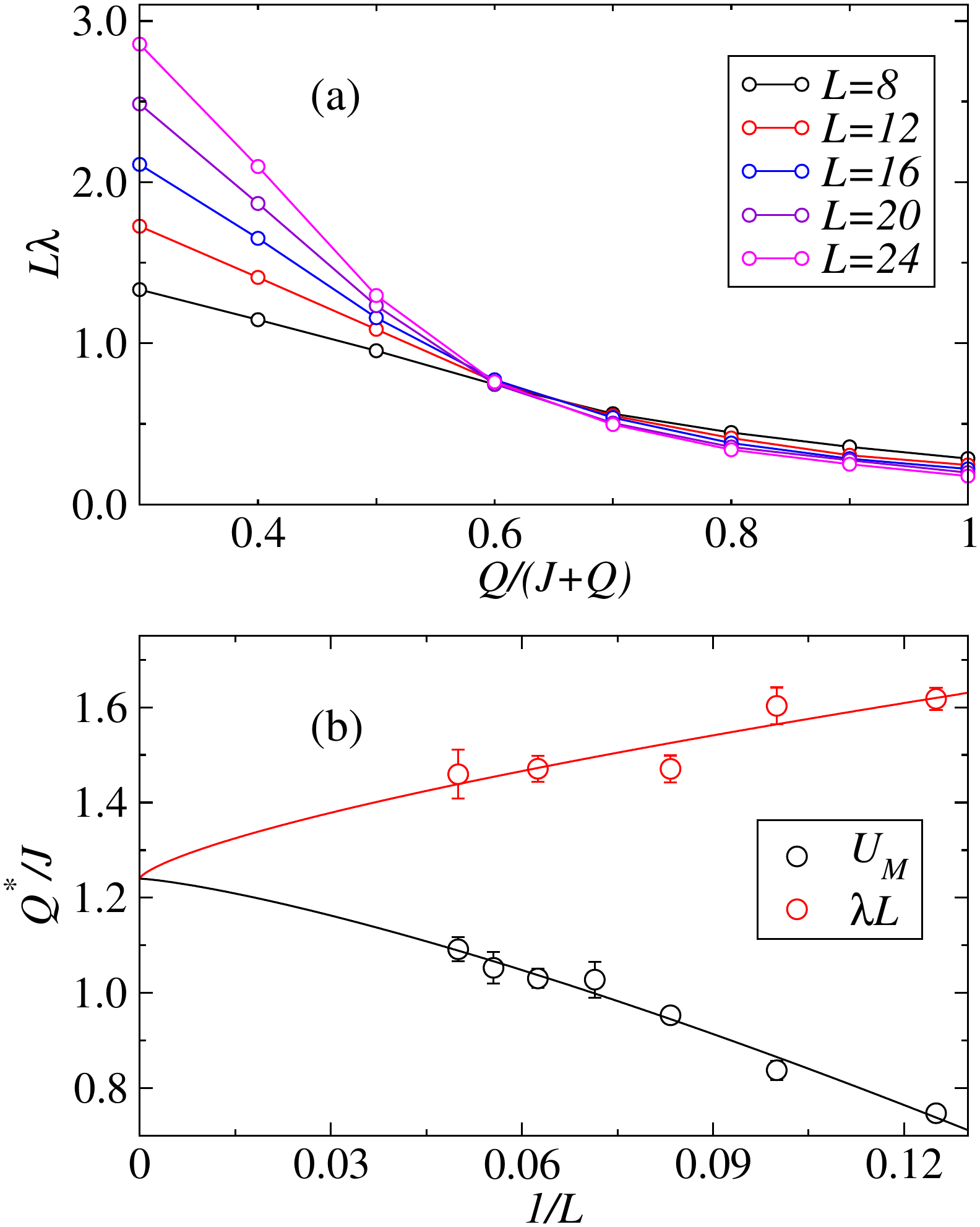}
\vskip-2mm
\caption{(a) String fraction multiplied by $L$ vs the coupling ratio $Q/(J+Q)$ of the random $Q$ model for several system sizes. In (b),
crossing points $Q^*/J$ extracted from system size pairs $(L,2L)$ of data sets such as those in (a) are graphed vs the inverse system size, along with
crossing points extracted from the Binder cumulant $U_M$ in Fig.~\ref{qdisum}(a). The curves are fits to a common constant (the critical value 
$Q_c/J=1.24 \pm 0.13$) with corrections $\propto L^{-\omega}$, where $\omega \approx 1.4$ and $0.8$ for the $U_M$ and $\lambda L$ crossings,
respectively.}
\label{crit}
\end{figure}

Interestingly, as shown in Fig.~\ref{crit}(a), when graphed versus the coupling ratio, $L\lambda$ for different system sizes exhibits crossing
points. This would not necessarily be expected when the behavior throughout the RS phase is $\lambda \sim L^{-1}$, but is still possible due to
scaling corrections; indeed, the fact that the crossings occur at smaller relative angles when $L$ increases and all the curves are close to each
other for large coupling ratios suggest that corrections to the dominant power law are responsible. While the crossing point is still quite well
defined and suggestive of a critical point, the weak size dependence inside the putative RS phase makes it hard to accurately extract the crossing
points between curves for, e.g., system sizes $L$ and $2L$ when $L$ is large. Nevertheless, we have extracted several crossing points and compare
them with the crossing points extracted from Binder cumulant data such as those in Fig.~\ref{qdisum}. As shown in Fig.~\ref{crit}(b), 
the size dependence is consistent with flows to a common value when $L \to \infty$ (the critical point), with power law corrections in $1/L$.
The two data sets approach the transition from different sides, which is helpful for locating the critical point. We do not have any physical
explanation for the different behaviors of the two different data set, but note that prefactors of scaling corrections are not universal and
there is no a-priori reason to expect that two different finite-size estimates of a critical point should approach it from the same side of
the transition.

Since the number of data points for both crossing quantities is rather small, and a common extrapolated $L \to \infty$ point appears visually very likely, 
in Fig.~\ref{crit}(b) we carried out a constrained fit with a common infinite-size point. This fit delivers $Q_c/J = 1.24 \pm 0.13$ (the error bar representing 
one standard deviation). An independent fit to only the $U_M$ points gives a fully compatible result, while a fit to only the $\lambda L$ points gives a slightly 
higher value, $1.4 \pm 0.1$. In the latter case the number of data points is very small (the number of degrees of freedom of the three-parameter fit is only 2) 
and the error bar is therefore not reliable. Considering the statistically sound joint fit, we take it as strong evidence that both $U_M$ and $L\lambda$ 
are valid indicators of a quantum critical point separating the AFM phase and a non-magnetic phase that we argue is an RS phase.

\subsubsection{Correlation functions}
\label{sec:corr}

\begin{figure}[t]
\includegraphics[width=65mm, clip]{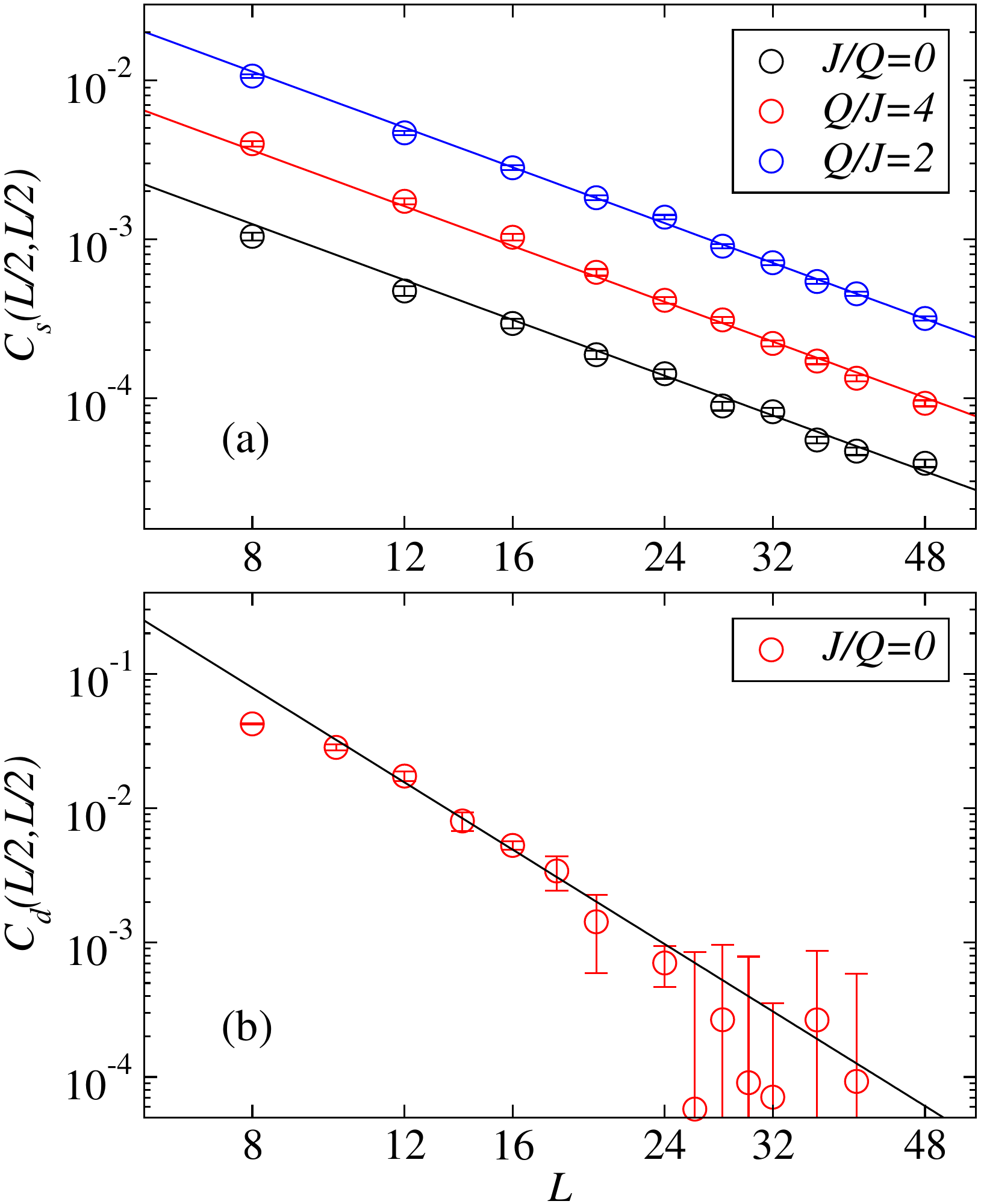}
\vskip-2mm
\caption{Absolute values of the mean long-distance spin (a) and dimer (b) correlations at three coupling ratios inside the RS phase of the random $Q$ model. Results
  are shown at the largest distance on the periodic $L\times L$ lattices. The three lines in (a) correspond to decay of the form $\propto L^{-2}$ and
  the line in (b) shows the form $\propto L^{-4}$.}
\label{corsd}
\end{figure}

Next, we consider the mean spin and dimer correlation functions. Fig.~\ref{corsd}(a) shows the spin correlations, Eq.~(\ref{csdef}), at the
largest distance on the periodic lattices, $r=L\sqrt{2}$, versus the system size $L$. For three different coupling ratios inside the RS phase,
we find the same behavior; a power-law decay corresponding to the distance dependence $C_s(r) \propto r^{-\alpha}$ with $\alpha=2$. Instead of carrying
out line fits to find $\alpha$, we here just show comparisons with the form with $\alpha=2$, but individual fits in all cases are also consistent
with this value. Interestingly, $C(r) \propto r^{-2}$ is also the form at the RS fixed point in 1D \cite{fisher94}, though in that case there
are apparently also multiplicative logarithmic corrections \cite{shu16} that we do not find here in 2D. In the case of the dimer correlations
defined in Eq.~(\ref{cddef}), Fig.~\ref{corsd}(b) shows results at the longest distance where we have extracted the relevant connected
piece of $C_d(r)$ as the difference between even and odd distances $r$, which produces less noisy results than the method of subtracting the mean
value in Eq.~(\ref{cddef}). Here the relative error bars are still rather large for the larger systems, and we only show consistency with
the form $C_d(r) \propto r^{-4}$, which again is the same form as in 1D (up to the log corrections found in 1D) \cite{shu16}.

\begin{figure}[t]
\includegraphics[width=70mm, clip]{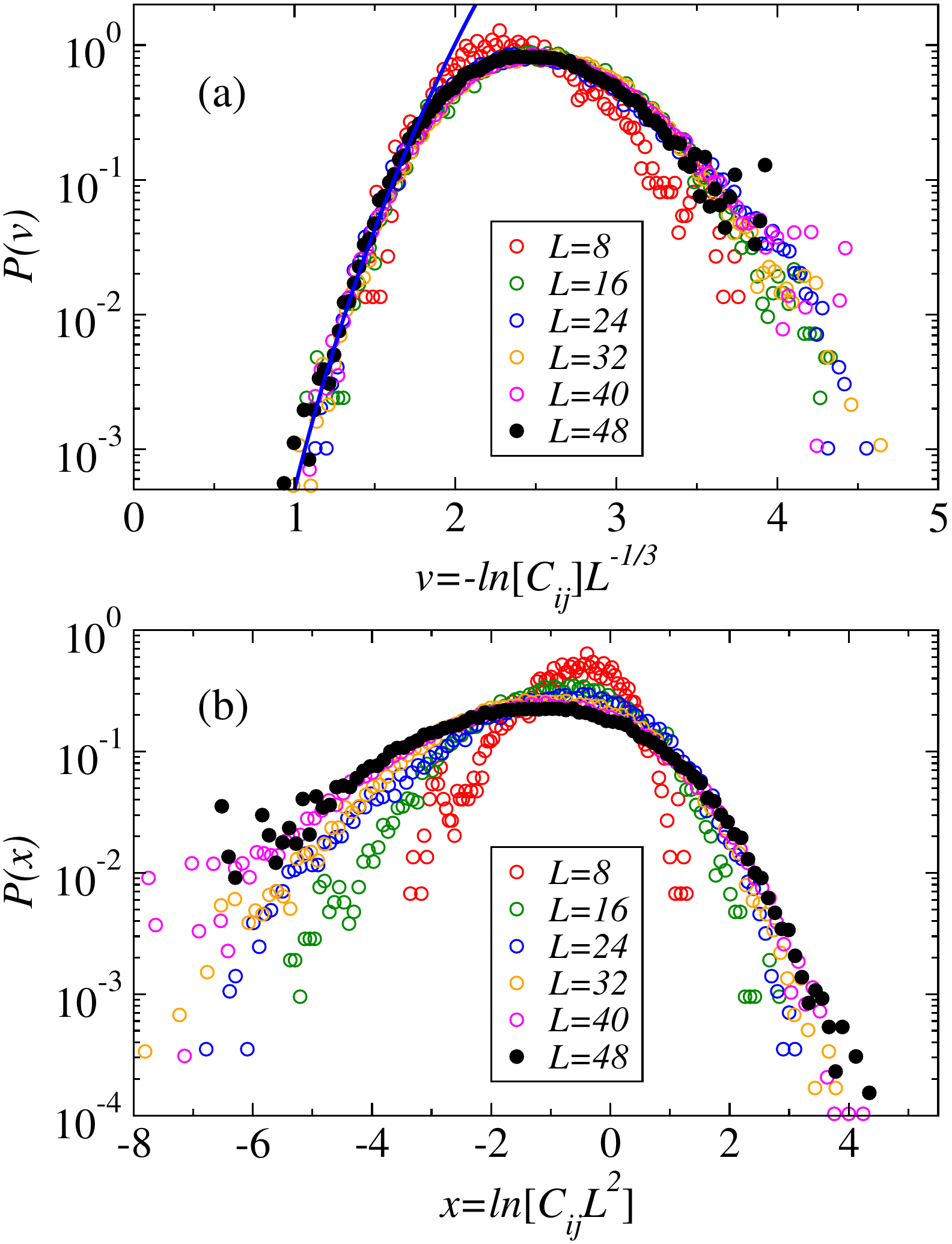}
\vskip-2mm
\caption{Two types of histograms showing the distribution of the spin correlations at distance ${\bf r}=(L/2,L/2)$. In (a) the exponent $a$
  in the variable $v$ in Eq.~(\ref{lamdef}) has been set to $a=1/3$, close to its optimal value for collapse of the data for the larger systems.
  The blue fitted curve on the left side of the distribution corresponds to the power-law behavior $P(v) \propto v^n$ with $n=11$.
  In (b) the scaling variable $x=\ln(C_{ij}L^2)$ is used.}
\label{corhist}
\end{figure}

It is also interesting to investigate the probability distribution of the values of the correlation functions in the spatially non-uniform system.
Here we again consider the longest distance $r_{ij}=L\sqrt{2}$ on the periodic square lattice and accumulate in histograms all the individual spin
correlations $C_{ij}=C(r_{ij})$ for spins at sites $i,j$ separated by this distance, with a large number of disorder realizations used to produce
reasonably smooth distributions. In this case it is important to run rather long simulations for
each individual disorder realization, so that the statistical errors do not influence the distributions significantly for the smaller instances of
$C(r_{ij})$ (in contrast to the mean disorder-averaged values, where one only has to make sure that the individual simulations are equilibrated, and
the final statistical error is dictated by the number of disorder instances). There will always be some problems with large relative errors for the
smallest correlations, and therefore we expect the distributions presented below to be most reliable at the upper end of the distribution.

To investigate scaling of the distributions, we first attempt a scaling variable similar to one applicable to end-to-end spin correlations of
the random transverse-field Ising chain, which realizes an IRFP \cite{fisher98},
\begin{equation}
v = -\ln(C_{ij})L^{-a},
\label{lamdef}      
\end{equation}  
and transform the histograms to the distribution $P(v)$. In Ref.~\onlinecite{fisher98} the exponent $a=1/2$, but here this does not work, and we therefore
consider $a$ as a fitting parameter. This indeed works quite well for the larger system sizes if $a\approx 1/3$, as shown in Fig.~\ref{corhist}(a).
We also need the resulting data-collapsed distribution to be consistent with the mean correlation function,
\begin{equation}
\langle C_{ij}\rangle = \int_0^\infty dv {\rm e}^{-vL^a} P(v),
\label{meanc}      
\end{equation}  
for which we previously found $\langle C_{ij}\rangle \propto L^{-2}$. We can obtain a power law if the behavior of the probability distribution $P(v)$
for the scaled variable $v$ follows a power law close to $0$; $P(v) \propto v^n$. It is easy to see that the contribution to the mean value from small $v$
then decays as $\langle C_{ij}\rangle \propto L^{-a(n+1)}$, and with $a=1/3$ we therefore need $n=5$. The behavior in Fig.~\ref{corhist}(a) is not consistent
with this value of $n$, instead giving an exponent $n$ more than twice as large (corresponding to $\langle C_{ij}\rangle \propto L^{-4}$), as shown with a
fitted curve in the figure. However, the part of the distribution away from the region where the power law applies still changes the scaling of the mean
value to the observed $L^{-2}$ form for the rather small systems we have access to, for which ${\rm e}^{-vL^a}$ in Eq.~(\ref{meanc}) is not yet very small
when $v \approx 2 \sim 3$. For large system sizes, the power law region would always dominate the integral and with the fitted form we would then obtain
an $L^{-4}$ decay. Since our data do not extend very close to $v=0$ we can not exclude that the distribution still changes and evolves into the 
$v^5$ form as $v \to 0$ and $\langle C_{ij}\rangle \propto L^{-2}$.

Considering the apparent inconsistencies arising with the scaling variable $v$ above, we explore an alternative form of the distribution.
Fig.~\ref{corhist}(b) shows distributions $P(x)$ with the scaling variable $x$ defined as
\begin{equation}
x=\ln(C_{ij}L^2).
\label{xdef}  
\end{equation}  
In this case any $P(x)$ trivially gives the desired $L^{-2}$ decay of the mean. Though the data collapse is not as good as in Fig.~\ref{corhist}(a),
the behavior does seem to improve with increasing $L$, especially at the high end of $x$.

A scaling variable of the form (\ref{lamdef}) and $P(v) \propto v^n$ for small $v$ implies different behaviors of the typical correlations (defined
conveniently by the peak of the distribution) and the conventional mean value; exponentially versus power-law decaying. At the IRFP, this behavior is a consequence
of the divergent dynamic exponent \cite{fisher98}. As we will show in Sec.~\ref{sec:tfinite}, the RS state in the
random $J$-$Q$ model has finite dynamic exponent, and the scaling with the variable in Eq.~(\ref{xdef}), which implies the same power-law decay of the
mean and typical values, may appear more plausible from this perspective. However, the scaling with the logarithmic variable in Fig.~\ref{corhist}(a)
works noticeable better and we cannot exclude that mean and typical values will scale differently even though $z$ is finite. It would clearly be useful to
study larger system sizes and further test the two scenarios for the distributions. The inverse-square distance dependence of the mean correlations 
already appears to be well-established by the good scaling for a wide range of system sizes and three different $Q/J$ values in Fig.~\ref{corsd}.

\subsection{Random $J$ model}
\label{sec:qrand}

\begin{figure}[t]
\includegraphics[width=65mm, clip]{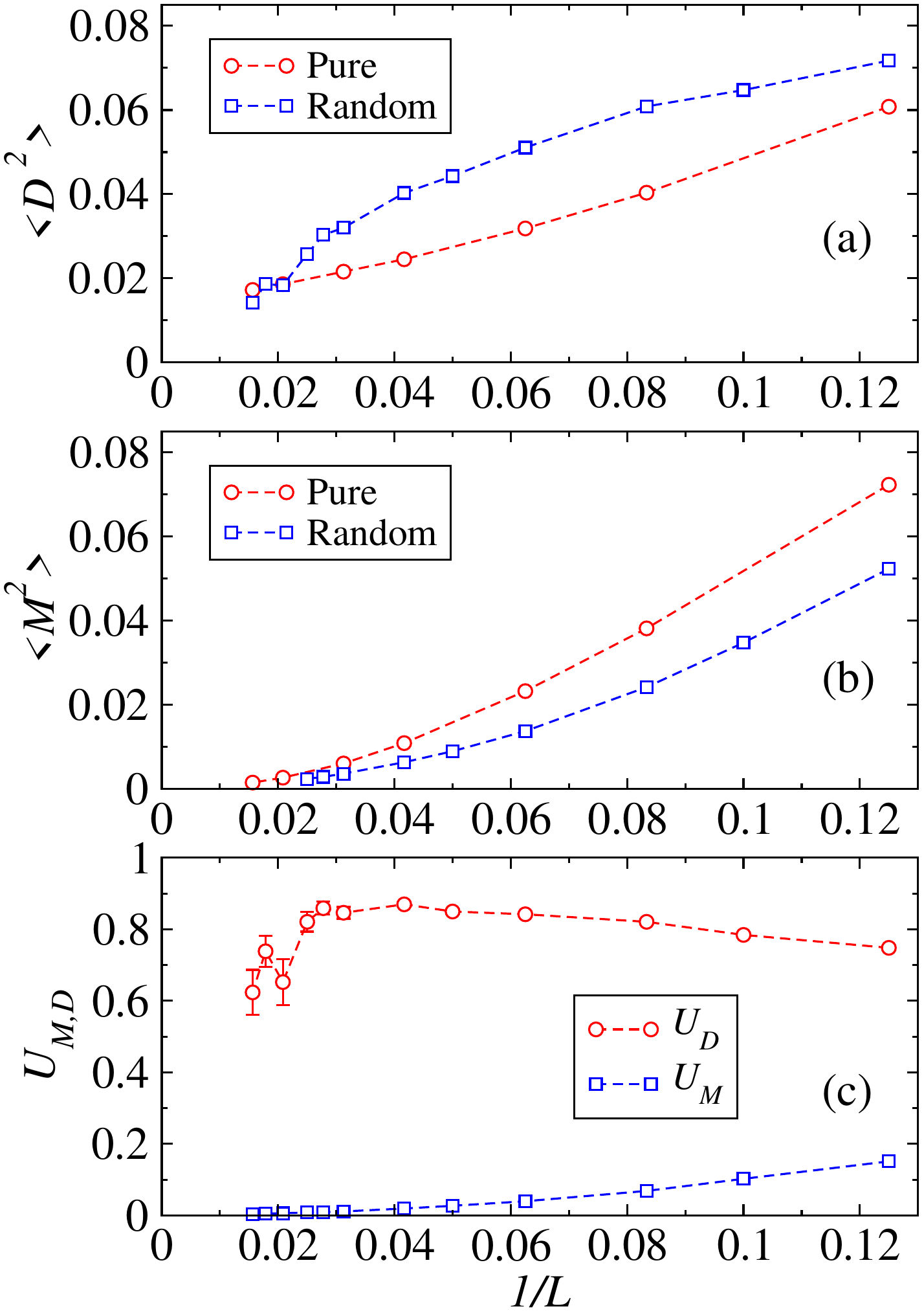}
\vskip-2mm
\caption{VBS (a) and AFM (b) order parameters and the corresponding Binder cumulants (c) versus the inverse system size for the model 
with bimodal $J$ couplings (50\% each of $J=0$ and $J=1$)  at $Q=2$. Results for the pure model with $J=1$ are shown for comparison in (a) and (b).}
\label{bimodalq2}
\end{figure}

In the random $J$ model, all $Q$ couplings are included and the $J$ couplings are drawn from a distribution. We have considered bimodal as well as
continuous distributions and find qualitatively the same kind of behaviors as above in the random $Q$ model. We therefore only provide a few
illustrative results showing these similarities.

Fig.~\ref{bimodalq2} shows results for the order parameters and Binder cumulants at $Q/J=2$ for the extreme bimodal case where half of the $J$
couplings are set to $0$ and the rest to $1$ (which we here take as the value of $J$ in the ratio $Q/J$). For reference we compare the size dependence
of these quantities with the corresponding pure system (all $J=1$). The results indicate that both order parameters vanish when $L \to \infty$, with the
VBS Binder cumulant showing a non-monotonic behavior with a drop toward zero starting when $L$ is of the order of the typical VBS domain size.
For $Q/J=2$ we conclude that the system is in the RS phase.

\begin{figure}[t]
\includegraphics[width=65mm, clip]{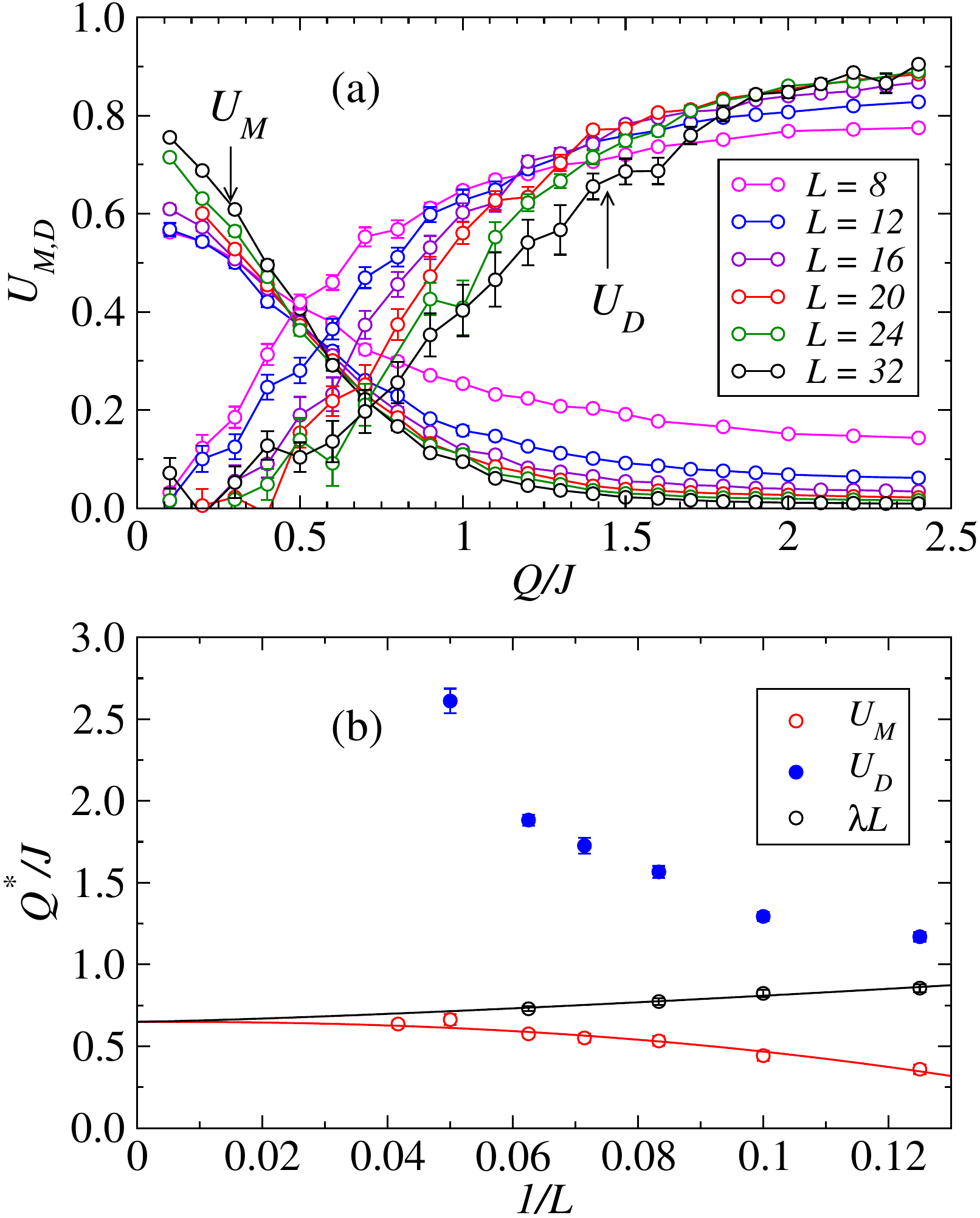}
\vskip-2mm
\caption{(a) Binder cumulants vs $Q/J$ for several system sizes of the bimodal random $J$ model. (b) Crossing points between
cumulants for system sizes $(L,2L)$ versus $1/L$. Crossing points of the size-normalized spinon string fraction $L\lambda$
(similar to those shown in Fig.~\ref{crit}) are also shown. Fits (the curves shown) to the latter data set and that for the $U_M$
crossing points were carried out using power-law corrections, $\propto L^{-\omega}$ (with $\omega\approx 1.5$ and $2.3$ for the $U_M$ and
$L\lambda$ set, respectively), with the constraint of the same value of the crossing point, $Q_c/J=0.65 \pm 0.02$, when $L \to \infty$.}
\label{bimodalqc}
\end{figure}

To confirm the existence of a critical point separating the AFM and RS phases, Fig.~\ref{bimodalqc}(a) shows scans for several system sizes of the Binder
cumulants versus $Q/J$ for the same bimodal $J$ distribution as in Fig.~\ref{bimodalq2}.
For $U_M$ we again see crossing points apparently converging toward a critical point,
similar to the behavior in the random $Q$ case in Fig.~\ref{qdisum}. The $(L,2L)$ crossing points are graphed versus the inverse system sizes in
Fig.~\ref{bimodalqc}(b), along with the crossing points of the scaled string fraction $L\lambda$. These two finite-size estimates of the critical
point again approach $Q_c$ from different directions. Requiring the fits with corrections $\propto L^{-\omega}$ to
have the same value of $Q_c/J$ but allowing for different values of $\omega$, we obtain $Q_c/J = 0.65 \pm 0.02$ and the exponents
$\omega\approx 1.5$ (for the cumulant crossings) and $2.3$ (for the string quantity). Given the rather small number of points and not very large
system sizes, the exponents should be regarded as ``effective exponents'' that are still influenced by  neglected higher-order corrections.
Since we only have four $\lambda L$ points in this case, the individual fit to this quantity is not reliable, but an individual fit to the
$U_M$ data gives results perfectly consistent with the joint fit.
Fig.~\ref{bimodalqc}(a) also shows the behavior of the VBS cumulants. It is clear that the crossing points here do not converge but flow to larger
$Q/J$ as the system size increases, as would be expected when arbitrary weak disorder destroys the VBS phase. The corresponding $(L,2L)$ crossing
points are graphed versus $1/L$ in Fig.~\ref{bimodalqc}(b).

Overall, with the results presented above and in other cases, we find very similar behaviors for the random $Q$ and random $J$ models, indicating
that the RS phase induced by these types of disorder is the same one. One notable aspect of the specific random $J$ model for which we have presented
results here is that the RS phase can arise not only out of the VBS phase of the pure model but also from the AFM state. The critical coupling extracted
in Fig.~\ref{bimodalqc} is at $Q/J \approx 0.65$, where the pure model with all $J=1$ Heisenberg couplings is still well inside the AFM phase (the
AFM--VBS transition of the pure system taking place at $Q/J \approx 1.50$). With the way we have defined the bimodal coupling strengths with $J=0$ and
$J=1$ at random locations, we can reach the RS from the AFM phase simply by removing some fraction of the $J$ interactions when $Q$ is between $0.65$
and  $1.50$. This random removal of $J$ couplings enhances the ability of the $Q$ terms to cause VBS formation, which in the random system only can take
the form of a domain-forming VBS. Thus, it seems very plausible that the same RS state will also be generated if the host system includes some frustrated
interactions that weaken the AFM order and favor local formation of VBS domains in a disordered system, instead of the $Q$ terms considered for that purpose
here. Such frustrated disordered systems can include the Heisenberg model on the triangular lattice, which is equivalent to the square lattice with half of
the diagonal couplings activated. It would then appear quite plausible that RS state we have identified here on the square lattice is actually the same
state as that discussed previously for frustrated systems. However, further characterization of the frustrated systems would be needed to confirm this.
We will discuss possible scenarios for RG fixed points and flows further in Sec.~\ref{sec:discussion}.

\subsection{Universality of the AFM--RS transition}

\begin{figure}[t]
\centering
\includegraphics[width=65mm, clip]{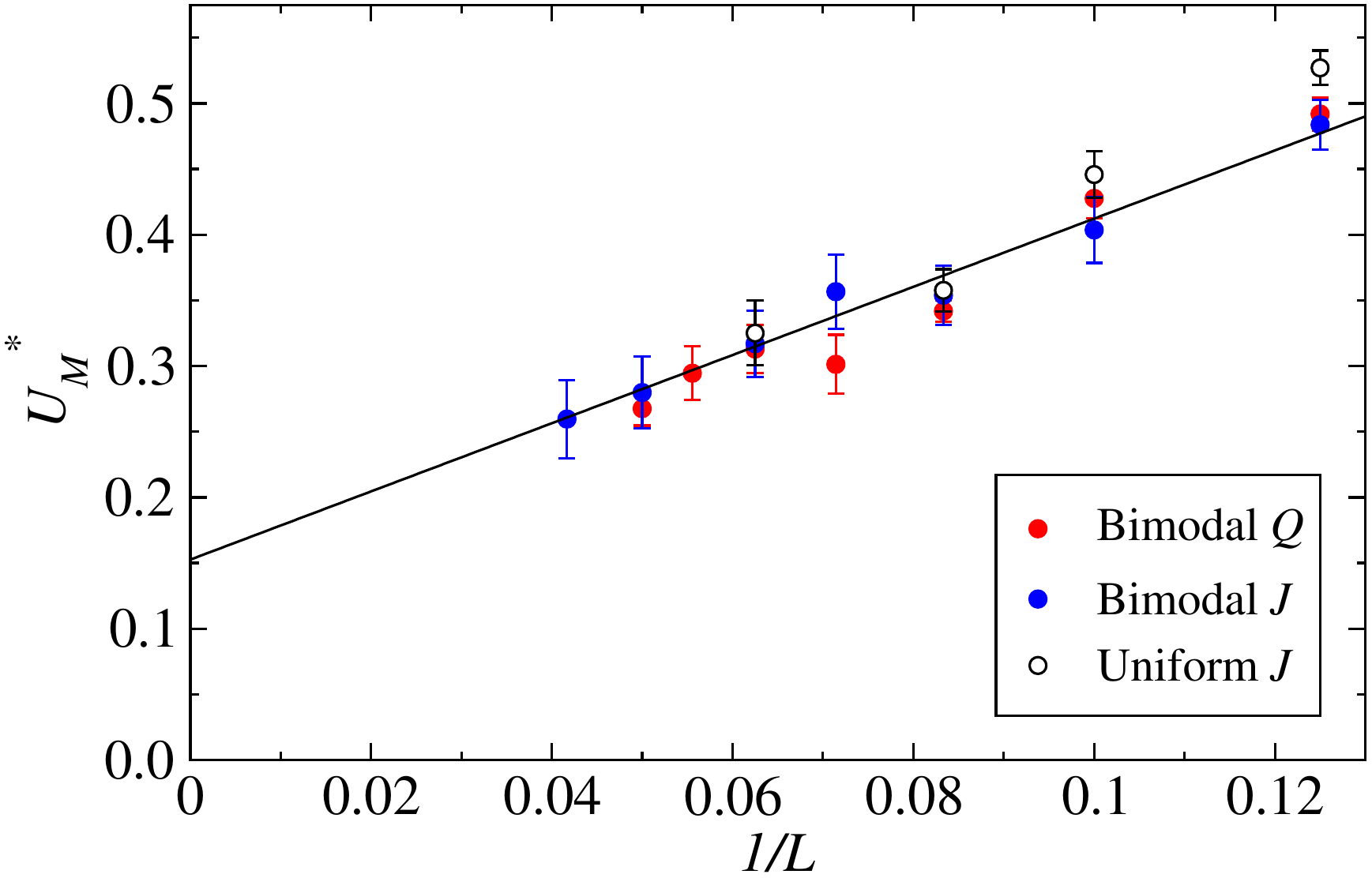}
\vskip-2mm
\caption{Binder cumulants vs the inverse system size at the $(L,2L)$ crossing points for systems with bimodal $Q$ and $J$ distributions
as well as a uniform distribution of $J$ from the range $[0,2]$. The line is a collective fit to the data for the two bimodal cases.}
\label{universal}
\end{figure}

Given our results presented above, it appears most likely that the AFM--RS transition is universal and that the RS phase itself has universal properties, such
as the $1/r^2$ power-law decay of the mean spin correlations (but we will show in Sec.~\ref{sec:tfinite} that the dynamic exponent is not universal inside
the RS phase but varies continuously---though it also is universal at the AFM--RS transition). An often used characteristic of a critical point is the
value of the Binder cumulant. This quantity is universal, in the sense that it is independent on microscopic details, but, unlike many other universal
quantities, such as critical exponents, it depends on boundary conditions and aspect ratios of the system \cite{kamienartz93,selke07,yasuda13}.
In the projector QMC method we effectively take the limit of the time-space aspect ratio $\beta/L \to \infty$ and the system geometry is also the
same for both the random $Q$ and random $J$ models. Thus, we have identical boundary conditions and aspect ratios, and would expect the same value
of the Binder cumulant at the AFM--RS transition point.

In Fig.~\ref{universal} we show results for three disorder types for which we have sufficient data to carry out meaningful studies of the scaling of the
AFM cumulant at the $(L,2L)$ crossing points; in addition to the bimodal $Q$ and $J$ cases we also show results for a continuous distributions of $J$,
with values drawn uniformly from the range $[0,2]$. Remarkably, the cumulants for all cases not only appear to flow to the same point in the limit of
infinite size, but even the leading correction in $1/L$, including the prefactor, seems to be the same. This correction
appears to be almost linear, and we analyze the data under this assumption, though it is possible that the form is $L^{-\omega}$ with $\omega$ just close to
$1$. For the two bimodal distributions all the data fall on the line as closely as would be statistically expected (with excellent goodness of fit),
while for the continuous distribution we see that the data for the smaller sizes deviate more significantly, indicating that the higher-order
corrections do depend on the kind of the disorder distribution. These results clearly lend further support to the existence of a universal AFM--RS
critical point, and, therefore, to the existence of the RS phase. 

The slope of the Binder cumulant, evaluated at the infinite-$L$ critical point or at crossing points, can be used to extract the critical
correlation-length exponent $\nu$,
\begin{equation}
\frac{dU}{dg}\Bigr |_{g=g_c} = aL^{1/\nu} + b L^{1/\nu-\omega} + \ldots ,
\end{equation}
where $g$ is the control parameter used, here $g=Q/J$, and $\omega$ is the exponent of the leading scaling correction (and $a,b$ are non-universal
constants). In practice, it is again convenient to use pairs of system sizes, e.g., $L_1=L$ and $L_2=2L$, and replace $g_c$ by the crossing point $g^*(L)$
of the two cumulants. Then one can show that (see, e.g., Ref.~\cite{shao16}):
\begin{equation}
\frac{1}{\nu^*(L)}=\frac{1}{\ln(2)} \ln \left (\frac{U'(2L)}{U'(L)} \right ) = \frac{1}{\nu} + cL^{-\omega} + \ldots ,
\label{nustar}
\end{equation}
where $U'$ denotes the derivative at the crossing point and $c$ is a non-universal constant. We here obtain the derivatives from the polynomial
fits used to interpolate the crossing points from data sets such as those in Fig.~\ref{qdisum}.

\begin{figure}[t]
\centering
\includegraphics[width=65mm, clip]{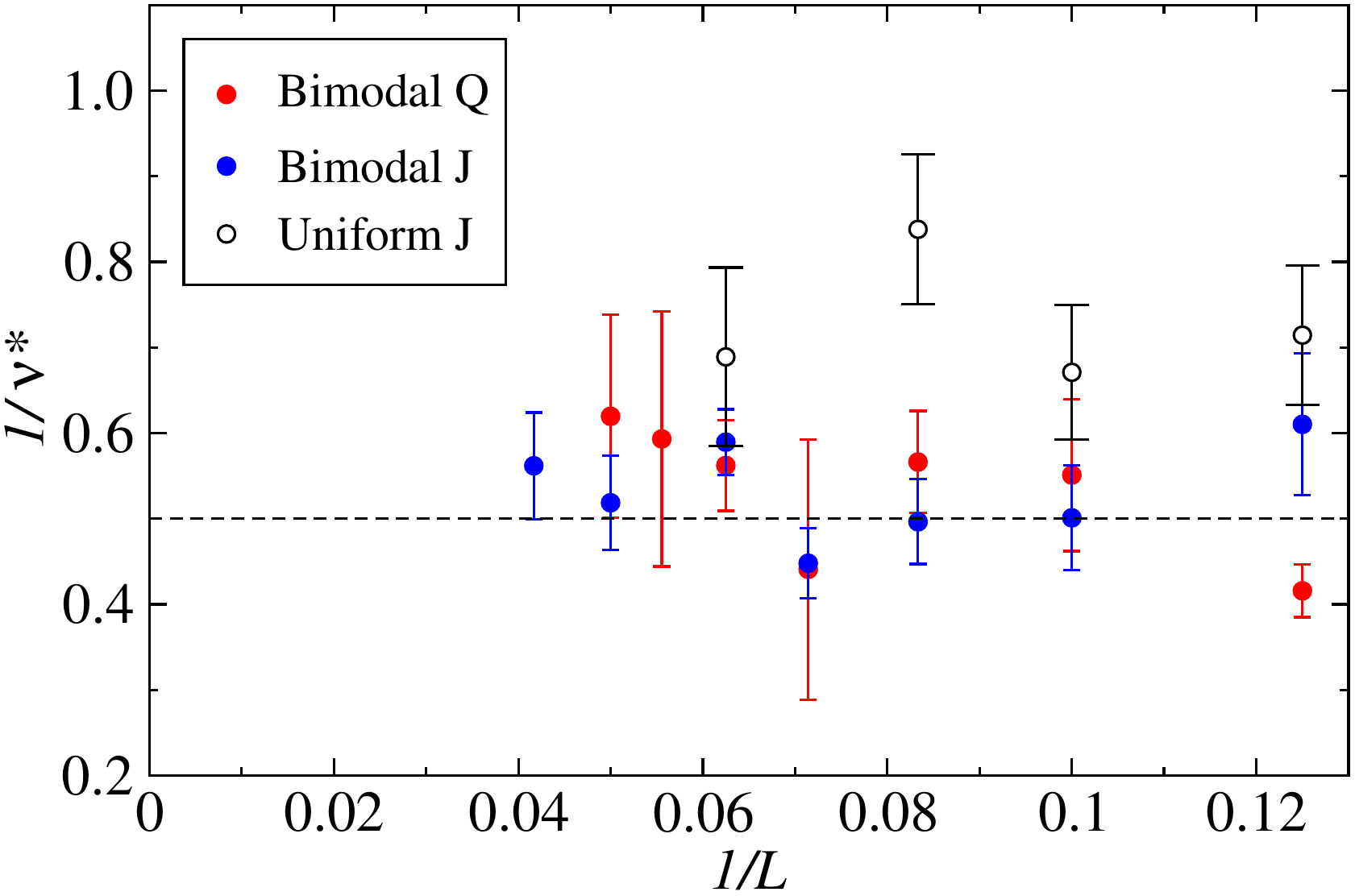}
\vskip-2mm
\caption{Inverse system size dependence of the correlation length exponent defined according to Eq.~(\ref{nustar}). The
disorder distributions are the same as those in Fig.~\ref{universal}. The conjectured exponent $\nu=2$ is indicated by
the dashed line.}
\label{nu}
\end{figure}

In Fig.~\ref{nu} we graph results for $1/\nu^*$ for the same disorder types as in Fig.~\ref{universal}. In the case of the bimodal $J$ and $Q$
distributions the size dependence is weak and the results indicate that $\nu \approx 2$. For the weaker, continuous $J$ distribution the values
of $1/\nu^*$ are overall larger. However, it is possible that only the bimodal distributions represent strong enough disorder for carrying out
reliable extrapolations to infinite size based on the current system sizes, i.e., the corrections to the asymptotic exponent may be larger for
the uniform $J$ distribution. An intriguing possibility is that all systems have $\nu=2$, which is also the universal value of this exponent at
the 2D superfluid to Bose-glass transition \cite{fisher89}, though the symmetries there are different and there is no a priori reason to expect the
exponents to be the same. Further work will be required to test this scenario.

\subsection{Site diluted $J$-$Q$ model}
\label{sec:jqdil}

In the site diluted $J$-$Q$ model, $J$ or $Q$ term in Eq.~(\ref{jqham}) acting on one or more vacancies are excluded from the Hamiltonian. We consider small 
vacancy concentrations $p$ and always remove an equal number of sites on the two sublattices. In the gapped VBS host, when $Q > Q_c$, with $Q_c/J \approx 1.50$ 
\cite{lou09}, we expect the vacancies to act as nucleation centers for VBS vortices \cite{kaul08}. Here no spinon will appear in the VBS vortex core as there
is an empty site. However, with the random distribution of the vacancies there will be local sublattice imbalance, i.e., unequal numbers of vortices on the two 
sublattices; within a group of $n$ vacancies there will be an imbalance of order $\sqrt{n}$ that makes impossible the short-distance pairing of all vacancy
vortices and antivortices. Therefore, additional vortices will form away from the vacancies, and these will source unpaired spins (spinons). There are reasons 
to believe that these spinons cannot be paired up into singlets in the way this happens in the RS state, because of the local imbalance between A and B 
spinons. The short-distance pairing mechanism responsible for the RS state is, thus, missing, and AFM order should form as in the $J_1$-$J_2$ model 
studied in Sec.~\ref{sec:j1j2}.

\begin{figure}[t]
\centering
\includegraphics[width=65mm, clip]{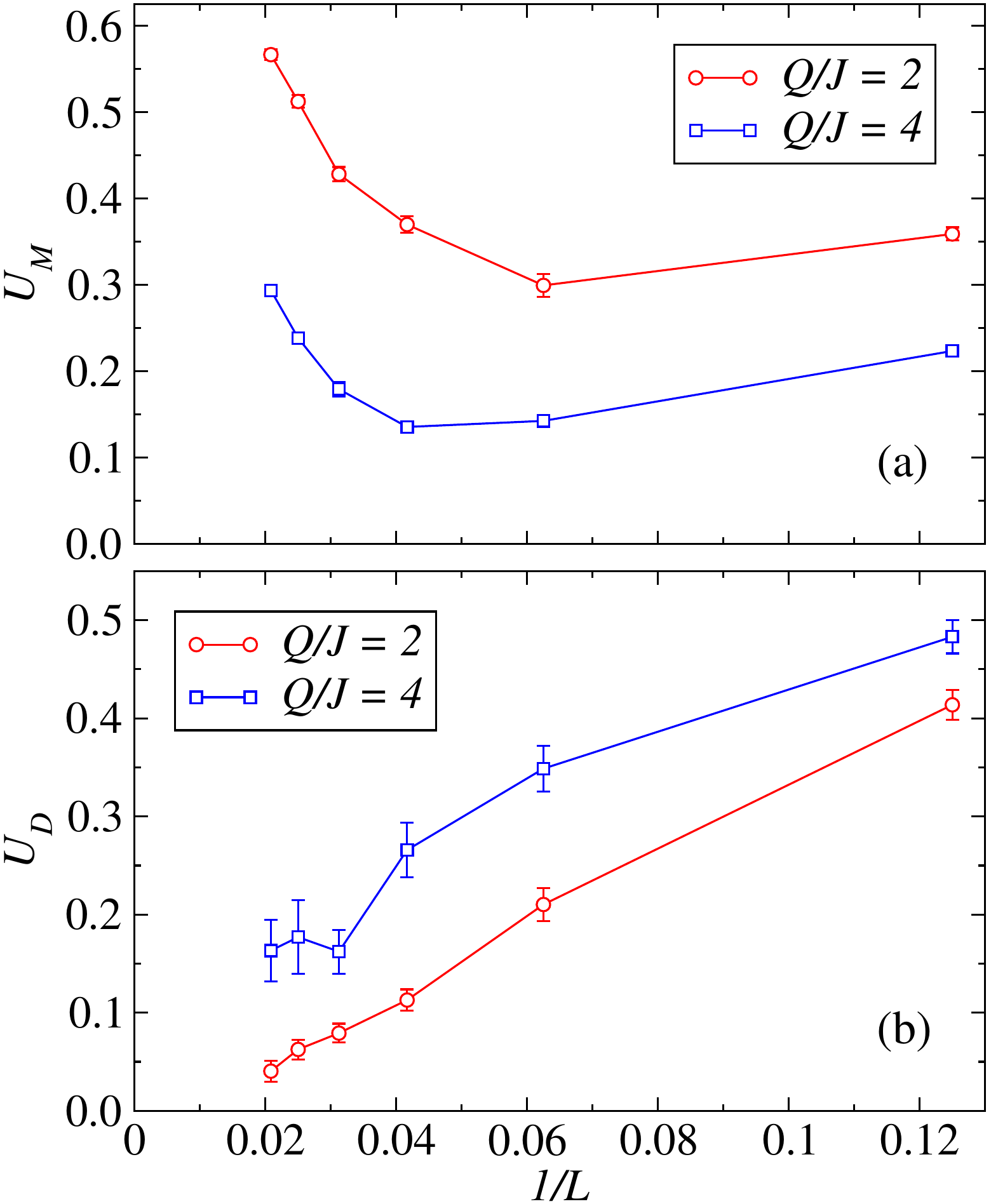}
\vskip-2mm
\caption{Binder cumulants of the AFM (a) and VBS (b) order parameters of the site-diluted $J$-$Q$ model at $Q/J=2$ and $4$,
graphed vs the inverse system size. The dilution fraction is $p=1/32$, with exactly half of the vacancies in each sublattice.}
\label{jqdil}
\end{figure}

Results for $p=1/32$ at two different values of the coupling ratio are shown in Fig.~\ref{jqdil}. Here, in Fig.~\ref{jqdil}(a), we can again see, as we did
in the case of the $J_1$-$J_2$ model in Fig.~\ref{j1j2dil}, how the AFM Binder cumulant first decreases with increasing system size but then starts to grow
when the number of moments becomes sufficient for AFM order to form. This cross-over occurs for larger sizes for the larger $Q/J$ value, which is again
similar to the behavior found for increasing coupling ratio $J_2/J_1$ in the $J_1$-$J_2$ model. Figure \ref{jqdil}(b) shows that the cumulant of the
dimer order parameter approaches zero with increasing $L$, as expected for a VBS breaking up into domains. These results lend support to a phase diagram
of the type in Fig.~\ref{phases}(a), with no phase transition for $\Lambda > 0$, just a cross-over between strong and weak AFM order.

\section{Finite temperature properties and the dynamic exponent}
\label{sec:tfinite}

Finite temperature properties are useful for extracting the dynamic exponent $z$ and may be the most direct route to connect to experiments.
We will here consider the uniform magnetic susceptibility,
\begin{equation}
\chi_{\rm u} = \frac{1}{TN}\langle m_z^2\rangle,~~~ m_z = \sum_{i=1}^N S^z_i,
\label{chiudef}
\end{equation}
and the local susceptibility at location ${\bf x}$ defined by the Kubo integral
\begin{equation}
\chi_{\rm loc}({\bf x}) = \int_0^{1/T} d\tau \langle S^z_{\bf x}(\tau)S^z_{\bf x}(0)\rangle,
\label{chilocdef}
\end{equation}
where $S^z_{\bf x}(\tau)$ is the standard imaginary time-dependent spin accessible in QMC simulations. We here use the SSE method and
refer to the literature, e.g., Ref.~\onlinecite{sandvik10a}, for further technical information. In this section, we average the local
susceptibility over all the sites ${\bf r}$ of the system (as well as over disorder realizations) and call this averaged quantity $\chi_{\rm loc}$.
In Sec.~\ref{sec:jeff} we will show an example of the spatial dependence of $\chi_{\rm loc}({\bf x})$ for a fixed disorder realization.

\subsection{Power-law behaviors}

At a quantum critical point, or in an extended quantum critical phase, since the magnetization is a conserved quantity
the susceptibility, Eq.~(\ref{chiudef}), should scale with the temperature as \cite{fisher89,chubukov94}
\begin{equation}
\chi_{\rm u} \propto  T^{D/z-1},
\label{chiutdep}
\end{equation}
where $D=2$ in our case. In contrast, the local susceptibility, Eq.~(\ref{chilocdef}), is sensitive to the fluctuations of the non-conserved critical order parameter. 
Generalizing the result by Fisher et al.~\cite{fisher89} for a critical point of a disordered boson system (the Bose-glass to superfluid point) to the critical RS phase, 
the mean spin-spin correlation function in imaginary time at zero spatial separation should have the form
\begin{equation}
\langle S^z_{\bf x}(\tau)S^z_{\bf x}(0)\rangle \propto \tau^{-(D+z-2+\eta)/z} = \tau^{-2/z},
\end{equation}
where in the equality we have used our finding that the equal-time correlation function $C(r) \propto r^{-(D+z-2+\eta)}$ \cite{fisher89}
always decays with distance as $1/r^2$, so that $\eta=2-z$. The local susceptibility (\ref{chilocdef}) is then predicted to take the following forms
\begin{equation}
\chi_{\rm loc} = \left \{ \begin{array}{ll} a + b\ln{(1/T)}, & {\rm for}~ z=2, \\ cT^{2/z-1}, & {\rm for}~z>2, \end{array} \right.
\label{chiloctdep}
\end{equation}
with nonuniversal constats $a,b,c$.
Here and in Eq.~(\ref{chiutdep}) it is interesting to note that the uniform and local susceptibilities should take the same divergent form
if $z=2$, while for $z=2$ the logarithmic divergence in $\chi_{\rm loc}$ is not present in $\chi_{\rm u}$, which instead should be
temperature independent (up to possible additive corrections).

\begin{figure}[t]
\centering
\includegraphics[width=65mm, clip]{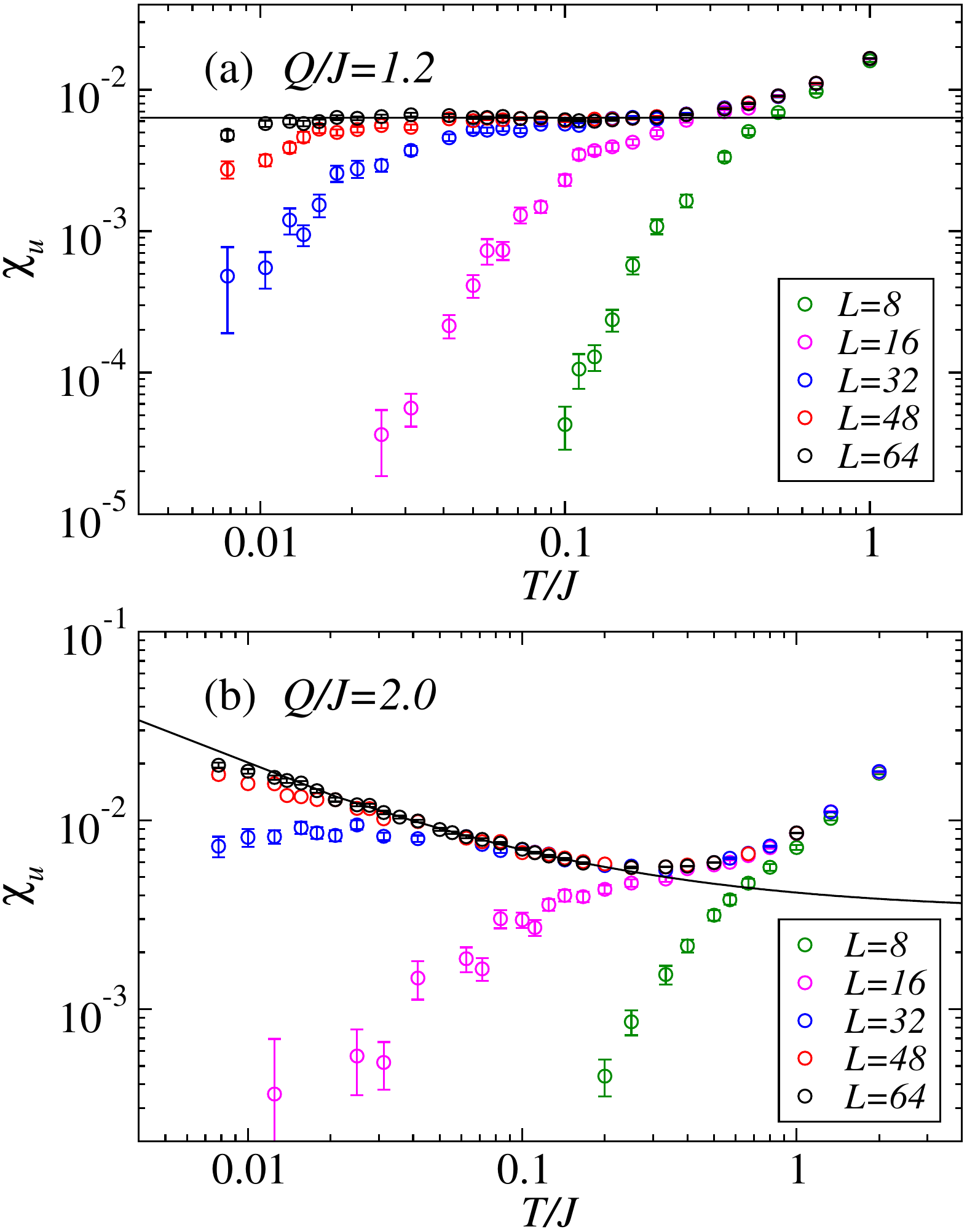}
\vskip-2mm
\caption{Temperature dependence of the uniform susceptibility of the random $Q$ model for several different system sizes. (a) shows results at
$Q/J=1.2$, within the error bars of the estimated AFM--RS critical point (Fig.~\ref{crit}), while the system in (b) is inside the RS phase, at $Q/J=2$. 
The horizontal line in (a) corresponds to the scaling expected if the dynamic exponent $z=2$. The curve in (b) shows a fit of the $L=64$ data to the
form $\chi_{\rm u}=c+bT^{-a}$ with the exponent $a = 0.60 \pm 0.08$, corresponding to $z=2/(1-a) \approx 5$.}
\label{qc_x}
\end{figure}

For the above forms of $\chi_{\rm u}$ and $\chi_{\rm loc}$ to be valid, we not only have to reach sufficiently low in $T$, but also the system 
size has to reach the range where there is no longer any size dependence left. This requirement limits the temperatures we can reach, as demonstrated 
in Fig.~\ref{qc_x} for the case of the uniform susceptibility of the random $Q$ model close to the critical point and inside the RS phase. We can still
clearly observe critical behaviors emerging for a range of low temperatures for the largest system sizes. In Fig.~\ref{qc_x}(a), at $Q/J=1.20$, which should be
very close to the AFM--RS transition according to the results in Fig.~\ref{crit}(b), we find very little temperature dependence (except for the lowest
temperatures, where there are still clearly some effects of finite size), indicating, by Eq.~(\ref{chiutdep}), that $z=D=2$ at the transition.
In principle one should also expect some corrections to the constant behavior, but apparently those are very small in this case.

\begin{figure}[t]
\centering
\includegraphics[width=65mm, clip]{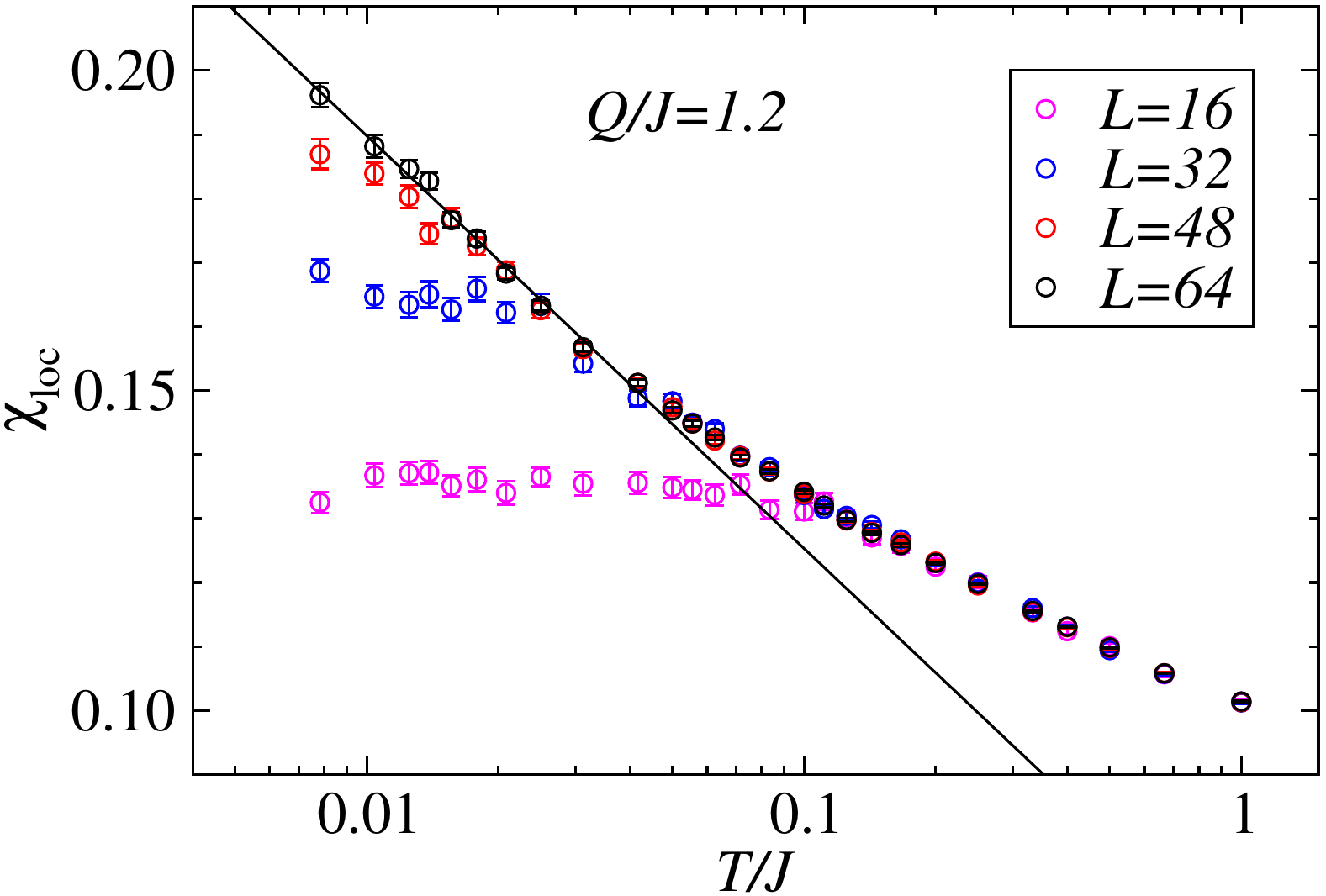}
\vskip-2mm
\caption{Temperature dependence of the local susceptibility of the random $Q$ model at its AFM--RS transition for several different system sizes. 
The line is a fit of the $L=64$ low-$T$ points to the logarithmically divergent form in Eq.~(\ref{chiloctdep}).}
\label{qc_xloc}
\end{figure}

If the dynamic exponent indeed takes the value $z=2$, then according to Eq.~(\ref{chiloctdep}) the local susceptibility should exhibit a 
logarithmic divergence. As shown in Fig.~\ref{qc_xloc}, this indeed appears to be the case. Here we have fitted the low-$T$ behavior to the first 
line in Eq.~(\ref{chiloctdep}), which already contains a constant (unlike the uniform susceptibility in Fig.~\ref{qc_x}, where we included a constant 
as a correction to the leading form). 

\begin{figure}[t]
\centering
\includegraphics[width=65mm, clip]{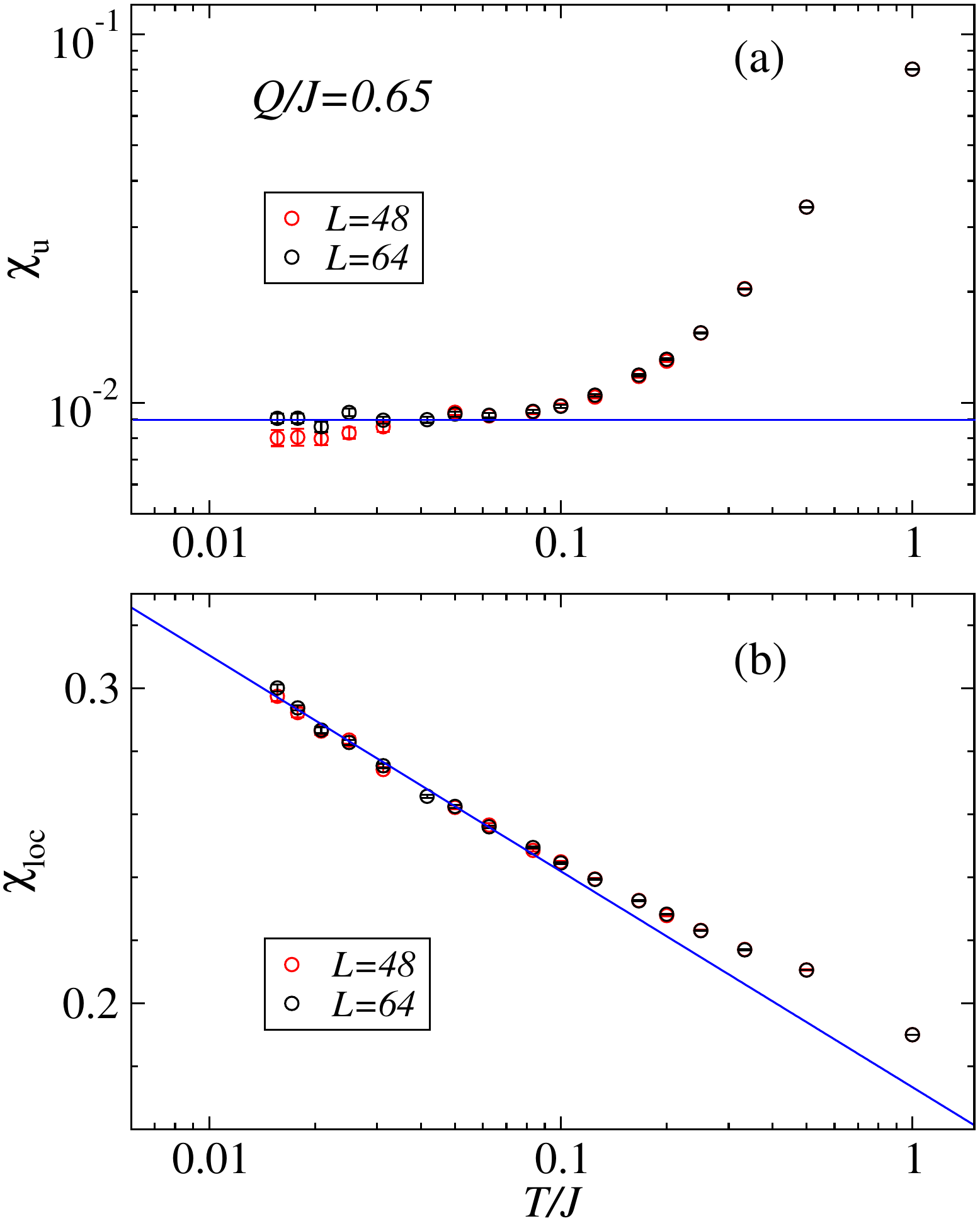}
\vskip-2mm
\caption{Temperature dependence of the uniform (a) and local (b) susceptibility of the bimodal random $J$ model at its estimated critical point
$Q_c/J \approx 0.65$ [see Fig.~\ref{bimodalqc}(b)]. The horizontal line in (a) indicates the size independent behavior expected from Eq.~(\ref{chiutdep}) if $z=2$,
while the line in (b) is a fit to logarithmic form in Eq.~(\ref{chiloctdep}).}
\label{xurjc}
\end{figure}

To test the universality of the scaling of the susceptibilities at the transition, in Fig.~\ref{xurjc} we show results for the bimodal
random $J$ model at its critical point extracted in Fig.~\ref{bimodalqc}(b). We include results for two different system sizes to demonstrate
that the thermodynamic limit should be reproduced for the larger size ($L=64$). We again see a significant low-$T$ regime where $\chi_u$ appears 
to be temperature independent, while the local susceptibility diverges logarithmically, supporting an AFM--RS transition with $z=2$ independently of 
model details.

Well inside the RS phase, at $Q/J=2$ in the random Q model, as shown in Fig.~\ref{qc_x}(b) we find a clearly divergent low-$T$ behavior of $\chi_u$.
Since the overall magnitude of the susceptibility originating from the localized spinons is still not very large at these temperatures, when fitting
to the expected power-law form we also include a constant, as a natural leading correction to the asymptotic divergent form. This works well and the 
value of the exponent given by the fit corresponds to $z\approx 5$. Thus, we find that $z$ increases as the RS phase is entered. 

\begin{figure}[t]
\includegraphics[width=62mm, clip]{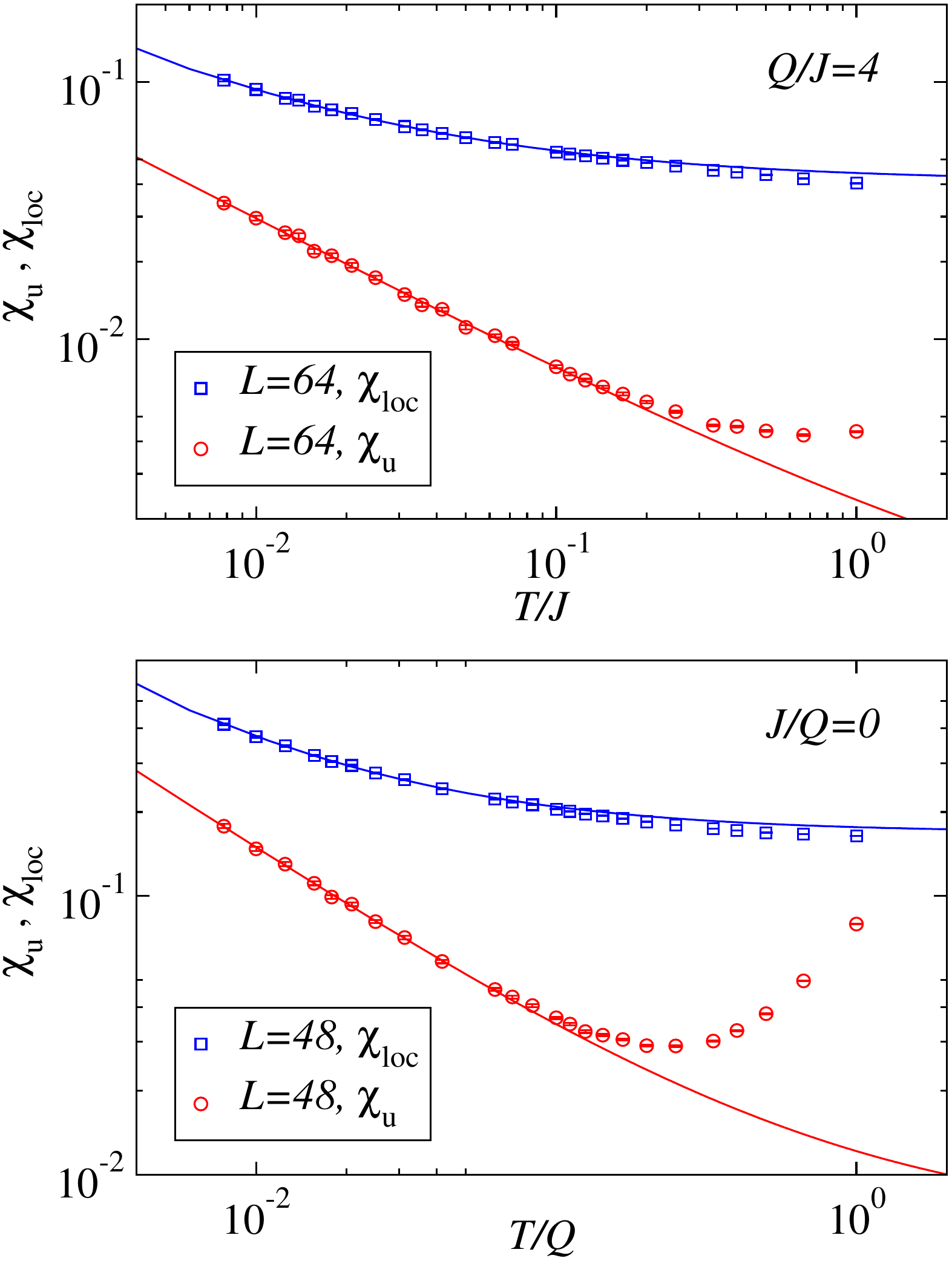}
\vskip-2mm
\caption{Temperature dependence of the uniform and local susceptibilities of the random $Q$ model deep insider the RS phase, at $Q/J=4$ and $J/Q=0$.
The curves show fits of the low-$T$ data to the predicted forms, Eq.~(\ref{chiutdep}) for $\chi_{\rm u}$ and the second line of Eq.~(\ref{chiloctdep}) for 
$\chi_{\rm loc}$, enforcing common exponents for the two quantities. The fits give $1-2/z = 0.60 \pm 0.10$ ($Q=1, J=1/4$) and and 
$0.72 \pm 0.06$ ($Q=1,J=0$).}
\label{qdis_x}
\end{figure}

Figure \ref{qdis_x} shows results even further inside the RS phase, along with fits such as those discussed above. Here we also show results for 
the local susceptibility, which at first sight appears to diverge slower though the ultimate power laws should be the same if $z>2$, according to
Eqs.~(\ref{chiutdep}) and (\ref{chiloctdep}). However, with independent constant corrections added, both $\chi_{\rm u}$ and  $\chi_{\rm loc}$ can be fitted 
with the same power laws (using joint fits). These fits give the exponent $1-2/z = 0.60 \pm 0.10$ and $0.72 \pm 0.06$ for $Q=1, J=1/4$ and $Q=1,J=0$, 
respectively, i.e., the dynamic exponents are $z\approx 5$ at $Q/J=4$ and $z\approx 7$ when $Q/J\to \infty$. 

In the case $J=0$, it should be noted that the bimodal disorder distribution that we have used here, where half the $Q$ couplings are set to zero, can 
lead to isolated spins that contribute $\propto 1/T$ to the susceptibility. However, we have avoided this issue by ``patching'' such rare isolated spins 
by adding a randomly chosen $Q$ interaction for each of those rare sites to connect them to the rest of the system. The differences between the patched and 
original systems are barely noticable in the disorder averaged quantities. 

The larger role of the correction (the constant terms used in the fits in
Fig.~\ref{qdis_x}) in $\chi_{\rm loc}$ than in $\chi_{\rm u}$ can likely be traced to the fact that the local susceptibility only contains a small 
fraction of the dominant staggered response at $q=(\pi,\pi)$ in momentum space, and therefore one may expect large corrections from all the other momenta, 
at which the response is weaker. An alternative way to detect the dominant dynamic response, but that we have not yet pursued, would be to compute the 
susceptibility at $q=(\pi,\pi)$. 

\subsection{Griffiths-McCoy singularities}

To properly classify the proposed RS state, we need to consider the fact that disordered systems generically have regions in parameter space called
Griffiths, or Griffiths-McCoy, phases. These phases or regions are characterized by spatial 'commingling' of two phases \cite{griffiths69,mccoy69}.
Fluctuations in the quenched disorder can favor a phase B within a limited part of a system that is overall in a phase A. Griffiths phases, which do not
always have well-understood RG fixed-point analogues (but sometimes they do \cite{grinstein76}), appear close to critical points and are
normally associated with weaker singularities than the actual critical points (for reviews, see Refs.~\onlinecite{vojta10} and \onlinecite{vojta13}).
The singularities arise from exponentially rare regions (e.g., large domains of phase B inside phase A) and have the most profound effects on dynamical
properties. 

In quantum systems, Griffiths phases typically have large but finite dynamic exponents, with associated divergent susceptibilities if $z>D$.
The large $z$ values (long time scales) motivate the often used term ``glass'' for these phases, though a Griffiths phase is not
normally associated with the multitude of thermodynamic states (by replica symmetry breaking and related phenomena) of classical and quantum spin
glasses (and it was also claimed that the valence-bond glass state undergoes replica symmetry breaking \cite{tarzia08} but this may be a consequence of a
classical treatment). Examples of Griffiths phases include the Bose glass in the Bose-Hubbard model with random potentials \cite{fisher89} and the
Mott glass in particle-hole symmetric boson systems where randomness is introduced in the hopping (and there are indications that this state
can also form with random potentials due to emergent particle-hole symmetry \cite{wang15}). The spin analogue of particle-hole symmetry is also present
in 2D random exchange Heisenberg antiferromagnets, where Mott-glass phases have been identified \cite{roscilde07,ma14}.

An important question is whether the RS state we identify in the random $J$-$Q$ model is also a Griffiths phase. We argue that it is not, because
equal-time correlations in Griffiths phases should decay exponentially with distance (a fundamental consequence of the rare-region mechanism), while we
have fond strong evidence for power-law decaying correlations.

There is a further strong argument against the RS phase being a Griffiths phase: If, in the language above, we consider the AFM as phase A, there is no obvious
phase B with which A can commingle to form the RS phase as a Griffiths phase. The RS phase is then actually that phase B, and in principle Griffiths singularities
could appear due to comingling of the AFM and RS phases close to the phase boundary. However, since the AFM and RS phases are both gapless, the Griffiths
singularities would be very hard to detect and would very unlikely be responsible for the power laws we have identified here. Most likely, they would only
cause scaling corrections and no separately identifiable Griffiths phase in addition to the AFM and RS phases.

RS states and Griffiths phases have been contrasted in detail in Heisenberg chains \cite{hyman96}.
In a chain with the same disorder distribution
on all links, the RS state forms generically. However, if there is furthermore an alternating strength of the mean couplings (static dimerization), in which
case the pure system is gapped and has exponentially decaying correlations, a critical disorder strength is required to induce the RS state. The weakly
disordered system is in a Griffiths phase, where rare RS regions in the otherwise gapped system imply gapless excitations of the system 
However, the spin correlations in this phase remain exponentially decaying. The RS state itself is not a Griffiths phase.

\section{Spinon interaction mechanism}
\label{sec:jeff}

The way the localized spinons interact with each other is a crucial ingredient in the formation of the RS state. In order for singlets to be gradually
``frozen out'' as the energy scale is reduced (as in the Ma-Dasgupta strong-coupling RG procedure \cite{dasgupta80,bhatt82}), and for AFM order not to form
on large length scales, in 2D it seems necessary that the spinon-spinon interactions are not completely random. This is evidenced by the fact that 2D SDRG
calculations on $S=1/2$ systems with various coupling distributions
have so far not been able to generate a phase similar to the RS phase identified here \cite{lin03,lin06}. In the $J$-$Q$ model, the observation that
spinons are created in pairs (spinons and antispinons) when VBS domains are formed already implies a correlation that favors closer typical distance between
a spinon and the nearest antispinon, because the domain walls will provide an effective potential due to domain-wall energy between a spinon and anti-spinon site
connected by a wall. There is, however, potentially also another effect, namely, the effective magnetic interactions between the spinons are
likely mediated mainly through the domain walls. The putative role of domain walls as mediators of spin correlations was mentioned in Ref.~\onlinecite{kimchi17}
but was not developed into an actual mechanism suppressing AFM order and causing the singlet formation in the RS state. Here we will provide evidence
for such a mechanism within our models on the square lattice. We note that the effective interactions should have the same bipartite nature as the
microscopic interactions in the pure system, as was discussed generically in Ref.~\cite{kimchi17} (and earlier in specific cases, e.g., in
Ref.~\cite{laflorencie04}).

\begin{figure}[t]
\includegraphics[width=45mm, clip]{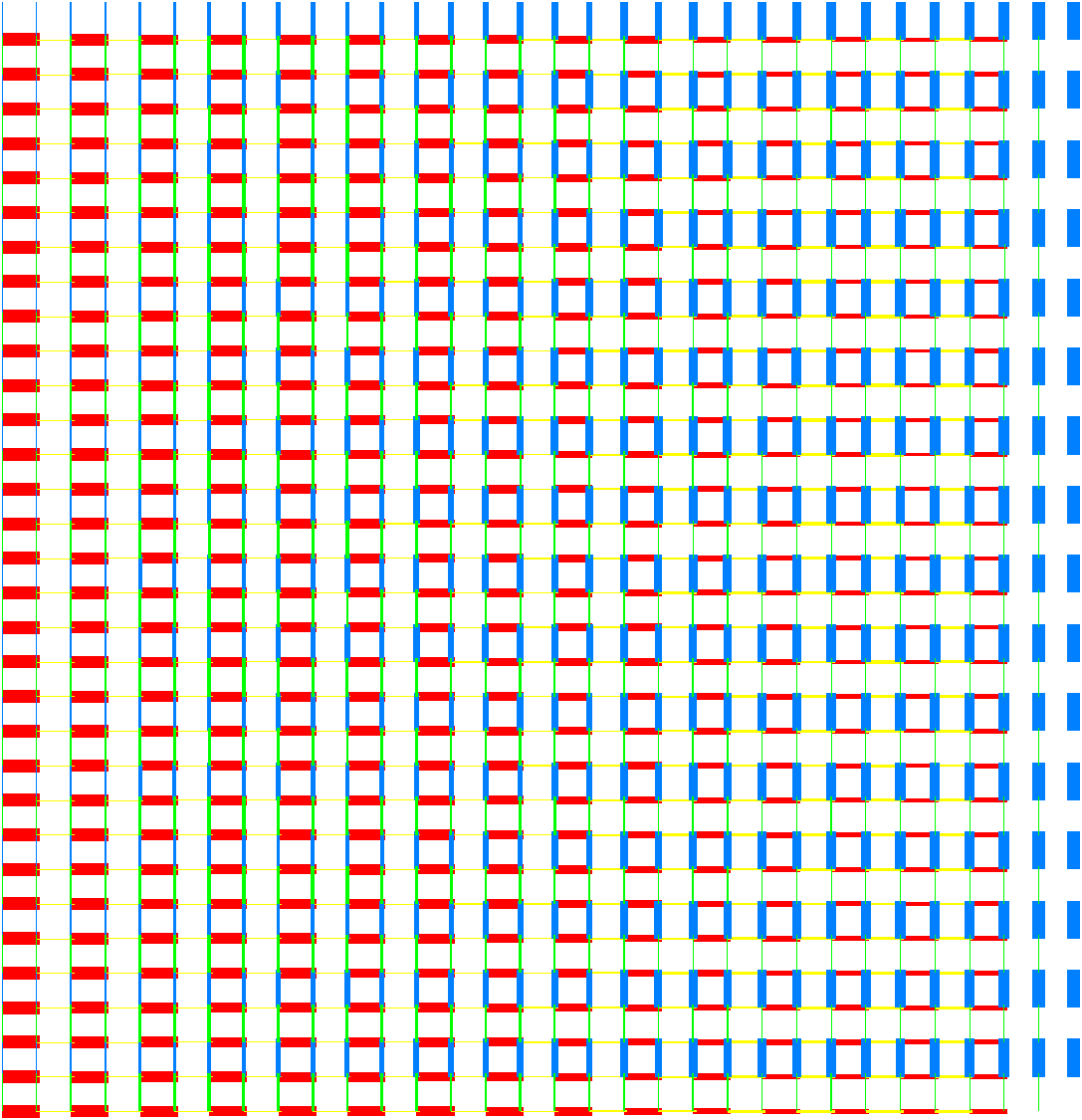}
\vskip-2mm
\caption{Domain wall on a $32\times 32$ lattice for a system with $J=0$, $Q=1$. The bonds are colored according to the convention in Fig.~\ref{spinon}
and the line thickness represents the expectation value $-\langle {\bf S}_i \cdot {\bf S}_j\rangle$. The left and right open boundaries have been
modified at the Hamiltonian level to lock in VBS patterns differing in angle by $\Delta\phi=\pi/2$. Note that a $2\times 2$ plaquette with equal
correlation on all edges, seen in the middle of the system, corresponds to $\phi=\pi/4$.}
\label{dw}
\end{figure}

\subsection{Uniform domain wall}

First, let us consider a uniform domain wall in the pure $J$-$Q$ model in its VBS state. According to the DQC theory \cite{senthil04a}, the thickness of a
domain wall between VBS domains, across which the angle $\phi$ defined in Fig.~\ref{spinon} changes by $\Delta\phi=\pi/2$, is not governed by the standard
correlation length $\xi$, but by a longer length scale $\xi'$ (i.e., this length diverges faster than $\xi$ as the DQC point is approached). This affects the
scaling of the energy density of the domain wall as the critical point is approached \cite{senthil04b,shao15}, which may also have a counterpart
at the AFM--RS transition. We here only mention this and do not explore the domain wall thickness further. Instead we discuss the spin gap of a domain
wall, i.e., the energy difference between the $S=0$ ground state and the lowest $S=1$ state in a system with a domain wall imposed by boundary
conditions.

Figure \ref{dw} shows an example of a domain wall, where the
bond thickness on a $32\times 32$ lattice corresponds to the magnitude of the spin correlation on that bond, and the colors of the bonds are
coded as in Fig.~\ref{spinon}. The boundary conditions are periodic in the vertical direction but in the horizontal direction the interactions
have been modified (see Ref.~\onlinecite{shao15}) so that the edges are locked into VBS realizations differing by the angle $\Delta\phi=\pi/2$ of an
elementary domain wall. Here it should be noted that the length scale over which the angle $\phi$ changes in Fig.~\ref{dw} is not the intrinsic domain
wall width, because the location of the wall also has quantum fluctuations that smear it out when expectation values are computed. The spin gap of the wall
is still a completely well defined quantity, as long as the $S=1$ excitation (observed, e.g., with the spinon strings illustrated in Fig.~\ref{strings})
is not repelled from the wall. We have confirmed that the excited spin is attracted to the domain wall (which by itself implies, by energy minimization, 
a smaller gap on the wall than in the bulk VBS away from the wall).

\begin{figure}[t]
\includegraphics[width=65mm, clip]{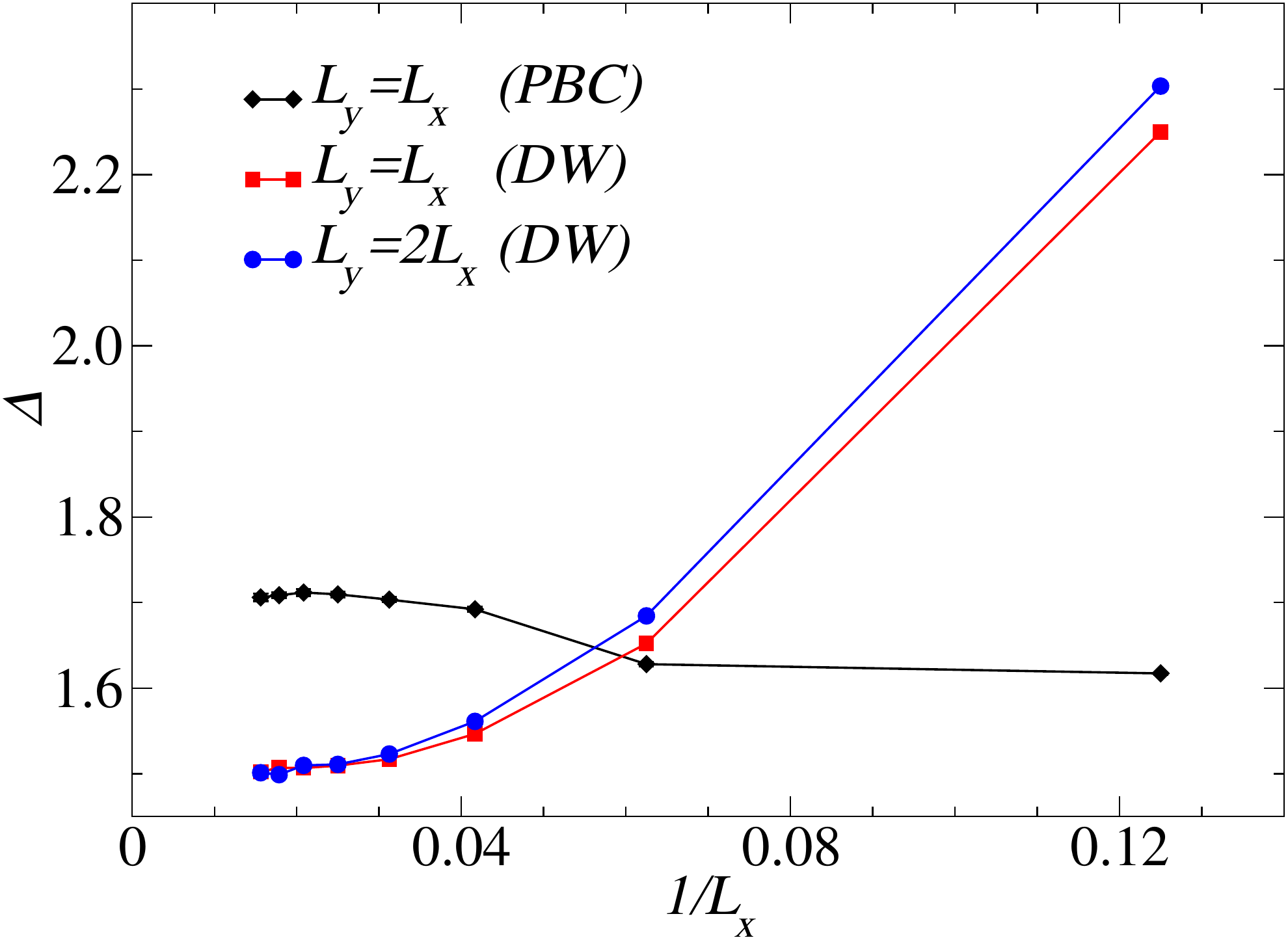}
\vskip-2mm
\caption{Singlet-triplet gaps of the non-random system at $J=0$, $Q=1$, with and without a domain wall.
  The bulk VBS gap corresponds to the $L \to \infty$ limit of the
results for system with periodic boundaries. For systems with a domain wall (illustrated in Fig.~\ref{dw}) induced by modified open boundaries
in the horizontal direction, two different aspect ratios $L_y/L_x$ are considered as a check of a unique intrinsic gap of the domain wall
when $L_x \to \infty$.}
\label{gap}
\end{figure}

The spin gap is obtained by simply taking the difference between total ground state energies computed in the two spin sectors. Fig.~\ref{gap} shows
results for the uniform system without domain wall (obtained with fully periodic $L \times L$ lattices) and with domain walls on lattices with two
different aspect ratios, as a test of the expected independence on the lattice geometry when $L \to \infty$. For small systems with a wall,
the gap is strongly influenced by the boundary modifications, which here extend three rows into the system on each side, and one should not draw
any conclusions on the differences between the system with and without the domain wall until $L$ is much larger and the wall has converged to its
intrinsic thickness. For large $L$, it is clear that the gap on the domain wall, $\Delta/Q \approx 1.5$, is significantly smaller than in the bulk,
$\Delta/Q \approx 1.7$, as one might have expected just from the fact that the domain wall has weaker order, i.e., more fluctuations, than the bulk VBS.
Thus, in a non-random system, a domain wall will be a more effective mediator of correlations, and thereby of effective interactions between
impurity spins, than in the bulk VBS. 

\subsection{Local susceptibility}

The above results for a pure infinitely long domain wall should only be taken as suggestive of enhanced spinon interactions along domain walls
in the disordered system. We can obtain further evidence by examining the spatial variations of the local susceptibility, Eq.~(\ref{chilocdef}),
for individual disorder realizations (see Ref.~\cite{wang10} for similar calculations for a diluted Heisenberg system). A large susceptibility can be
taken as a sign of a small local gap, through the sum rule (here written only for $T=0$)
\begin{equation}
\chi_{\rm loc}({\bf r}) = 2\int_0^\infty d\omega \omega^{-1}S_{\rm loc}({\bf r},\omega),
\label{chilocint}
\end{equation}
where $S_{\rm loc}({\bf r},\omega)$ is the local dynamic spin structure factor, which satisfies another sum rule,
\begin{equation}
\int_0^\infty d\omega S_{\rm loc}({\bf r},\omega)=S_{\rm loc}({\bf r})=\langle S^z_{\bf r}S^z_{\bf r}\rangle = \frac{1}{4}.
\end{equation}
For any finite system, the spectral weight in $S_{\rm loc}({\bf r},\omega)$ does not extend all the way down to $\omega=0$, and in a single-mode
approximation, where there is only a single $\delta$-function at $\omega=\Delta$, we can extract the local gap as
$\Delta({\bf r}) = 2 S_{\rm loc}({\bf r})/\chi_{\rm loc}({\bf r}) = [2\chi_{\rm loc}({\bf r})]^{-1}$. In the realistic case where there is a broader
distribution of spectral weight, $\chi_{\rm loc}({\bf r})$ can still be regarded as a proxy for the typical local low-energy scale, and it should
then also be a measure of the local ability of a region of the system to mediate effective spin-spin interactions.

\begin{figure}[t]
\centering
\includegraphics[width=80mm, clip]{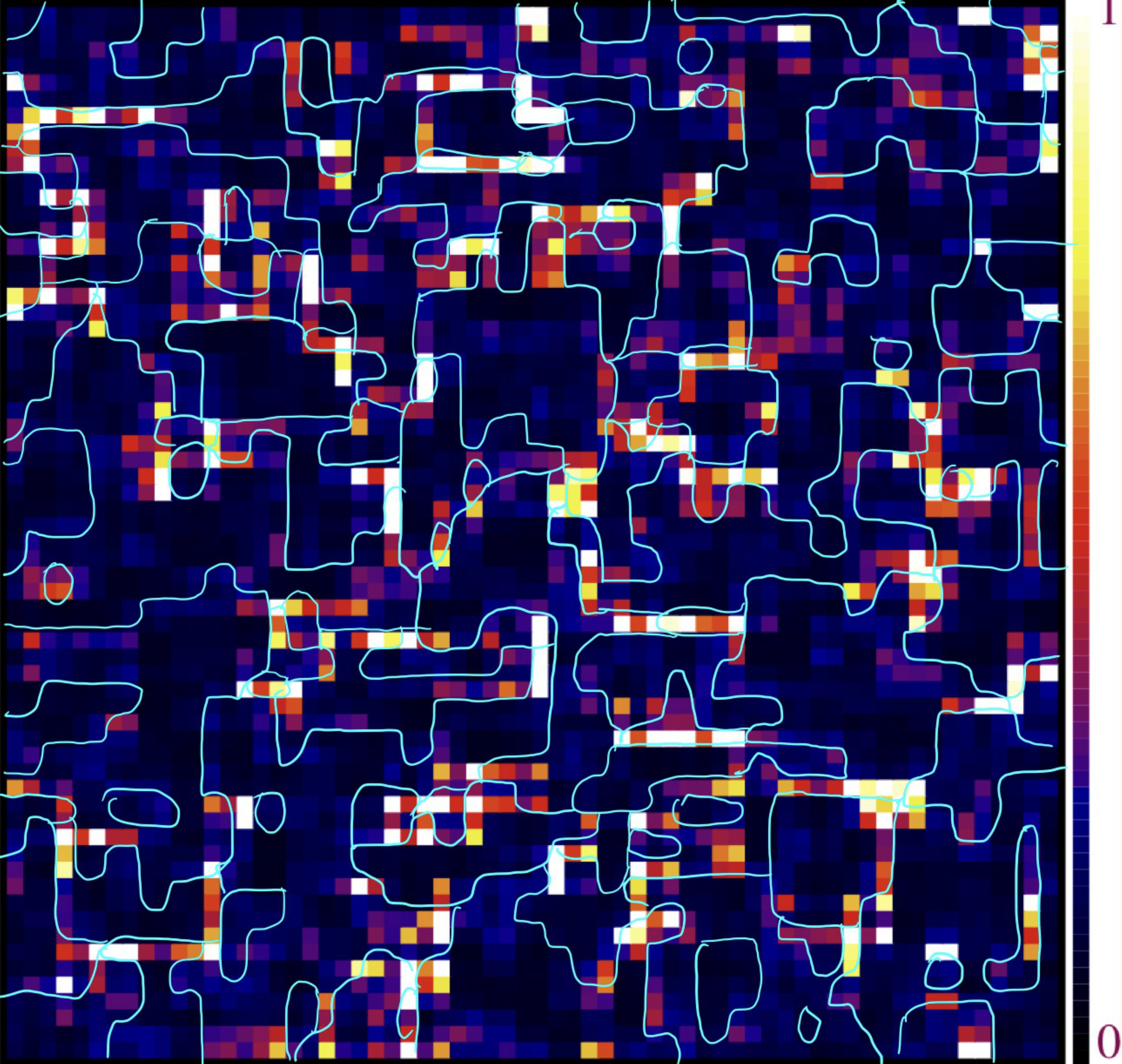}
\caption{Visualization of the local susceptibility for the same coupling realization of the random $Q$ model for which the VBS domains are
illustrated in Fig.~\ref{vbsdom} and with the same hand-drawn domain boundaries (turquoise curves). The values of the susceptibility defined
in Eq.~(\ref{chilocdef}) have been rescaled so that the maximum is $1$, and the color coding is shown on the bar on the right side. Note that,
in some cases a nexus of four domain wall is ``dark'' because it does not correspond to a spinon. This can be seen by comparing with Fig.~\ref{vbsdom},
where it can be seen that the cases in question only involve two different domains (thus no spinon is present).}
\label{loc}
\end{figure}

In Fig.~\ref{loc} we show the spatial dependence of the local susceptibility for the same $Q$ disorder realization as in the illustration
of VBS domains in Fig.~\ref{vbsdom}. Several bright spots on the susceptibility map can be observed, and many of them can be matched with
meeting points of four VBS domain walls, where spinons should localize. Naturally, the sites on which the spinons reside should have enhanced
susceptibility (and note that a single spinon will be spread out over several sites due to quantum fluctuations). There are also bright regions
in Fig.~\ref{loc} along many of the domain walls, while in the bulk of large VBS domains there are no bright spots. These observations support
the notion that the domain walls act as mediators of spinon-spinon interactions, which should play an important role in the formation of the RS state.

\subsection{Spinon strings}
\label{subsec:strings}
           
As seen in Eq.~(\ref{chilocint}), the local static susceptibility represents an inverse-frequency weighted average over a local dynamic
response function. By selection rules the relevant excited states have total spin $S=1$. We can also access specifically the lowest $S=1$ state,
by projector QMC simulations in the extended valence-bond basis with two unpaired spins, as discussed in Sec.~\ref{sec:spinonstring}. Previously we
used the mean length of the spinon strings in this basis (see Fig.~\ref{strings}) as a means to detect the AFM--RS transition. The strings can also
provide spatial information in the form of the site-dependent string density, which for a string with $n$ sites in a given configuration (transition graph) 
is $\rho_r=1/n$ for sites $r$ covered by the string and $0$ else \cite{stringnote}. Averaging over the simulation we obtain the mean 
string density $\rho(x,y)$, which should reflect the spatial structure of the lowest $S=1$ wave function. A quantity similar to the string density
was previously studied in a site-diluted Heisenberg model at the percolation point, and it gave useful information on the nature of 
quasi-localized moments in that case \cite{wang10}.

In the picture we have outlined for the RS state, the lowest $S=1$ state should be formed mainly from the localized $S=1/2$ spinons, corresponding
to breaking some of the singlets formed among groups of spinons. In a large system, the lowest excitation may not involve all the spinons, but
in the relatively small systems we can access here there is typically some string density in all regions identifiable as spinons. We expect a
given string to be mainly confined to a region corresponding to a localized spinon, but the strings will also migrate between spinons (those
involved significantly in the lowest $S=1$ state), and this should lead to elevated string density also on the domain walls. This migration of
spinons should take place mainly within the same sublattice, i.e., a sublattice-X spinon, X$\in\{{\rm A,B}\}$, will migrate predominantly between
spinons located on the X sublattice (with interesting violations of this rule to be discussed in Sec.~\ref{dynamicspinon}). Fluctuations of the
strings between nearby A and B spinon regions should also take place along domain walls connecting them, reflecting the anticipated role of
the domain walls in mediating antiferromagnetic spinon-spinon interactions.

\begin{figure}[t]
\centering
\includegraphics[width=60mm, clip]{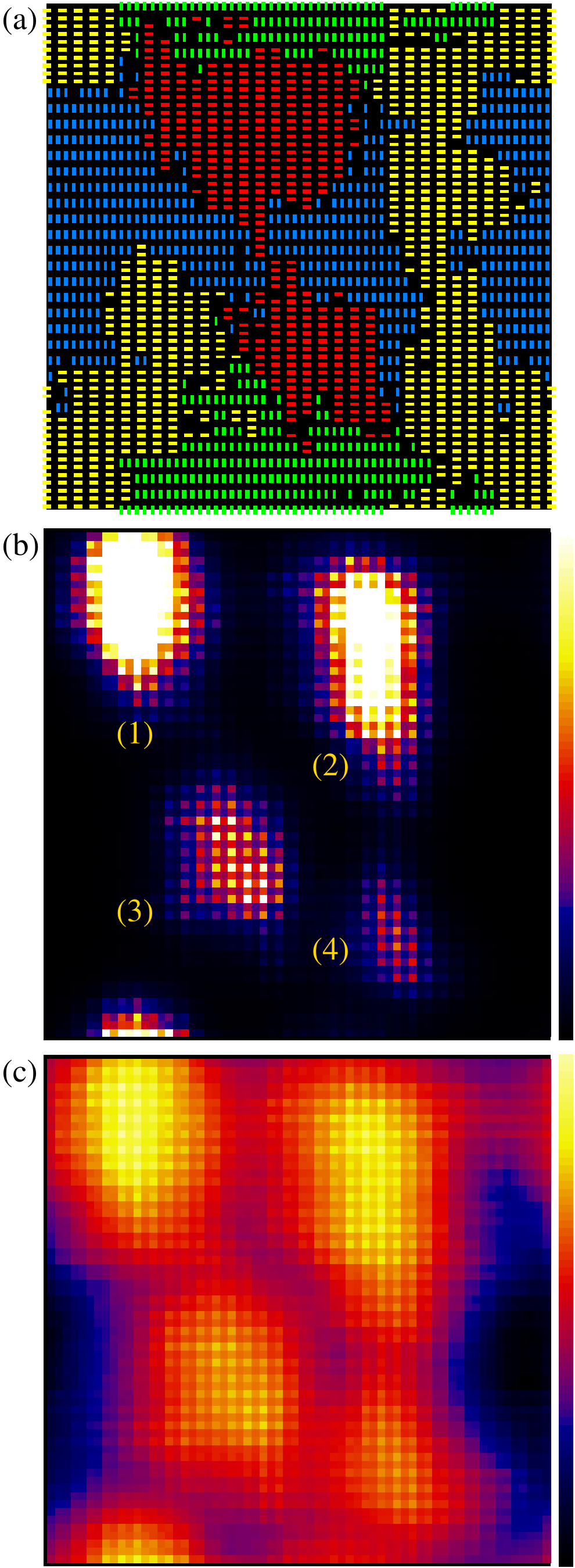}
\caption{Example of domain structure and spinon string density for an instance of the bimodal random-$J$ model. In (a) the bonds are
color coded as in Fig.~\ref{vbsdom}. The mean spinon string density $\rho(x,y)$ is graphed on a linear scale in (b), with
the color bar corresponding to $\rho \in [0,0.002]$. The actual maximum value is $\rho\approx 0.017$, but all  $\rho >0.002$ have 
been shown as white in order to make the weaker features better visible. In (c), the color scale is for $\ln(\rho) \in [-15,-3]$ 
(the full range of the computed values). The labels 1-4 for the spinon regions in (b) are referenced in Fig.~\ref{states4}.}
\null\vskip-3mm
\label{all3}
\end{figure}

In order to have a clear example of the string density in regions of spinons and domain walls, we show a case of the random-$J$ model with a small
number of spinons in Fig.~\ref{all3}. Here the domains are larger than in the previous random-$Q$ instance considered (Figs.~\ref{vbsdom} and \ref{loc})
and in Fig.~\ref{all3}(a) one can clearly identify four cases of meeting points of four domain walls. Accordingly, in Fig.~\ref{all3}(b) there
are four islands of high string density (where the periodic boundary conditions should be noted). The two upper islands have much
higher integrated string density than the other two, indicating that the lowest excitation mainly corresponds to breaking up a singlet formed between
those two spinons. 

On the linear color scale in Fig.~\ref{all3}(b) one cannot easily detect any structure corresponding to the
domain walls. When considering instead a logarithmic scale, Fig.~\ref{all3}(c), we can see channels formed in the regions corresponding to domain
walls in Fig.~\ref{all3}(a), though these channels are clearly much more spread out than the domain walls visualized as in Fig.~\ref{all3}(a). Note again,
however, that the representation in Fig.~\ref{all3}(a) is based on mean values in which the spatial fluctuations of the domain walls are not apparent.
There are domain walls also within the completely dark regions, but those walls do not represent short paths connecting spinons.
The spinon-spinon interactions should be carried mainly along the shortest domain walls connecting spinons.

In Figs.~\ref{all3}(b,c), checker-board patterns can be observed in the spinon regions. The pattern arises from the sublattice imbalance of the
strings, which always cover an odd number of sites and the end points of which always stay on the same sublattice. Thus, the string density originating
from an X-sublattice spinon, ${\rm X} \in \{{\rm A},{\rm B}\}$, is higher on the X sublattice. 

\subsection{Dynamic spinons}
\label{dynamicspinon}

\begin{figure}[t]
\centering
\includegraphics[width=75mm, clip]{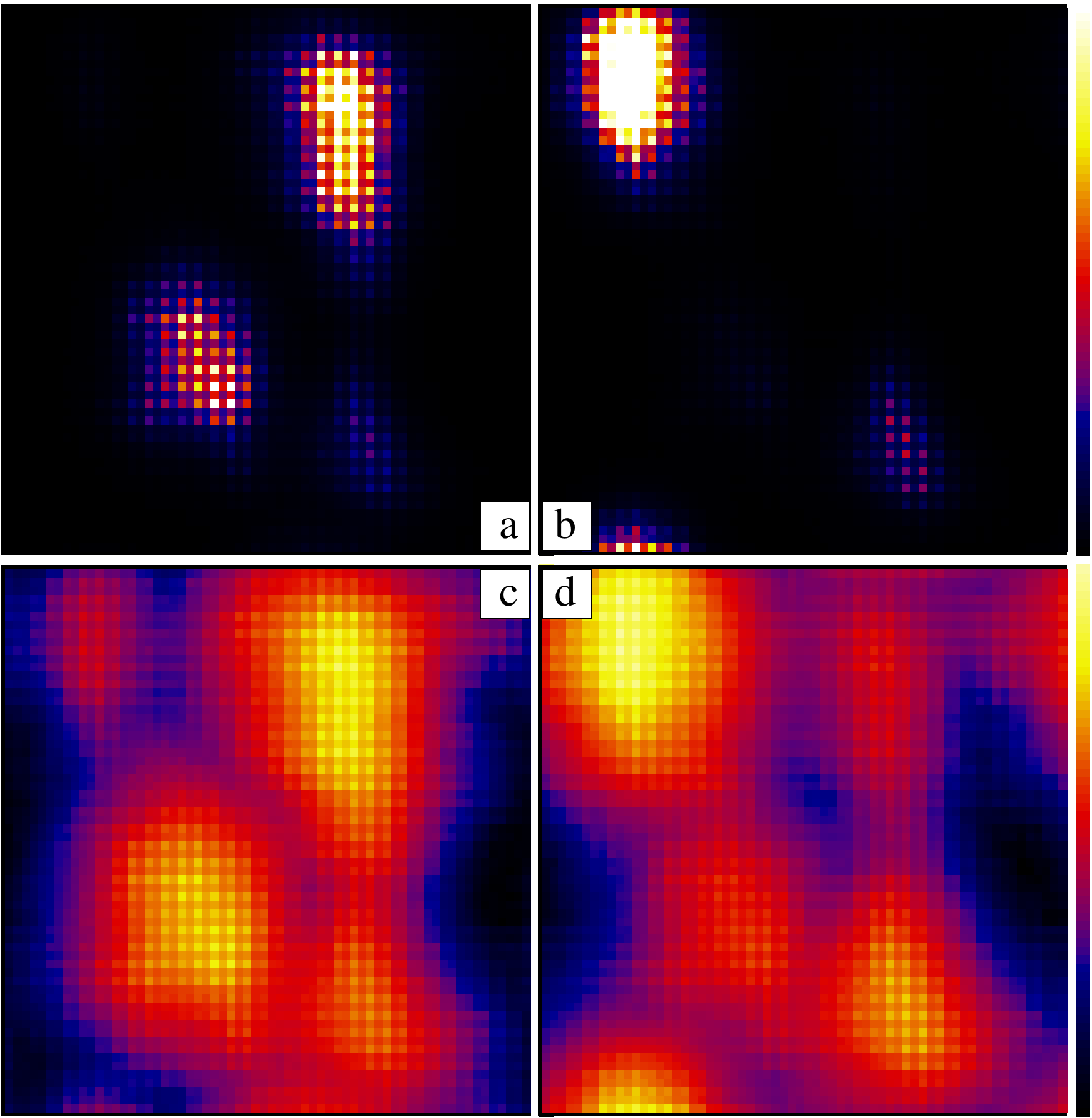}
\caption{The total string density $\rho$ in Fig.~\ref{all3}(b) resolved into contributions from the individual A [$\rho_A$ in (a) and (c)] and 
B [$\rho_B$ in (b) and (d)] strings. The scale is linear in (a) and (b), spanning the range $[0,0.003]$ (and white is used for larger values, 
up to $0.0097$ for $\rho_A$ and $0.017$ for $\rho_B$). In (c) and (d) the same data are shown on a logarithmic scale, with the color bar 
corresponding to $\ln(\rho_A),\ln(\rho_B) \in [-15,-3]$.}
\label{dst}
\end{figure}

The string density shown in Figs.~\ref{all3}(b,c) is the total density, with equal contributions from the A and B sublattice strings.
Figure \ref{dst} shows the A and B densities separately. Here we can see clearly that the A and B strings are attracted predominantly, but not
exclusively, to different regions of the lattice. The small but noticable coexistence of the two spinon strings within the same regions requires
dynamical aspects of the spinons that are not captured within the completely static picture of the localized spinons that we have had in mind
throughout the discussion of the results so far. In particular, the results in Fig.~\ref{dst} imply that one cannot consider a region of high
string density as occupied by either a spinon or an antispinon, but there is some degree of both spinon and antispinon in each vortex
region. 

As apparent from the basic
picture of a spinon in Fig.~\ref{spinon}, we can attach sublattice labels A and B to spinons and antispinons, respectively, as we have
frequently done. Thus, a spinon can be characterized by two labels, the spin-$z$ component $\sigma \in {\uparrow,\downarrow}$ as well as
the sublattice label ${\rm X} \in \{{\rm A},{\rm B}\}$ (spinon, antispinon). If we consider two regions, $1$ and $2$, in which spinons
can exist, a singlet can be written as
\begin{eqnarray}
  \Psi_S = && a_{\rm R}(\uparrow_{1{\rm A}}\downarrow_{2{\rm B}}-\downarrow_{1{\rm A}}\uparrow_{2{\rm B}}) \\
           &&~~~+ a_{\rm C}(\uparrow_{1{\rm B}}\downarrow_{2{\rm A}}-\downarrow_{1{\rm B}}\uparrow_{2{\rm A}}), \nonumber
\label{twospinonpsi}
\end{eqnarray}
where $\uparrow_{1A}$ means that the spinon in region $1$ is on sublattice A with spin $\uparrow$, etc., and we demand that the two spinons
cannot simultaneously occupy the same region. If region $1$ is predominantly occupied by the spinon (i.e., it sits on sublattice A) and region 2 is
occupied by the antispinon, then we consider the first term as the regular (R) term and the second (C) term is the cross term, and
$|a_{\rm R}| \gg |a_{\rm C}|$. Naturally it is the details of the disordered couplings that determine which of the terms is the regular one; essentially
as in the static spinon picture. The static description implies that $a_{\rm C}=0$, but what we see in Fig.~\ref{dst} is that cross terms
in fact must exist, not only in the simple two-spinon system but also in systems with more than two spinons. Thus, it is not
completely correct to say that disorder leads to localized spinons of type A or B as dictated by the VBS votices and antivortices, but the physics
is more complicated and involves quantum fluctuations that exchange not only the spins forming singlets but also the sublattice labels of the
different spinon regions, with the constraint of a total spin singlet and equal numbers of $A$ and $B$ labels.

\begin{figure}[t]
\centering
\includegraphics[width=60mm, clip]{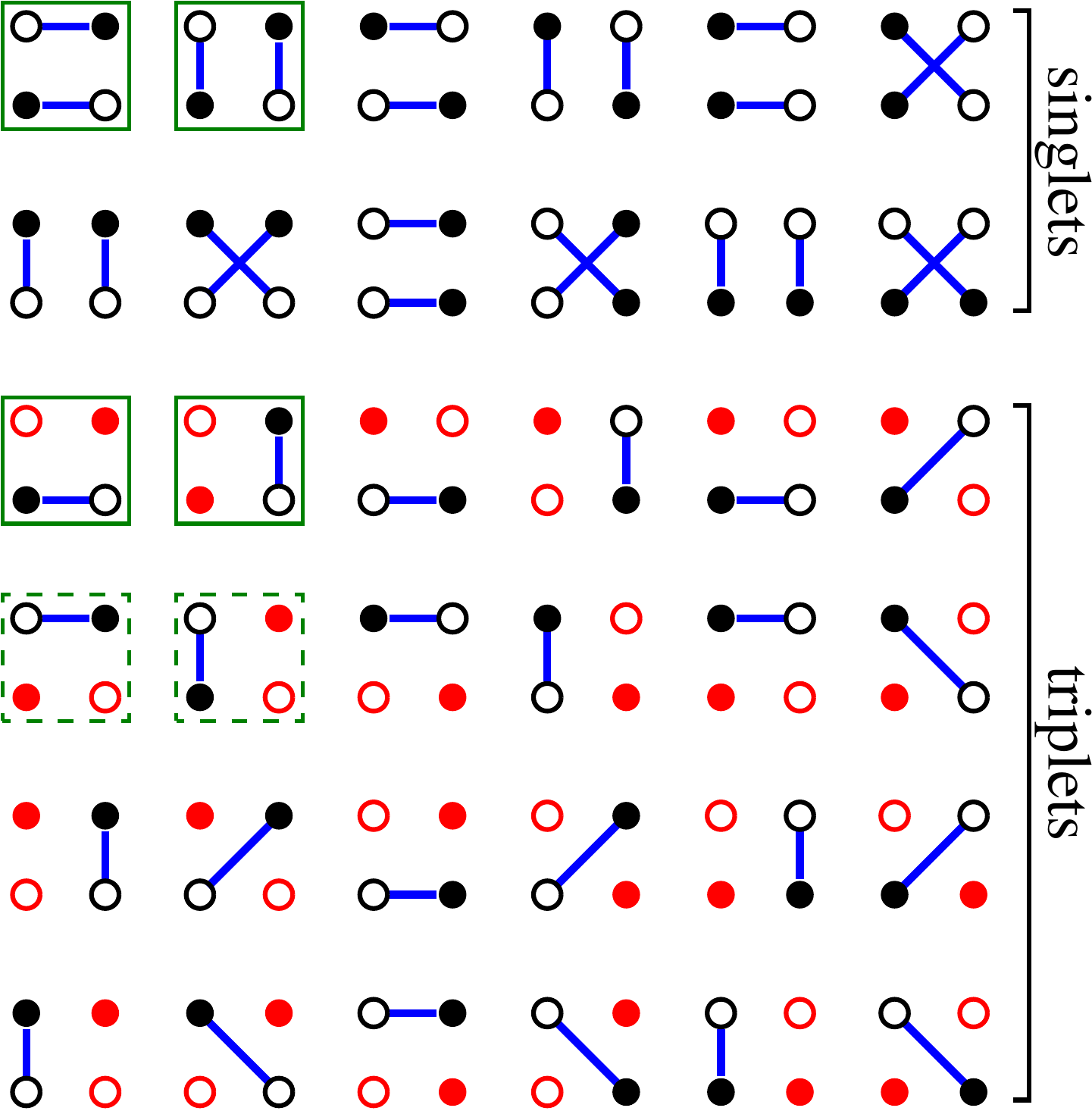}
\caption{Spinon wave function components for singlet and triplets of two spinons and two antispinons, with the circles in order
$~^1_3~^2_4$ corresponding to the bright (spinon) regions in Fig.~\ref{all3}.
Spinons (on sublattice A) and antispinons (on sublattice B) are denoted by solid and open circles, respectively, and singlet pairs
are indicated by lines connecting circles. Unpaired red circles correspond to up spins (or other triplet states of the pairs). The
components enclosed by solid green squares represent the dominant components of the wave functions as discussed in connection with
Fig.~\ref{dst}, and the dashed green boxes indicate further regular components in the triplet sector. All
other configurations correspond to cross terms existing when spinons and antisponons can trade places.}
\label{states4}
\end{figure}

In the simple picture of a spinon as a vortex in the VBS structure as drawn in Fig.~\ref{spinon}, the unpaired spin is tied to the core of the 
VBS vortex. It seems unlikely that the VBS vortices in a disordered system can migrate substantially, since they are formed due to the local random 
environment. The unpaired spins, however, can migrate, and this is made easier owing to the presence of valence bonds longer than the shortest 
bonds in the simple picture in Fig.~\ref{spinon}. The longer valence bonds also connect only spins on different sublattices, and in the case of
a single VBS vortex on an infinite lattice the unpaired spin cannot dissociate completely from the vortex core, as long as there is VBS order and
the probability of very long bonds decays exponentially with the bond length. Thus,  an X-type spinon, ${\rm X} \in \{{\rm A},{\rm B}\}$,
is the composite object of a VBS X-type vortex and an unpaired spin on sublattice X. However, in the case of a disorder-induced vortex-antivortex
pair, there will be some probability of an unpaired spin on the A sublattice to migrate to the antivortex associated with the B sublattice, and
vice versa. Thus, the unpaired spins are not completely tied to the VBS vortices on their own native sublattices. We will here consider the
unpaired spins as the spinons (on sublattice A) and antispinons (on sublattice B), and refer to the VBS vortices and antivortices as separate
objects. A spinon (antispinon) is still predominantly associated with a VBS vortex centered on sublattice A (antivortex centered on
sublattice B). We disregard the presumably low probability of a spinon and antispinon existing simultaneously at the same VBS vortex or
antivortex.

It is important to note that there is no symmetry analogous to the SU(2) symmetry of the spins in the sublattice
labels---instead, one should think of the spinons and antispinons as moving in different random potentials (likely with some repulsive interactions
between spinons and antispinons in the same vortex region, though we have not quantified this). It is remarkable that these potentials, as is apparent 
in Fig.~\ref{dst}, both have minimums at roughly the same locations (the VBS vortices), instead of the spinons being repelled by the VBS antivortices 
and antispinons being repelled by the VBS vortices. This suggests a picture of the excited spinons as itinerant particles that can tunnel through 
channels corresponding to the domain walls between regions attracting both A and B spinons, but with a typically large difference in the depth of 
the A and B potential wells (A and B spinons being more attracted to type A and B VBS vortices, respectively). The excitations (of which we here have only 
studied the lowest one) should be localized, in the sense that a given spinon only migrates substantially between a limited subset of the VBS vortices.

It is instructive to construct a general wave function for four spinons satisfying the above constraint, and to analyze the A and B
string densities in Fig.~\ref{dst} within that formal framework. Figure \ref{states4} shows all possible singlet and triplet components
of wave functions for two spinons and two antispinons. Here the four circles have been arranged to correspond closely to the 
four regions of elevated string density in Fig.~\ref{dst} (which form roughly the corners of a square). The single dominant bright spot
occupied by the B string and the two bright spots (one of which is dominant) containing the $A$ string can be achieved in the triplet channel if
the predominant wave function components are the two enclosed by the solid-line boxes in Fig.~\ref{states4}. The dashed-line boxes enclose
the two other triplet components corresponding to the same A, B arrangements but with the dimmer A-string spot in Fig.~\ref{dst}(b)
having one of the unpaired spins instead of the brighter spot. Based on these triplet components we can deduce that the dominant singlet
components are those two indicated by squares. All other components, in both the singlet and the triplet sectors, are analogues of the cross
term in the two-spinon state, Eq.~(\ref{twospinonpsi}), and they have smaller amplitudes that can, in principle, be roughly estimated
from the integrated densities within the different spots in Fig.~\ref{dst}. Doubly occupied vortices should also have some contributions 
in the wave function, but for simplicity we have neglected those here.

It is not completely clear what the consequences are of these dynamical spinon effects. Most likely, the fluctuations are secondary
effects and the RS fixed point can be realized even with frozen A, B labels. However, this still needs to be tested. It is possible that
the A-B exchange processes actually contribute to the effective spinon-spinon interactions and, thus, further reinforce the scenario of
spinon-antispinon singlet pairing as the mechanism responsible for the RS phase.

\section{Conclusions and Discussion}
\label{sec:discussion}

\subsection{Summary}

Using the $J$-$Q$ model, we have demonstrated that an RS phase can be induced by disorder in a quantum spin system even though all microscopic
interactions are bipartite, lacking the geometric frustration that so far was believed to be a necessary ingredient for this type of 2D state. The
RS phase is characterized by algebraically decaying mean spin and dimer correlations, with distance dependence $\propto r^{-2}$ and $\propto r^{-4}$,
respectively. For the continuous AFM--RS transition, we have sufficient evidence to conjecture universal critical exponents; $\nu=2$, $\eta=0$,
and $z=2$ (with $\nu$ affected by the largest uncertainty). The $r^{-2}$ form of the spin correlation function also applies inside the RS phase, 
and this implies the exponent relation $\eta=2-z$ based on standard definitions of critical correlation functions \cite{fisher89}. The dynamic exponent $z$ 
increases from $2$ as the RS phase is entered, according to our results for a corresponding power-law divergent uniform and local magnetic susceptibilities.
The observed consistency of the scaling forms for $\chi_{\rm u}(T)$ and $\chi_{\rm loc}(T)$ [Eqs.~(\ref{chiutdep}) and (\ref{chiloctdep})], with the 
latter derived under the condition of the exponent relation $\eta=2-z$ obtained at $T=0$, provides strong evidence for a critical RS phase in which 
the standard quantum-critical scaling laws apply.

The key physical mechanism underlying the RS phase, argued in the case of the triangular lattice by Kimchi et al.~\cite{kimchi18} and observed
directly here in our simulations of the square lattice, is the pairwise creation of localized spinons (spinon-antispinon pairs) as the VBS is broken
up into domains in the presence of disorder (which is similar also to the previously observed RS state arising out of the dimerized phase of the $J$-$Q$
chain \cite{shu16}). This correlated spinon distribution leads to a network of weakly interacting spinon pairs and no long-range AFM order, in sharp
contrast to the case of completely randomly distributed un-frustrated magnetic moments. We have also argued that the VBS domain walls connecting spinons
act as channels mediating the effective spinon-spinon interactions within the pairs. Moreover, we have found a dynamical effect whereby the individual
spatial regions containing the spinons cannot be associated purely with spinons or anti-spinons, but there is some mixing originating from migration
of unpaired spins between the VBS vortices and antivortices. A given VBS vortex can still be classified either as predominantly a spinon or an
antispinon according to the vortex type, and the mixing may not be necessary in describing the RS fixed point.

Because of the presence of spinons and their dominant role in the physical properties of the RS state, this state should not be referred to as a
valence-bond glass, which is a term normally reserved for a state with random arrangements of short valence bonds, with critical dimer correlations but no
liberated spinons \cite{tarzia08} (though in the literature other kinds of states have also been referred to using the same term, e.g., in
Refs.~\onlinecite{watanabe14,kawamura14,shimokawa}). The spin correlations in the valence-bond glass were not discussed explicitly in 
Ref.~\onlinecite{tarzia08}, but they should decay exponentially in such a 2D state with only short valence bonds.

It is interesting to compare the critical exponents we have obtained here at the AFM--RS transition
with those at the transition between a superfluid and a Bose-glass in
the Bose-Hubbard model with random potentials \cite{fisher89}. Though the symmetries are different, the superfluid breaking U(1) symmetry and the AFM
state considered here breaking O(3) symmetry, the exponents that we have obtained here appear to be the same or satisfy the same bounds. At the Bose-glass
transition in $D$ dimensions the dynamic exponent $z=D$, the same as $z=2=D$ that we have found here. Also, the correlation length exponent $\nu=2$ at the Bose-glass
transition, which we have not been able to fully confirm in the case of the AFM--RS transition but was conjectured based on the results shown in Fig.~\ref{nu}.
Moreover, in the Bose-glass case the anomalous dimension should satisfy the bound $\eta \ge 0$. The mean spin correlations decaying as $1/r^2$, in the
RS phase and at the AFM--RS phase boundary, corresponds to $\eta=0$ and the bound is satisfied. Apart from the obviously different symmetries, the BG phase
is a Griffiths phase, which we have argued is not the case for the RS phase. The two systems should therefore not be expected to belong to the same
universality class. The fact that the exponents nevertheless appear to be the same is intriguing and deserves further study.

It appears most likely that the RS state identified here is the same one, in the RG sense, as those previously conjectured in frustrated
systems \cite{watanabe14,kawamura14,shimokawa,uematsu17,kimchi17,kimchi18,wu18}, though the lack of definitive quantitative results in the previous works
(e.g., exponents governing various power-law behaviors) makes it difficult to definitely ascertain this at the moment. For example, it was argued that
the low-$T$ susceptibility follows a Curie form in the frustrated honeycomb Heisenberg model in the RS phase \cite{uematsu17}, while we
have here demonstrated a $T^{-a}$ behavior with varying $a <1$ in the random $J$-$Q$ model (and $a \to 0$ as the AFM phase is approached).
Clearly ED studies of lattices with only up to $\approx 20$ sites cannot be used to reliably address the detailed form of the divergence, as we have
seen even with much larger systems here. In the work of Kimchi et al.~\cite{kimchi17} as well, it was not possible to obtain quantitative
values of most of the exponents pertaining to the RS phase in the triangular lattice, though we note a more recent work in which scaling
forms for the heat capacity (which we have not yet investigated) were obtained under various conditions and compared with experiments \cite{kimchi18}.
We note, in particular, that the previous works have not discussed any details of the AFM--RS phase transitions, for which we have obtained specific
results here on power laws both at $T=0$ and $T>0$. In any case, there are no apparent contradictions between our RS state and that of Kimchi et al.,
and, given that the proposed mechanisms underlying the formation of these states is similar, a common RS fixed point appears plausible.
We still discuss our results in the context of other possible scenarios.

\subsection{Fixed points}

In the case of the triangular lattice, it was pointed out that the RS phase may eventually, at the longest length scales, be unstable to the
formation of a spin glass state \cite{kimchi18}. Similarly, the square-lattice random $J$-$Q$ model might possibly be unstable to the formation of weak
AFM order, though we have seen no signs of this up to the largest lattice studied here ($L=64$). The fact that we observe such good scaling up to
these system sizes at the very least implies that an RS fixed point exists (in the models studied here or outside but close to the present model space)
and is responsible for the observed behaviors. The question then is 
whether there is one or two fixed points---one for bipartite interactions and one for frustrated interactions (perhaps above some critical
strength of the frustration). We discuss possible RG scenarios for either case.

i) There is a single RS fixed point. Let us call this the bipartite RS fixed point (BRSFP), even though it may attract also frustrated systems. 
A random system flowing to this fixed point has a true RS ground state. One possibility is that the BRSFP is stable for bipartite interactions
but unstable when frustrated interactions are included---in that case an interesting question is whether the fixed point is unstable in the presence of
arbitrarily weak frustration or only above a critical frustration strength. Adapting the picture of Kimchi et al.~\cite{kimchi18} to this scenario,
the flow away from the BRSFP would eventually lead to a spin glass fixed point if the frustration is sufficiently strong
\cite{deltanote}. Another possibility is that the BRSFP is also unstable in many bipartite systems (i.e., reaching it would require fine-tuning of
parameters), and in that case the flow would be toward an AFM ordered fixed point. Even then, provided that the length scale at which the flow deviates
from the BRSFP is sufficiently large, there will still be experimentally observable consequences of the proximity to the BRSFP fixed point
(e.g., the temperature dependent susceptibility studied here).

ii) There are two fixed points; the BRSFP discussed above as well as a frustrated-RS fixed point (FRSFP). In this case, frustrated interactions,
arbitrarily weak or above a critical strength, would cause a flow toward the FRSFP. These fixed points (both of them or only one of them) could also
require fine-tuning, in principle. As in scenario i), even unstable fixed points would lead to experimentally observable consequences if the
flows lead sufficiently close to the fixed points. The BRSFP and FRSFP must have some differences in their operator contents, and there should
then be some ways to distinguish them in numerical model studies and in experiments. As mentioned above, so far there are no explicit indications
of two fixed points based on existing numerics.

In addition to the several fixed points mentioned above---the BRSFP, FRSFP, AFM, and the spin glass---there could also be various other ``random spin liquid''
fixed points in a wider space of disordered frustrated and bipartite quantum magnets. Experimentally, the question of how to distinguish between a spin
glass and a random spin liquid has attracted considerable attention \cite{li15,ma18} and the issue is as of now unresolved. On general grounds
one would expect gapped (topological) spin liquids to be stable to weak disorder, while gapless (algebraic) spin liquids may generically flow
to RS fixed points (the BRSFP or the FRSFP).

If indeed the BRSFP encompasses the $J$-$Q$ model as well as the multitude of frustrated quantum magnets, the
ability to study the former with large-scale unbiased QMC simulations has significant consequences in the context of experiments. It will
then be possible to relate observed power laws directly to unbiased calculations, e.g., to test relationships between the power laws for
different physical observables. Although the $J$-$Q$ model does not represent the correct microscopic interactions of specific materials, its
phases can still contain the experimentally relevant low-energy physics. This is in the spirit of ``designer Hamiltonians'' \cite{kaul13}, which
are tailored to realize collective quantum states and quantum phase transitions while at the same time being amenable to numerical calculations,
especially sign free QMC simulations, on large scales without approximations. Given some of they key results that we have obtained here, such as the
$r^{-2}$ form of the decay of the mean spin correlation functions and the temperature independent magnetic susceptibility at the AFM--RS transition
(and the divergent behavior with varying exponent inside the RS phase), targeted calculations aiming at these specific universal
characteristics can hopefully soon be carried out also for the frustrated models. One promising calculational route here is tensor network
states tailored specifically to disordered spin models \cite{goldsborough14,lin17}. Though such calculations are certainly challenging, it may 
still be possible to reach larger system sizes than in the previous exact ED and DMRG studies.

A crucial question is whether and how the RS fixed point(s) can be obtained in SDRG calculations. The key physical ingredients underlying the RS 
phase---VBS domains and localized spinons---are unlikely generated correctly in the initial (high energy) stages of the SDRG procedure applied directly 
to microscopic bipartite Heisenberg Hamiltonians; in 1D the method can only partially reproduce VBS domains \cite{shu16}. It is furthermore very 
difficult to apply the SDRG approach to more complicated interactions like the six-spin $Q$ terms used here (which are difficult to deal with even 
in 1D systems \cite{shu16}) since many kinds of effective couplings can be generated. It may be more fruitful to consider SDRG calculations carried 
out on a suitably constructed effective subsystem of the localized spinon-antisponon pairs with their domain-wall mediated interactions. With the 
A-B sublattice correlation effect built into such an effective model, in combination with suitable inter- and intra-pair coupling distributions, 
singlets should gradually 'freeze out' one-by-one in an SDRG procedure. Like in 1D, a rare-event mechanism \cite{dasgupta80,fisher94} would likely be 
responsible for some pairing over larger distances, which is required for obtaining power-law correlations. It would be interesting to carry out 
SDRG calculations on effective models of randomly located spinons with different degrees of pair formation among A and B sublattice spins.

\subsection{Experiments}

A promising system for realizing a square-lattice RS state is the quasi-2D material
Sr$_2$CuTe$_{1-x}$W$_{x}$O$_6$, which was initially synthesized at $x=0,0.5$, and $1$ \cite{mustonen18a}, and more
recently also for several other values of $x \in [0,1]$ \cite{mustonen18b,watanabe18}. The corresponding isostructural compounds Sr$_2$CuTeO$_6$ and
Sr$_2$CuWO$_6$ have dominant nearest- and next-nearest-neighbor spin interactions, respectively, owing to the different orbital properties of the plaquette
centered Te and W ions. With random distribution of these ions, it was argued in Ref.~\cite{mustonen18b} that an RS-type state forms in a sizable
region of $0 < x < 1$, though detailed comparisons with specific RS predictions were still lacking. In Ref.~\cite{watanabe18}, it was instead argued
that the state of the random system is a valence-bond glass with a singlet gap. Within our scenario, the RS state on the square lattice could form
as a consequence of couplings locally favoring VBS domains, and this could possibly be the case when the $J_1$ and $J_2$ couplings are mixed at random,
even though the pure $J_1$ and $J_2$ systems are magnetically ordered (with N\'eel and stripe AFM order for $J_1$ and $J_2$ couplings, respectively).
Note that the uniform frustrated Heisenberg model with variable $J_2/J_1$ has a VBS phase in its phase diagram \cite{gong14,morita15,wang17}.

\begin{figure}[t]
\includegraphics[width=84mm, clip]{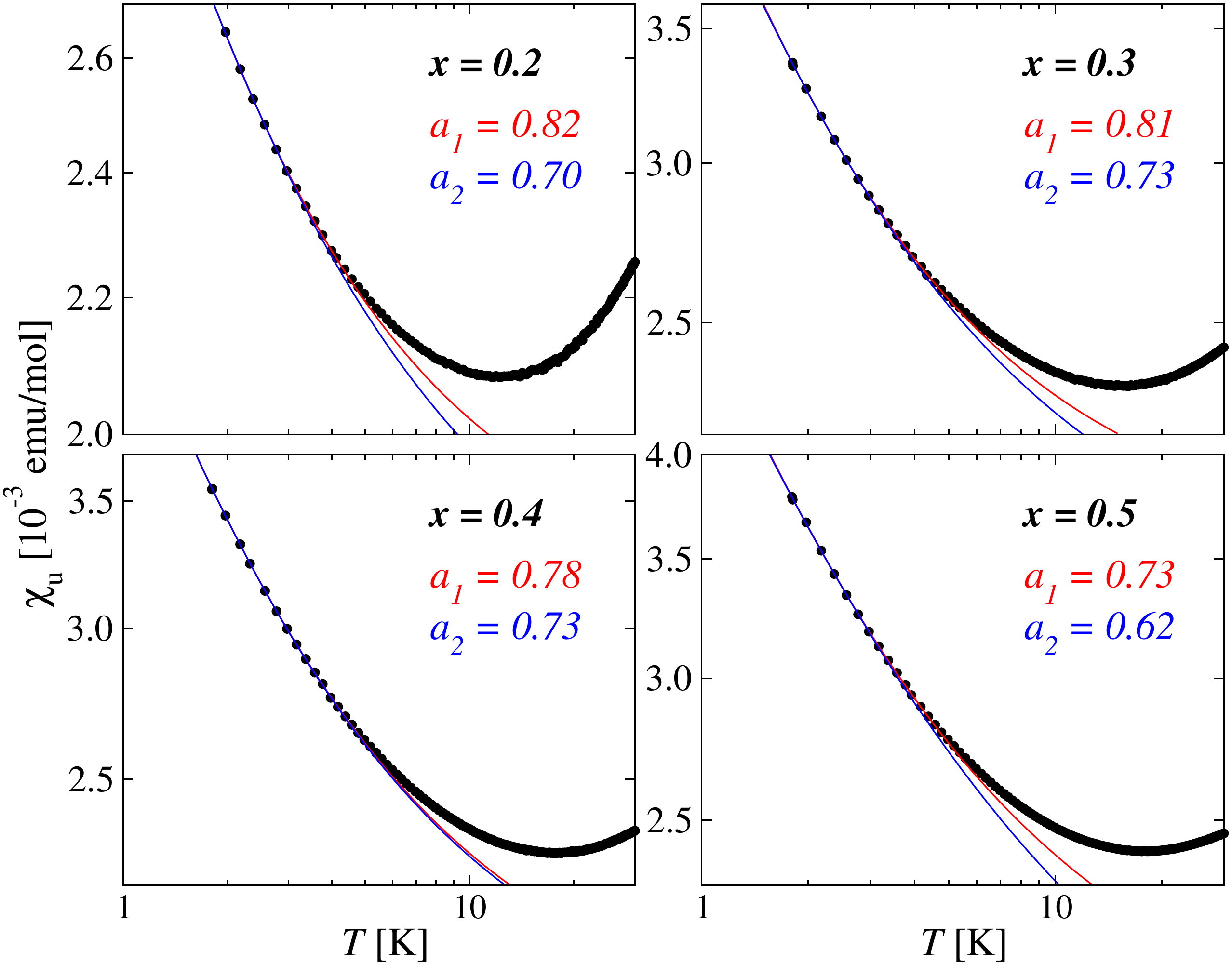}
\vskip-2mm
\caption{Experimental susceptibility data (black circles) for the random quantum magnet Sr$_2$CuTe$_{1-x}$W$_{x}$O$_6$ from Ref.~\cite{watanabe18}.
The panels correspond to W fractions $x=0.2,0.3,0.4,0.5$, as indicated. Logarithmic scales are used for all axes.
The curves are of the form $\chi_u = \chi_0 + cT^{-a}$, with red and blue corresponding, respectively, to $T < 4$K and $T < 3$K data used
in the fits. The temperature exponents ($a_1$ for $T < 4$ K and $a_2$ for $T < 3$K) obtained from the fits are also indicated in the panels.}
\label{expfig}
\end{figure}

An intriguing observation \cite{mustonen18b,watanabe18} is a divergent low-$T$ susceptibility for $x \in [0.2,0.5]$. 
This divergence was interpreted as a Curie tail originating from isolated magnetic moments in the random systems.
In light of the findings we have presented here for the susceptibility in the RS state, we have re-analyzed the susceptibility data of
Ref.~\cite{watanabe18} (Fig.~2) in the regime of Te-W mixing $x$ where RS physics may pertain.
Figure \ref{expfig} shows the low-temperature susceptibility for the W fraction $x$ in the range $0.2-0.5$ fitted
to the form $\chi_u = \chi_0 + cT^{-a}$. We have used two different temperature windows for these fits, $T<4$K (shown as red curves) and $T < 3$K
(blue curves). In all cases, we find that the divergence is slower than the Curie law. For the $T<4$K fits we obtain exponents $a$ in
the range $0.73-0.82$, while the range is $0.62-0.73$ when the lower cutoff is used. An important observation is that the exponent
consistently decreases when the temperature cutoff is reduced. This  makes it seem unlikely that the low-$T$ susceptibility follows the Curie law,
though we note that reasonable fits can also be obtained with $a=1$---these fits work approximately up to considerably higher temperatures than 
those shown in Fig.~\ref{expfig}, but the low-$T$ data are not as closely matched as with the fits in Fig.~\ref{expfig}. 

While the data fits are not conclusive, the findings motivate further experimental studies and analysis based on the concrete RS predictions we have 
reported here. Experiments at still lower temperatures would be desirable in this regard. It would be particularly interesting to test our prediction of 
a temperature independent low-$T$ susceptibility at the AFM--RS transition.

\subsection{Future extensions}

Many interesting QMC calculations are called for as extensions of the initial study of the random $J$-$Q$ model presented here. For example, the
evolution of the RS phase as a function of an external magnetic field (which was recently studied in the triangular lattice \cite{kimchi18}) 
is very interesting theoretically and also from the experimental perspective. The field can be included in SSE
simluations of the $J$-$Q$ model \cite{iaizzi17,iaizzi18}. Dynamical signatures,
e.g., the dynamic spin structure factor, can also be studied using SSE supplemented by analytic continuation techniques \cite{shao17}, and it
will be interesting to compare the 2D RS phase with the random exchange Heisenberg chain, which was also recently
studied with the above mentioned techniques \cite{shu18}. We also note that the specific heat played a major role in the experiments in
Ref.~\cite{watanabe18}, and also theoretically in the context of other materials in Ref.~\cite{kimchi18}. We did not report specific
heat QMC results here, because they require significantly more computational resources than the susceptibility. We plan to calculate
the specific heat in future work.

The diluted $J$-$Q$ model also deserves further studies. Here we have merely confirmed that it does not have an RS ground state but hosts weak AFM order.
However, the system mixes aspects of vacancy induced moments and RS physics, and potentially it could exhibit clear RS behaviors on intermediate length scales,
e.g., it may show an anomalous divergent susceptibility similar to that of a system with RS ground state. This kind of behavior was indeed suggested
recently in the context of random spin liquids in frustrated quantum magnets, where experimentally one may expect the presence of more than one type of
impurity moments, including localized spinons, and these different interacting moments may collectively cause a non-Curie susceptibility \cite{riedl18}.
Given the similarities we have discussed here between the $Q$ terms and frustrated interactions, it would be interesting to investigate the susceptibility
and other experimentally relevant $T>0$ properties of the diluted $J$-$Q$ model.

Other variants of the $J$-$Q$ model can be studied in order to test the universality of the RS--AFM transition and the RS state. Preliminary
results for the $J$-$Q_2$ model [i.e., two instead of three singlet projectors in Eq.~(\ref{jqham}) and Fig.~\ref{terms}] with random $J$ and $Q_2$ look
very similar to the random $J$-$Q_3$ model. 
For example, in the system with bimodal random $Q_2 \in \{0,2Q\}$ we observe behaviors like in Fig.~\ref{qc_x}, 
with the susceptibility at $Q/J \approx 2.5$ converging toward a $T$-independent form with increasing system size, while a slowly divergent $T \to 0$ behavior 
is seen for larger $Q$. These results support a universal $z=2$ at the AFM--RS transition and $z>2$ inside the RS phase.

\begin{acknowledgments}
We would like to thank K. Beach, I. Kimchi, T. Okubo, S. Sachdev, T. Senthil, J. Takahashi, and L. Wang for valuable discussions. We are grateful to
the authors of Ref.~\onlinecite{watanabe18} for sending us the experimental susceptibility data displayed in Fig.~\ref{expfig}. W.G. was supported by
NSFC under Grants No.~11734002 and No.~11775021. H.S. was supported by the China Postdoctoral Science Foundation under Grants No.~2016M600034 and
No.~2017T100031. Y.-C.L. was supported by the MOST (Taiwan) and the National Center for Theoretical Sciences (NCTS). A.W.S was supported by the NSF 
under Grant No.~DMR-1710170 and by a Simons Investigator Award, and he also gratefully acknowledges travel support from Beijing Normal University under 
YinZhi project No.~C2018046, as well as from the NCTS.  Some of the numerical calculations were carried out on Boston University's Shared Computing Cluster.
\end{acknowledgments}



\begin{thebibliography}{999}
\bibitem{bondnote}
In a frustrated system, non-bipartite bonds, if present, do not vanish upon time evolution, but it should be noted that the bipartite valence bond
basis is always (over)complete; thus, on the square-lattice one can also use a basis without frustrated bonds even when frustrated interactions 
are present. However, it is only when the interactions are bipartite that the ground state can be expressed as a positive-definite superpositions 
of valence bond configurations.
\bibitem{udnote}
  A sign-free QMC simulation corresponds to a related classical statistical-mechanics problem, the properties of which
  reflect those of the quantum system of interest. Any phase transition taking place in the classical system must
  also correspond to a phase transition in the quantum system, and vice versa. Thus, for the purpose of locating a phase transition
  in the quantum system, one does not necessarily have to study a property with a strict relationship to a true quantum mechanical
  expectation value, but one can choose some quantity that is convenient to evaluate in the simulation. As an example, in valence-bond
  projector QMC simulations $\langle D^n\rangle$ for $n>2$ is difficult to evaluate as an exact quantum expectation value, but by examining the
  relevant expressions \cite{beach06}, one can see that the finite-size scaling, at least to leading order, should be the same for
  $\langle (\tilde D)^n\rangle$, where $\tilde D$ is the easily computed real number corresponding to the operator $D$ in
  a QMC transition graph [and $(\tilde D)^n$ differs only slightly from the number $\tilde D^n$ giving the exact $\langle D^n\rangle$].
  This is what we use in defining the Binder cumulant of the dimer order parameter.
\bibitem{stringnote}
Another possible definition of the string density is to use $\rho=1$ for the covered sites, or one can just consider the locations of the unpaired 
spins (which carry the net spin) as defining the spinon density \cite{tang11}.
\bibitem{deltanote}
In principle, the stability of the BRSFP to weak frustration can be studied in the RS phase by computing the scaling dimension of the operator used to
define the frustrated terms on the square lattice (i.e., next-nearest-neighbor exchange interactions), in practice using
the decay of the corresponding correlation function. Even without doing such a calculation, it seems most likely that the state we have studied here should
not be unstable to weak frustration, because the scaling dimension of an added interaction such as next-nearest-neighbor exchange is not dependent on the
sign of the interactions. Since next-nearest-neighbor interactions with negative (ferromagnetic) sign are not frustrated, we do not see any reason for
instability in this case, even when the sign is positive (frustrated interactions). QMC simulations with frustrated interactions explicitly included are 
essentially impossible, due to the exponential increase in the severity of the sign problem (the small mean value of the $\pm 1$ sign factor of the
sampled configurations) with increasing $L$ or $\beta$, even for weak frustration.
\end{thebibliography}
\end{document}